\documentclass[equations,11pt]{article}
\usepackage{amsfonts}
\usepackage{a4,tikz}
\usepackage{geometry}

\makeatletter
\newcommand\ackname{Acknowledgements}
\if@titlepage
  \newenvironment{acknowledgements}{%
      \titlepage
      \null\vfil
      \@beginparpenalty\@lowpenalty
      \begin{center}%
        \bfseries \ackname
        \@endparpenalty\@M
      \end{center}}%
     {\par\vfil\null\endtitlepage}
\else
  \newenvironment{acknowledgements}{%
      \if@twocolumn
        \section*{\abstractname}%
      \else
        \small
        \begin{center}%
          {\bfseries \ackname\vspace{-.5em}\vspace{\z@}}%
        \end{center}%
        \quotation
      \fi}
      {\if@twocolumn\else\endquotation\fi}
\fi
\makeatother

\usepackage{amsmath}
\usepackage{amssymb}
\usepackage{amsthm}
\usepackage{eucal}
 \usepackage[usenames,dvipsnames]{pstricks}
 \usepackage{epsfig}
 \usepackage{pst-grad} 
 \usepackage{pst-plot} 
\renewcommand{\theequation}{\arabic{equation}}
\usepackage{amssymb,amsmath}
\usepackage{color}
\usepackage{accents}
\usepackage{texdraw}
\usepackage{eucal}
\usepackage{epic,epsfig}
\usepackage{graphicx}

\theoremstyle{definition}

\numberwithin{equation}{section}
\DeclareMathAccent{\wtilde}{\mathord}{largesymbols}{"65}
\DeclareMathAccent{\what}{\mathord}{largesymbols}{"62}

\def\m@th{\mathsurround=0pt}
\mathchardef\bracell="0365
\def\upbrall{$\m@th\bracell$}
\def\undertilde#1{\mathop{\vtop{\ialign{##\crcr
    $\hfil\displaystyle{#1}\hfil$\crcr
     \noalign
     {\kern1.5pt\nointerlineskip}
     \upbrall\crcr\noalign{\kern1pt
   }}}}\limits}
\def\m@th{\mathsurround=0pt}
\mathchardef\bracell="0365
\def\upbrall{$\m@th\bracell$}
\def\underhat#1{\mathop{\vtop{\ialign{##\crcr
    $\hfil\displaystyle{#1}\hfil$\crcr
     \noalign
     {\kern1.5pt\nointerlineskip}
     \upbrall\crcr\noalign{\kern1pt
   }}}}\limits}
\usepackage{pgf}
\usepackage{tikz}
\usepackage{verbatim}
\usepackage[utf8]{inputenc}
\usetikzlibrary{arrows,automata}
\usetikzlibrary{positioning}

\def\theequation{\arabic{section}.\arabic{equation}}

\newcommand{\wh}{\widehat}
\newcommand{\wt}{\widetilde}

\def\hypotilde#1#2{\vrule depth #1 pt width 0pt{\smash{{\mathop{#2}
\limits_{\displaystyle\widetilde{}}}}}}
\def\hypohat#1#2{\vrule depth #1 pt width 0pt{\smash{{\mathop{#2}
\limits_{\displaystyle\widehat{}}}}}}

\newcommand{\bblu}{\begin{color}{blue}}
\newcommand{\bred}{\begin{color}{red}}
\newcommand{\ecl}{\end{color}}

\newcommand{\bE}{\boldsymbol{E}}

\newcommand{\bI}{\boldsymbol{I}}

\newcommand{\bK}{\boldsymbol{K}}
\newcommand{\bL}{\boldsymbol{L}}
\newcommand{\bM}{\boldsymbol{M}}
\newcommand{\bN}{\boldsymbol{N}}

\newcommand{\bU}{\boldsymbol{U}}

\newcommand{\bsX}{\boldsymbol{X}}
\newcommand{\bsY}{\boldsymbol{Y}}

\newcommand{\bLam}{\boldsymbol{\Lambda}}

\newcommand{\pl}{\partial}

\newcommand{\be}{\begin{equation}}
\newcommand{\ee}{\end{equation}}
\newcommand{\bea}{\begin{eqnarray}}
\newcommand{\eea}{\end{eqnarray}}
\newcommand{\bse}{\begin{subequations}}
\newcommand{\ese}{\end{subequations}}
\newcommand{\nn}{\nonumber}

\begin{document}

\def\theequation{\arabic{section}.\arabic{equation}}

\newtheorem{thm}{Theorem}[section]
\newtheorem{lem}{Lemma}[section]
\newtheorem{defn}{Definition}[section]
\newtheorem{ex}{Example}[section]
\newtheorem{rem}{Remark}
\newtheorem{criteria}{Criteria}[section]
\newcommand{\ra}{\rangle}
\newcommand{\la}{\langle}

\title{\textbf{Discrete-time Calogero-Moser system and Lagrangian 1-form structure}}
\author{\\\\Sikarin Yoo-Kong$^1$, Sarah Lobb$^2$, Frank Nijhoff$^3$ \\
\small \emph{School of Mathematics, Department of Applied Mathematics, University of Leeds,}\\
\small \emph{United Kingdom, LS2 9JT}\\
\small  $^1$syookong@gmail.com,
\small $^2$sarahlobb@gmail.com,
\small $^3$nijhoff@maths.leeds.ac.uk}
\maketitle


\abstract
We study the Lagrange formalism of the (rational) Calogero-Moser (CM) system, both in discrete time as well as in continuous time, as a first 
example of a Lagrange 1-form structure in the sense of the recent paper \cite{SF1}. The discrete-time model of the CM system 
was established some time ago arising as a pole-reduction of a semi-discrete version of the KP equation, and was shown to lead to an exactly integrable 
correspondence (multivalued map). In this paper we present the full KP solution based on the commutativity of the discrete-time flows in 
the two discrete KP variables. The compatibility of the corresponding Lax matrices is shown to lead directly to the relevant closure relation 
on the level of the Lagrangians. Performing successive continuum limits on both the level of the KP equation as well as of the CM system, we 
establish the proper Lagrange 1-form structure for the continuum case of the CM model. We use the example of the three-particle case 
to elucidate the implementation of the novel least-action principle, which was presented in \cite{SF1}, for the simpler case of Lagrange 1-forms.


\section{Introduction}
\setcounter{equation}{0}
The Calogero-Moser (CM) model \cite{C1,C3} is an integrable one-dimensional many-particle system with long-range interactions, originally given as 
a continuous system with pairwise inverse square potential, but which also has been generalized to the trigonometric \cite{C2} and 
the elliptic case \cite{C4}, cf. also \cite{R0}, and later to a relativistic model (the Ruijsenaars-Schneider model \cite{R1,R2}). The model has 
been extensively studied in both the classical and quantum cases \cite{OP}, and is integrable on both levels. 
From a physical perspective the emphasis is on the dynamics of real particles subject to pairwise repulsive potentials. However, in recent years the study 
of CM models has attained a wider mathematical significance, from the seminal paper \cite{KKS} considering the symplectic geometry 
associated with the model, to more recently its role in the representation theory of Lie and quantum algebras \cite{Eting}. In this context, 
the investigation of CM systems with attractive potential, allowing particles to collide, has led to interesting new perspectives from the point of view of
algebraic geometry, cf. e.g. \cite{Wilson}. 

An integrable discrete-time version of the (rational) CM model was presented in \cite{F2}, where it was obtained from 
a semi-discrete Kadomtsev-Petviashvili (KP) type equation with two discrete and one continuous independent variable. 
The construction, which we summarize in Section 2, follows closely Krichever's pole-reduction of the continuous KP equation leading 
to a connection with the CM system \cite{Kri}. In the discrete case, this leads to a rather complicated system of ordinary difference equations 
(O$\Delta$Es) which constitutes an integrable Lagrangian correspondence, i.e., a  
multivalued symplectic map \cite{V1}. The construction provides a Lax pair, whilst the classical $R$ matrix is esentially the 
same as for the continuous case (as given by \cite{AT}), from which the Liouville integrability (in the sense of \cite{V1}) follows. The structure of 
the Lax matrices was used to obtain the solution of the 
initial-value problem in an adaptation of the standard way, cf. \cite{OP,KKS}. It was shown that a ``naive'' continuum limit of the 
discrete equations reduces them to the equations of motion of the continuous CM system in the case of an attractive, rather than repulsive, potential. In spite of the 
fact that this makes the discrete-time CM model perhaps less relevant from a physical perspective, it is certainly of interest from  
the mathematical point of view in that it provides us an insight into the basic structures underlying discrete integrability. Thus, in the same 
vein, we will use the model here to understand better the novel insights from our recent paper \cite{SF1} of the Lagrangian multi-form 
structure underlying systems which are integrable in the sense of \textit{multidimensional consistency}. 

Our focus in the present paper on Lagrangian structures has many motivations, most notably the possibility of quantizing the discrete-time 
model via a path integral formalism. Furthermore, whilst many integrable discrete systems admit a Lagrangian description, in the discrete-time case the 
Hamiltonian no longer provides such a natural framework as it does in the continuous-time case. 
Recently, two of the authors observed a new fundamental property of Lagrangians for integrable (in the sense of multidimensionally consistent) systems 
which reflects the multidimensional consistency of the lattice systems \cite{SF1}. This is given by a \emph{closure relation} which holds for Lagrangians 
when the system is embedded in a higher-dimensional lattice (meaning that the same equations hold in a multitude of sublattices of the multidimensional 
lattice together with their corresponding Lagrangians), and leads to the interpretation of the Lagrangians as closed forms on the extended lattice. 
This observation, based on the elaboration of various explicit examples, prompted the proposal of a new variational principle for integrable systems 
which involves explicitly the geometries in the space of independent variables. In \cite{SF1} we focused on establishing and interpreting  
the closure relation of Lagrangian 2-forms for cases of integrable discrete equations with two independent variables, as well as of their continuous 
analogues, for lattice equations in the ABS list \cite{ABS}. Subsequently, the closure property was also established for Lagrangians describing 
multicomponent lattice systems in the so-called lattice Gel'fand-Dikii hierarchy, \cite{SF2}, and for Lagrangian 3-forms in the case of the discrete 
bilinear KP equation in \cite{SFQ}, whilst in \cite{BS} it was shown that all equations in the ABS list possessed this property (cf. also \cite{XNL} 
for a ``universal'' Lagrange multi-form description for quadrilateral affine linear equations and their continuous counterparts).   
An important case which has not yet been covered is that of Lagrangian 1-forms, which applies to the case of integrable ODEs. Natural candidates for exhibiting 
such a structure are integrable finite-dimensional many-body systems, such as the Calogero-Moser systems. 

Thus, in this paper we set out to establish the Lagrange 1-form structure of the discrete-time CM system in the rational 
case. Whilst some results demonstrating that the considerations extend to the hyperbolic/trigonometric and elliptic case as well are given in 
Appendix \ref{ET}, we prefer to work with this simplest case, which is most transparent, and for which we have exact solutions at our disposal. 
Starting with the discrete-time case, we will establish the Lagrange structure also for the continuous case by continuum limits, rather than using the 
(known) Hamiltonians for the higher CM flows. This is necessary because, as it turns out, the proper Lagrangians cannot be obtained by performing 
Legendre transformations separately on each of the Hamiltonians. This leads, beyond the second-order flow, to Lagrangians containing rather complicated  
algebraic expressions of the higher-time-derivatives. Instead, we obtain a hierarchy of mixed polynomial Lagrangians in terms of these higher-order 
derivatives from the discrete-time case by systematic expansions (indicating \textit{en passant} that Legendre transformations should probably not be  
implemented order-by-order in the flows but rather altogether in one stroke). In all these considerations, the link with the underlying KP system is instrumental 
in deciding how to perform the higher-order continuum limits by systematic expansions and in terms of the proper parameters. 

The organization of the present paper is as follows. In Section 2, we present the full solution of the semi-discrete KP equation,
which requires commuting flows in different discrete-time directions, and thus generalizes the one-dimensional discrete-time CM 
flow in \cite{F2}. In Section 3, we establish the closure relation of the discrete-time Lagrangian 1-form of the discrete-time CM 
model through the compatibility of the time-part Lax matrices, which are of Cauchy type. In Section 4, we investigate the semi-continuum 
limit, or skew limit, of the semi-discrete KP equation. This limit leads to an equation which we refer to as the semi-continuous KP equation 
(defined in terms of two continuous and one discrete independent variable), from which we obtain by pole-expansion the corresponding 
differential-difference system which we coin the \textit{semi-continuous CM system}. In Section 5, the latter, which acts as a generating system for the 
CM hierarchy, is a Lagrangian system in its own right obeying a closure relation of differential-difference type with the original 
discrete-time Lagrangian of the CM system. In Section 6, the full continuum limit is performed, and we thus recover the usual fully continuous 
CM hierarchy in mixed form, together with continuous Lagrangians exhibiting a closed 1-form structure in Section 7. We concentrate in Section 8 on the 3-particle case, 
governed by the CM flow and its first higher-order counterpart, in order to implement the geometric 
variational principle which was formulated in \cite{SF1}. Thus, we show how the principle leads in this case to the derivation both of
the closure relation as well as the pertinent Euler-Lagrange (EL) equations. We end the paper with some discussion of open problems and possible 
extensions.

\section{Pole reduction of the semi-discrete KP equation}
\setcounter{equation}{0}
In this section we review the connection between the semi-discrete KP equation, following on from \cite{F0,F1}, and the discrete-time CM model as in 
\cite{F2,FP}. This gives us an occasion to introduce appropriate notation which we will use throughout the paper. Furthermore, it helps us to identify the 
commuting flows needed to establish the corresponding nontrivial solution of the semi-discrete KP equation. 
\\
\\
The semi-discrete KP equation that was used in \cite{F2} is
\begin{equation}\label{eq:2.1}
\partial _\xi  (\widehat u - \widetilde u) = (p - q + \widehat u - \widetilde u)(u + \widehat{\widetilde{u}} - 
\widehat u - \widetilde u)\;,
\end{equation}
where $p$ and $q$ are two lattice parameters, $u$ is the classical field, and $\widetilde u$ represents the discrete shift of $u$ 
corresponding to a transition in the ``time'' direction, while $\widehat u$ represents the discrete shift of $u$ 
corresponding to a translation in the ``spatial'' direction. There are two discrete variables $n,m$ to which
 the lattice parameters $p,q$ and shifts $\widetilde u ,\widehat u$ correspond respectively, and one continuous variable $\xi$.
\\

We observe that eq. (\ref{eq:2.1}) is a consequence of the compatibility condition of a scalar Lax pair which reads
\begin{subequations}
\begin{eqnarray}
\widetilde\phi  &=& \phi _\xi   + (p + u - \widetilde u)\phi\; ,\label{eq:2.2}\\
\widehat\phi  &=& \phi _\xi   + (q + u - \widehat u)\phi\; .\label{eq:2.3}
\end{eqnarray}
\end{subequations}
Taking $u$ to have the form
\begin{equation}\label{eq:2.1a}
u = \sum\limits_{i = 1}^N {\frac{1}{{\xi  - x_i (n,m)}}}\; , 
\end{equation}
we find that the corresponding solution of the linear Lax equations is
\begin{equation}\label{eq:2.4}
\phi  = \left( {1 - \frac{1}{k}\sum\limits_{i = 1}^N {\frac{{b_i (n,m)}}{{\xi  - x_i (n,m)}}} } \right)\left( {p + k} \right)^n
 \left( {q + k} \right)^m e^{k\xi }\;,
\end{equation}
where $k$ is a new spectral parameter and the $b_i$ are yet to be determined. Inserting expressions \eqref{eq:2.1a} and \eqref{eq:2.4} for $u$ and $\phi$ into the first Lax equation \eqref{eq:2.2}, we obtain the relations
\begin{eqnarray}
(p + k)b_i  &=& k+\left( {\sum\limits_{l = 1}^N {\frac{1}{{x_i  - \widetilde{x }_l}} - \sum\limits_{\mathop {l = 1}\limits_{l \ne i} }^N 
{\frac{1}{{x_i  - x_l }}} } } \right)b_i  - \sum\limits_{\mathop {j = 1}\limits_{j \ne i} }^N {\frac{{b_j }}{{x_i  - x_j }}}\; ,\label{eq:2.5}\\
(p + k)\widetilde {b}_i  &=& k - \sum\limits_{j = 1}^N {\frac{{b_j }}{{\widetilde x_i  - x_j }}} ,\begin{array}{*{20}c}
   {} & {i = 1,2,...,N} \\
\end{array}.\label{eq:2.6}
\end{eqnarray}
Introducing the vectors $\boldsymbol{ b}=(b_1,b_2,...,b_N)^T$,  and $\boldsymbol{e}=(1,1,...,1)^T$, and the Lax matrices
\begin{eqnarray}
\boldsymbol{L} &=& \sum\limits_{i,j = 1}^N {\frac{E_{ii}}{x_i-\widetilde x_j}}  - \sum\limits_{\mathop {i,j = 1}
\limits_{j \ne i} }^N \frac{E_{ii}+E_{ij}}{x_i-x_j}\;, \label{eq:2.7}\\
\boldsymbol{M} &=&  - \sum\limits_{i,j = 1}^N {\frac{{E_{ij} }}{{\widetilde x_i  - x_j }}}\; ,\label{eq:2.8}
\end{eqnarray}
where $E_{ij}$ are the generators of $GL_N$, i.e., matrices with entries $(E_{ij})_{kl}=\delta_{ik}\delta_{jl}$, we can now rewrite equations 
\eqref{eq:2.7} and \eqref{eq:2.8} as
\begin{subequations}\label{eq:Lax} 
\begin{eqnarray}\label{eq:2.9}
(p + k)\boldsymbol{b} &=& k\boldsymbol{e} + \boldsymbol{L}\boldsymbol{b}\;,\label{eq:2.9a}\\
(p + k)\widetilde {\boldsymbol{b}} &=& k\boldsymbol{e} + \boldsymbol{M}\boldsymbol{b}\;,\label{eq:2.9b}
\end{eqnarray}
\end{subequations}
which forms a $N\times N$ matricial Lax pair. The compatibility of equations \eqref{eq:2.9a} and \eqref{eq:2.9b} gives us the relation 
\begin{eqnarray*}
&& \left(\wt{\boldsymbol{L}}\,\boldsymbol{M}-\boldsymbol{M}\,\boldsymbol{L}\right)\boldsymbol{b}
 +k\left(\wt{\boldsymbol{L}}-\boldsymbol{M}\right)\boldsymbol{e}=  \\ 
 && = \sum_{i,j=1}^N \left[\sum\limits_{l = 1}^N \left( \frac{1}{\wt{x}_i  - \widetilde{\wt{x}}_l } +\frac{1}{\wt{x}_i - x_l}  \right)
 - 2\sum\limits_{\mathop {l = 1}\limits_{l \ne i} }^N {\frac{1}{{\wt{x}_i  - \wt{x}_l }}} \right]
 \left(\frac{k}{N} E_{ij}\boldsymbol{e}-\frac{E_{ij}}{\wt{x}_i-x_j}\,\boldsymbol{b}\right) = 0  
 \end{eqnarray*}
and since this must holds for arbitrary value of $k$ we deduce the discrete-time equations of motion of an $N$-particle system:
\begin{equation}\label{eq:2.11}
\sum\limits_{j = 1}^N \left( \frac{1}{x_i  - \widetilde x_j } +\frac{1}{x_i - {\hypotilde 0 x}_j}  \right)
 - 2\sum\limits_{\mathop {j = 1}\limits_{j \ne i} }^N {\frac{1}{{x_i  - x_j }}=0\  ,} \begin{array}{*{20}c}
   \mbox{~where~} & {i = 1,...,N}\; , \\
\end{array}
\end{equation}
in which the under-shift ${\hypotilde 0 x}_i$ is short-hand for a shift in the variable $n$ over one unit in the 
nagative direction, i.e., ${\hypotilde 0 x}_i(n,m)=x_i(n-1,m)$. 
Furthermore, if \eqref{eq:2.11} holds, both the following matrix relations are satisfied: 
\begin{equation}\label{eq:2.10}
\widetilde{\boldsymbol{L}}\boldsymbol{M} = \boldsymbol{M}\boldsymbol{L}\;,\begin{array}{*{20}c}
   \mbox{~and~} & {(\widetilde{\boldsymbol{L}} - \boldsymbol{M})\boldsymbol{e} = 0}\;,  \\
\end{array}
\end{equation}
and consequently the inhomogeneous Lax system \eqref{eq:Lax} leads to an isospectral problem in terms of the Lax matrices 
$\bL$ and $\bM$. 
As was shown in \cite{F2} eq. \eqref{eq:2.11} is an integrable symplectic correspondence, in the sense of Liouville, \cite{V1}, 
whose exact solutions $x_i(n,m)$, which denote the position of the particles $x_i$ at the $n$th time step (with fixed $m$), 
are obtained by a linearization procedure implementing the Lax pair (for details, see Appendix \ref{Solution}). The resulting solution of the 
initial value problem, imposing initial values $x_i(0,m)$ and $x_i(1,m)=\widetilde x_i(0,m)$, can be obtained by solving the 
secular problem given by the characteristic equation of the matrix 
\begin{equation}\label{eq:2.12}
\boldsymbol{Y}(n,m) = \boldsymbol{\Lambda}_{L}^{-n}\boldsymbol{Y}(0,m)\boldsymbol{\Lambda}_{L}^{n} -n\boldsymbol{\Lambda}_{L}^{-1}\;.
\end{equation}
subject to the constraint on the initial value matrix
\[ [\bsY(0,m)\,,\,\bLam_L]=\bI+ {\rm rank\,1}\  , \] 
(in which $\bI$ denotes the $N\times N$ unit matrix). 
The eigenvalues of $\bsY(n,m)$ given by \eqref{eq:2.12} are the particle positions $x_i(n,m)$, and they are determined up to permutations of the 
particles, which accounts for the multivaluedness of the system of O$\Delta$Es. Here the $\boldsymbol{\Lambda}_L$ 
denotes the diagonal matrix of eigenvalues of the Lax matrix $\boldsymbol{L}$ and the matrix $\boldsymbol{Y}(n,m)$ is related to the 
the diagonal matrix $\bsX(n,m)=\sum_{i = 1}^N x_i(n,m) E_{ii}$ of the particle positions by a similarity transformation. 
It follows from \eqref{eq:2.10} that the matrix $\boldsymbol{\Lambda}_L$ is invariant under time shifts, i.e., 
$\widetilde {\boldsymbol{\Lambda}}_L=\boldsymbol{\Lambda}_L$.
\\

We note that \eqref{eq:2.12} resolves only the $n$-dependence of the solution of the KP equation, and to provide the full solution we 
need also to consider the dependence on $m$. Obviously, the latter should come from the consideration of \eqref{eq:2.3}, which can be treated in 
much the same way, and lead to the following relations:
\begin{subequations}
\begin{eqnarray}\label{eq:2.13}
(q + k)\boldsymbol{b} &=& k\boldsymbol{e} + \boldsymbol{K}\boldsymbol{b}\;,\label{eq:2.13a}\\
(q + k)\wh{ \boldsymbol{b}} &=& k\boldsymbol{e} + \boldsymbol{N}\boldsymbol{b}\;,\label{eq:2.13b}
\end{eqnarray}
\end{subequations}
where $\boldsymbol{K}$ and $\boldsymbol{N}$ take exactly the same form as $\boldsymbol{L}$ and $\boldsymbol{M}$, with 
just the ``$\;\,\widetilde{}\;\;$'' shift replaced by the ``$\;\,\widehat{}\;\;$'' shift. The compatibility condition leads to 
\begin{equation}\label{eq:2.14}
\widehat{\boldsymbol{K}}\boldsymbol{N} = \boldsymbol{N}\boldsymbol{K}\;,\begin{array}{*{20}c}
   \mbox{~and~} & {(\widehat{\boldsymbol{K}} -  \boldsymbol{N})\boldsymbol{e} = 0}  \\
\end{array},
\end{equation}
and the equations of the motion can be directly obtained from \eqref{eq:2.11} by replacing the ``$\;\,\widetilde{}\;\;$'' shift by 
the ``$\,\;\widehat{}\;\;$'' shift, 
\begin{equation}\label{eq:2.11a}
\sum\limits_{j = 1}^N {\left( {\frac{1}{{x_i  - \widehat x_j }} + \frac{1}{{x_i  - {\hypohat 0 {x}}_j }}} \right)} 
 - 2\sum\limits_{\mathop {j = 1}\limits_{j \ne i} }^N {\frac{1}{{x_i  - x_j }}=0\ ,} \begin{array}{*{20}c}
   \mbox{~where~} & {i = 1,...,N} , \\
\end{array}
\end{equation}
in which the under-shift ${\hypohat 0 x}_i$ is short-hand for a shift in the variable $m$ over one unit in the 
negative direction, i.e., ${\hypohat 0 x}_i(n,m)=x_i(n,m-1)$
The exact solution of the system of O$\Delta$Es \eqref{eq:2.11a} can be obtained in the same way as \eqref{eq:2.12}, namely by considering 
\begin{equation}\label{eq:2.15}
\boldsymbol{Y}(n,m) = \boldsymbol{\Lambda}_{K}^{-m}\boldsymbol{Y}(n,0)\boldsymbol{\Lambda}_{K}^{m} -m\boldsymbol{\Lambda}_{K}^{-1}\;.
\end{equation}
subject to the constraint on the initial value matrix
\[ [\bsY(n,0)\,,\,\bLam_K]=\bI+ {\rm rank\,1}\  , \] 
where the matrix $\boldsymbol{\Lambda}_K$ is invariant under time shift, so that $\widehat{\boldsymbol{\Lambda}}_K=\boldsymbol{\Lambda}_K$.
\\
\\
The $n$- and $m$-parts of the solutions of course can only be combined if the corresponding flows are compatible. In particular this requires that 
the Lax matrices $\bL$ and $\bK$ commute, which in turn suggests that these matrices can be simultaneously diagonalized. Furthermore, it requires 
the compatibility of the flows in the two lattice directions, which leads to the 
relations 
\begin{eqnarray}
(p-q)\boldsymbol{b}&=& (\bL-\bK)\boldsymbol{b}\  , \label{eq:KL}\\ 
(p-q)k\boldsymbol{e}&=& (\wh{\bM}-\wt{\bN})k\boldsymbol{e}+ (\wh{\bM}\,\bN-\wt{\bN}\,\bM)\boldsymbol{b}\  .   \label{eq:MN} 
\end{eqnarray} 
Eq. \eqref{eq:KL} leads to the relation
\begin{equation}\label{eq:con}
 p-q=\sum_{l=1}^{N}\left(\frac{1}{x_i-\widetilde{x}_l}-\frac{1}{x_i-\widehat{x}_l}\right)\;,\quad i=1,\dots,N\;, 
\end{equation}
whereas \eqref{eq:MN} yields subsequently 
$$ 
\sum_{i,j=1}^N \left[ p-q + \sum_l \left( \frac{1}{\wh{\wt{x}}_i-\wh{x}_l}- \frac{1}{\wh{\wt{x}}_i-\wt{x}_l}\right)\right] 
\left( \frac{k}{N}E_{ij}\boldsymbol{e}-\frac{E_{ij}}{\wh{\wt{x}}_i-x_j}\,\boldsymbol{b}\right)=0 \  ,  
$$ 
which yields in addition
\begin{equation}\label{eq:con2}
 p-q=\sum_{j=1}^{N}\left(\frac{1}{\wh{\wt{x}}_i-\widetilde{x}_j}-\frac{1}{\wh{\wt{x}}_i-\widehat{x}_j}\right)\;,\quad i=1,\dots,N\; . 
\end{equation}
Furthermore, the relation 
\begin{equation}\label{eq:3.9a}
\sum_{l=1}^{N}\left( \frac{1}{\widehat{\widetilde{x}}_i-\widehat x_l}-\frac{1}{\widehat{\widetilde{ x}}_i-\widetilde x_l}\right)
=\sum_{l=1}^{N}\left( \frac{1}{\widetilde{x}_j-x_l}-\frac{1}{\widehat x_j- x_l}\right)\  , \quad i,j=1,\dots,N\ , 
\end{equation}
which is a consequence of \eqref{eq:con} and \eqref{eq:con2}, guarantees that the zero-curvature condition 
~$\widehat{\boldsymbol{M}}\boldsymbol{N}=\widetilde{\boldsymbol{N}}\boldsymbol{M}$~ holds, which in turn implies that 
~$(\wh{\bM}-\wt{\bN})\boldsymbol{e}=(p-q)\boldsymbol{e}$~. 

Equations \eqref{eq:con} and \eqref{eq:con2}, which we will refer to as the constraint equations, guarantee that the 
discrete flows in the variables $n$ and $m$ commute, and hence that the corresponding linear equations can be simultaneously solved. 
In fact, the downward shift of \eqref{eq:con2} yields 
\[   
 p-q=\sum_{l=1}^{N}\left(\frac{1}{x_i-{\hypohat 0 x}_l}-\frac{1}{x_i-{\hypotilde 0 x}_l}\right)\;,\quad i=1,\dots,N\;, 
\] 
which implies 
\begin{equation}\label{eq:con3} 
\sum\limits_{j = 1}^N \left( \frac{1}{x_i  - \widetilde x_j } +\frac{1}{x_i - {\hypotilde 0 x}_j}  \right)= 
\sum\limits_{j = 1}^N \left( \frac{1}{x_i  - \widehat{x}_j}  + \frac{1}{x_i  - {\hypohat 0 x}_j } \right)\;, 
\end{equation}
which is also a consequence of \eqref{eq:2.11} and \eqref{eq:2.11a}, and which expresses this compatibility with the sets of O$\Delta$Es. 
It is, thus, the consistency of the equations \eqref{eq:2.12} and \eqref{eq:2.15}, that lead us to the complete solution of the 
semi-discrete KP equation. In particular, from eq. \eqref{eq:con}, it follows that the diagonal parts of the matrices $\bL$ and 
$\bK$ differ by $(p-q)\bI$, and hence we can identify 
\[ \bLam_L=p\bI+\bLam\quad,\quad \bLam_K=q\bI+\bLam\  , \] 
with common diagonal matrix $\bLam$. Using this identification we can now combine the solution given by \eqref{eq:2.12} for the CM system of 
O$\Delta$Es in the $n$-direction and the solution \eqref{eq:2.15} in the $m$-direction into a simultaneous solution which satisfies also the 
commutativity constraints \eqref{eq:con} and \eqref{eq:con2}, leading to the following statement: 

\paragraph{Proposition:} \textit{ The eigenvalues $x_1(n,m), \dots, x_N(n,m)$ of the $N\times N$ matrix 
\begin{subequations}\begin{equation}
\bsY(n,m)=(p\bI+\bLam)^{-n}(q\bI+\bLam)^{-m}\bsY(0,0)(p\bI+\bLam)^{n}(q\bI+\bLam)^{m}-n(p\bI+\bLam)^{-1}-m(q\bI+\bLam)^{-1}\  
\end{equation} 
in which the initial value matrix $\bsY(0,0)$ is subject to the condition 
\begin{equation}
[\bsY(0,0)\,,\,\bLam]=\bI + {\rm rank\,1} \  , 
\end{equation}
\end{subequations} 
obey both the discrete-time Calogero-Moser systems given by eqs. \eqref{eq:2.11} and \eqref{eq:2.11a} as well as the systems of constraint 
equations given by \eqref{eq:con} and \eqref{eq:con2} }.

\paragraph{} 
The proof is presented in Appendix \ref{Solution}. In order to make a connection with an initial value problem, we mention that the initial value matrix 
$\bsY(0,0)$ can be obtained from the diagonal matrix of initial values $\bsX(0,0)$ by a similarity transformation with a matrix $\bU(0,0)$ 
which is an invertible matrix diagonalizing the initial Lax matrices $\bL(0,0)$ and $\bK(0,0)$. To find the latter, we need the initial values 
$x_i(0,0)$, $x_i(1,0)$ and $x_i(0,1)$, ($i=1,\dots,N$). We note that the secular problem can, hence, be reformulated as one for the following 
matrix 
\begin{equation}\label{eq:2.18a}
\boldsymbol{\mathcal X}(n,m)=\bsX(0,0)-n\bL^{-1}(0,0)-m\bK^{-1}(0,0)\  ,  
\end{equation} 
and hence the solution is provided by the roots of the characteristic equation:
\begin{equation}\label{eq:2.18b}
p_{\boldsymbol{\mathcal X}}(x)=\det(x \boldsymbol I- \boldsymbol{\mathcal{X}}(n,m))=\prod\limits_{i=1}^N(x-x_i(n,m))\;. 
\end{equation}
%

Returning now to the solution of the semi-discrete KP equation \eqref{eq:2.1}, inserting the roots obtained from \eqref{eq:2.18b} into 
the expression \eqref{eq:2.1a} we obtain the required pole solutions. Furthermore, for the solution of the corresponding eigenfunction 
$\phi(n,m)$ as given in \eqref{eq:2.4} for the eigenvalue problem, the vector $\boldsymbol{b}(n,m)$ must be determined. This is 
obtained, for arbitrary value of the spectral parameter $k$, from \eqref{eq:2.9a}, or equivalently \eqref{eq:2.13a}, and is given by
\begin{equation}
 \boldsymbol{b}=\left( p+k-\boldsymbol{L}\right)^{-1}k\boldsymbol{e}\;,\;\mbox{or\ \ equivalently}\;\;
 \boldsymbol{b}=\left( q+k-\boldsymbol{K}\right)^{-1}k\boldsymbol{e}\;.
\end{equation}
%
Plugging all the results into 
\eqref{eq:2.4}, we obtain the explicit form of the eigenfunction $\phi(n,m)$ solving eqs. \eqref{eq:2.2} and \eqref{eq:2.3}. 
Furthermore, eliminating the function $u$ from the latter, we obtain the following nonlinear differential-difference equation for $\phi$ itself:
\begin{equation}\label{eq:lattMKP}
\phi ( \wh{\phi}\;\wt{\phi}_\xi-\wt{\phi}\;\wh{\phi}_\xi)= 
(\wt{\phi}\;\wh{\phi}-\phi\;\wh{\wt{\phi}})\,(\wt{\phi}-\wh{\phi})\  , 
\end{equation}
whose pole-solutions are given by the construction outlined above. 

%
%

%
%
%
%
%
\section{The Lagrangian 1-form and its closure relation}
\setcounter{equation}{0}
In this section, in order to derive the Lagrangian $1$-form structure, we first establish a direct connection between 
the Lagrangian of the discrete-time CM model, as was presented in \cite{F2}, and the temporal part of the Lax representation. 
We will show that this connection, through the compatibility of the matrices $\boldsymbol M$ and $\boldsymbol N$ of the previous section, 
will immediately lead to a closure relation for the corresponding Lagrangians. 
\\
 
Let $\boldsymbol{x}=(x_{1},x_{2},\dots,x_{N})$. Equation \eqref{eq:2.11} can be computed from the variation of a discrete 
action given in \cite{F2} as
\begin{eqnarray}\label{eq:3.1}
\mathcal S_{(n)}[\boldsymbol x(n,m)] = \sum\limits_n {\mathcal L_{(n)}(\boldsymbol{x},\widetilde{\boldsymbol{x}})} 
 =\sum\limits_n \left( - \sum\limits_{i,j = 1}^N {\log \left| {x_i  - \widetilde x_j } \right|}  + \sum\limits_{\mathop {i,j = 1}
\limits_{i \ne j} }^N {\log \left| {x_i  - x_j } \right|}\right)  ,
\end{eqnarray}
where $\mathcal L_{(n)}(\boldsymbol{x},\widetilde{\boldsymbol{x}})$ is the Lagrangian corresponding to the ``$\;\,\widetilde {}\;\;$'' direction, and 
the sum over $n$ represents the sum over all
 discrete-time ``$\;\,\widetilde {}\;\;$'' iterates.  
Similarly, the action corresponding to the ``$\;\,\widehat{}\;\;$'' direction takes the form
\begin{eqnarray}\label{eq:3.3}
\mathcal S_{(m)}[\boldsymbol x(n,m)] = \sum\limits_m {\mathcal L_{(m)}(\boldsymbol{x},\widehat{\boldsymbol{x}})} 
 =\sum\limits_m \left( - \sum\limits_{i,j = 1}^N {\log \left| {x_i  - \widehat x_j } \right|}  + \sum\limits_{\mathop {i,j = 1}
\limits_{i \ne j} }^N {\log \left| {x_i  - x_j } \right|}\right)  ,
\end{eqnarray}
where $\mathcal L_{(m)}(\boldsymbol{x},\widehat{\boldsymbol{x}})$ is the Lagrangian corresponding to the ``$\;\,\widehat {}\;\;$'' direction, and 
the sum over $m$ represents the sum over all
 discrete-time ``$\;\,\widehat {}\;\;$'' iterates.
\\

The matrices $\boldsymbol{M}$ and $\boldsymbol{N}$ are Cauchy-type matrices, and so the exact forms of the determinant can be written as: 
\begin{eqnarray}
\det (\boldsymbol{M}) &=& \frac{{\prod\nolimits_{i < j} {(x_i  - x{}_j)(\widetilde x_i  -
 \widetilde x{}_j)} }}{{\prod\nolimits_{i,j} {(x_i  - \widetilde x{}_j)} }},\label{eq:3.5}\\
\det (\boldsymbol{N}) &=& \frac{{\prod\nolimits_{i < j} {(x_i  - x{}_j)(\widehat x_i  -
 \widehat x{}_j)} }}{{\prod\nolimits_{i,j} {(x_i  - \widehat x{}_j)} }},\label{eq:3.6}
\end{eqnarray}
and we also have that
\begin{subequations}\label{eq:3.7}
\begin{eqnarray}
\log \left| {\det (\boldsymbol{M})} \right| &=& \sum\limits_{\mathop {i,j = 1}
\limits_{i < j} }^N {\left( {\log \left| {x_i  - x{}_j} \right| 
+ \log \left| {\widetilde x_i  - \widetilde x_j} \right|} \right) - \sum\limits_{i,j=1}^N {\log \left| {x_i  - \widetilde x{}_j} \right|} } 
\;,\\
\log \left| {\det (\boldsymbol{N})} \right| &=& \sum\limits_{\mathop {i,j = 1}
\limits_{i<j} }^N {\left( {\log \left| {x_i  - x{}_j} \right| 
+ \log \left| {\widehat x_i  - \widehat x{}_j} \right|} \right) - \sum\limits_{i,j=1}^N {\log \left| {x_i  - \widehat x{}_j} \right|} }
\; .
\end{eqnarray}
\end{subequations}
Remarkably, the action \eqref{eq:3.1} can be obtained by considering the infinite chain product of the matrix $\boldsymbol{M}$ in the following way
\begin{eqnarray}\label{eq:3.9c}
\mathcal S_{(n)} &=&
	\log\left|\det\left(\prod_{n=+\infty}^{\curvearrowleft}\boldsymbol {M}(n)\right)\right|
\nonumber\\
&=&\sum\limits_n {\left( {\sum\limits_{\mathop {i,j = 1}
\limits_{i < j} }^N {\left( {\log \left| {x_i  - x_j} \right| + \log \left| {\widetilde x_i  - \widetilde x_j} \right|} \right) - \sum\limits_{i,j=1}^N {\log \left| {x_i  - \widetilde x_j} \right|} } } \right)} 
 \nonumber\\
&=&\sum\limits_n {\left( { - \sum\limits_{i,j = 1}^N {\log \left| {x_i  - \widetilde x_j } \right|}  + \sum\limits_{\mathop {i,j = 1}
\limits_{i \ne j} }^N {\log \left| {x_i  - x_j } \right|} } \right)} 
=\sum\limits_n {\mathcal L_{(n)}(\boldsymbol{x},\widetilde{\boldsymbol{x}})}\; .
\end{eqnarray}
The same procedure can performed on the matrix $\boldsymbol N$ in order to get the action \eqref{eq:3.3}:
\begin{eqnarray}\label{eq:3.9}
\mathcal S_{(m)}&=&
	\log\left|\det\left(\prod_{m=+\infty}^{\curvearrowleft}\boldsymbol {N}(m)\right)\right|
\nonumber\\
&=&\sum\limits_m {\left( {\sum\limits_{\mathop {i,j = 1}
\limits_{i<j} }^N {\left( {\log \left| {x_i  - x_j} \right| + \log \left| {\wh{x}_i  - \wh{x}_j} \right|} \right) - \sum\limits_{i,j=1}^N {\log \left| {x_i  - \wh{x}_j} \right|} } } \right)} 
\nonumber\\
&=&\sum\limits_m {\left( { - \sum\limits_{i,j = 1}^N {\log \left| {x_i  - \wh{x}_j } \right|}  + \sum\limits_{\mathop {i,j = 1}
\limits_{i \ne j} }^N {\log \left| {x_i  - x_j } \right|} } \right)}  
=\sum\limits_m {\mathcal L_{(m)}(\boldsymbol{x},\wh{\boldsymbol{x}})}\; .
\end{eqnarray}
We note that the connection between the time-part Lax matrix and the
Lagrangian also exists in the cases of the elliptic and trigonometric discrete-time CM systems (see Appendix \ref{ET}).
\\
\\
We now consider the compatibility relation $\widehat{\boldsymbol{M}}\boldsymbol{N}=\widetilde{\boldsymbol{N}}\boldsymbol{M}$, which we recall 
is satisfied if both \eqref{eq:con} and \eqref{eq:con2} hold, and we rewrite this as  
\begin{equation}
\log | {\det (\widehat {\boldsymbol{M}})}| + \log \left| {\det (\boldsymbol{N})} \right|  -
 \log | {\det (\widetilde{\boldsymbol{N}})} | - \log \left| {\det (\boldsymbol{M})} \right| =0\;,
\end{equation}
or
\begin{equation}\label{eq:3.10}
 \widehat{\mathcal L_{(n)}(\boldsymbol{x},\widetilde{\boldsymbol{x}})} - \mathcal L_{(n)}(\boldsymbol{x},\widetilde{\boldsymbol{x}}) -
 \widetilde{\mathcal L_{(m)}(\boldsymbol{x},\widehat{\boldsymbol{x}})} + \mathcal L_{(m)}(\boldsymbol{x},\widehat{\boldsymbol{x}}) = 0\;,
\end{equation}
where we define $\Xi=\sum_{i=1}^Nx_i$ to be the centre of mass and
\begin{subequations}\label{eq:3.10q}
\begin{eqnarray}
\mathcal L_{(n)}(\boldsymbol{x},\widetilde{\boldsymbol{x}})&=&\log|\det(\boldsymbol M)|+p(\Xi-\wt {\Xi})\;,\label{eq:3.10qa}\\
\mathcal L_{(m)}(\boldsymbol{x},\widehat{\boldsymbol{x}}) &=& \log|\det(\boldsymbol N)|+q(\Xi-\wh {\Xi})\;.\label{eq:3.10qb}
\end{eqnarray}
\end{subequations}
The last terms in \eqref{eq:3.10qa} and \eqref{eq:3.10qb} are the total derivative terms involving centre of mass motion which can be separated
from the relative motion. The extra equation comes directly from \eqref{eq:3.10} 
\begin{equation} 
\wt{\Xi}+\wh{\Xi}=\Xi+\wh{\wt{\Xi}}\;,
\end{equation}
which holds on the solutions.
\\
\\

We would now like to interpret \eqref{eq:3.10} as the closure relation for a Lagrangian $1$-form. 
Let $\boldsymbol{e}_{i}$ represent the unit vector in the lattice direction labeled by $i$ and let any position in the 
lattice be identified by the vector $\boldsymbol{n}$, so that an elementary shift in the lattice can be created by the
 operation $\boldsymbol{n}\mapsto\boldsymbol{n}+\boldsymbol{e}_{i}$. Since the Lagrangian depends on $\boldsymbol{x}$ and 
its elementary shift in one discrete direction, it can be associated with an oriented vector $\boldsymbol{e}_i$ on a curve 
$\Gamma_i(\boldsymbol{n})=(\boldsymbol{n},\boldsymbol{n}+\boldsymbol{e}_i)$, and we can treat these Lagrangians as defining 
a discrete 1-form $\mathcal L_i(\boldsymbol{n})$
\begin{equation}\label{eq:3.12}
\mathcal L_i(\boldsymbol{n})=\mathcal L_i(\boldsymbol{x}(\boldsymbol{n}),\boldsymbol{x}(\boldsymbol{n}+\boldsymbol{e}_{i})).
\end{equation}
\\
\textbf{Proposition}: \emph{As functions on the two-dimensional lattice, the discrete-time Lagrangians given in eq. \eqref{eq:3.10q}
 satisfy the following relation
\begin{eqnarray} 
&& \mathcal{L}_{i}(\boldsymbol{x}(\boldsymbol{n}+\boldsymbol{e}_{j}),
                                                      \boldsymbol{x}(\boldsymbol{n}+\boldsymbol{e}_{i}+\boldsymbol{e}_{j}))
                                     -\mathcal{L}_{i}(\boldsymbol{x}(\boldsymbol{n}),\boldsymbol{x}(\boldsymbol{n}+\boldsymbol{e}_{i}))\nn\\
                                   &&\;\;\;\;\;\;\;\;-\mathcal{L}_{j}(\boldsymbol{x}(\boldsymbol{n}+\boldsymbol{e}_{i}),
                                                      \boldsymbol{x}(\boldsymbol{n}+\boldsymbol{e}_{j}+\boldsymbol{e}_{i}))
                                     +\mathcal{L}_{j}(\boldsymbol{x}(\boldsymbol{n}),\boldsymbol{x}(\boldsymbol{n}+\boldsymbol{e}_{j}))=0\;.
\end{eqnarray}
}

Equation \eqref{eq:3.10} represents the closure relation of the Lagrangian $1$-form. The point is that the closure relation \eqref{eq:3.10} is derived from 
the Lax equation (the compatibility of the matrices $\boldsymbol M$ and $\boldsymbol N$), together with the property of determinants of Cauchy matrices. 
In contrast, in those cases of Lagrangian $2$-forms and $3$-forms in \cite{SF1, SF2, SFQ}, the equation of the motion must be invoked in order to show that 
the closure relation holds. Also in those cases the Euler-Lagrange equation stemming from the variational principle was not exactly the equation of motion 
needed to verify the closure relation; rather it was a discrete derivative, or a sum of copies, of the original equation of motion.

\begin{figure}[h]
\begin{center}
\begin{tikzpicture}[scale=0.5]
 \draw[->] (0,0) -- (8,0) node[anchor=west] {$n_{i}$};
 \draw[->] (0,0) -- (0,8) node[anchor=south] {$n_{j}$};
 \fill (1,1) circle (0.2);
 \draw (1,1)  -- (1,2) -- (3,2) -- (3,3) -- (4,3) -- (4,4) -- (5,4) -- (5,6) -- (6,6);
 \fill (6,6) circle (0.2);
 \draw (4,4)node[anchor=north west] {$\Gamma$};
\end{tikzpicture}
\begin{tikzpicture}[scale=0.5]
 \draw[white] (0,0) -- (6.5,0);
 \draw[white] (0,0) -- (0,6);
 \draw[->,thick] (2,4) -- (4,4);
\end{tikzpicture}
\begin{tikzpicture}[scale=0.5]
 \draw[->] (0,0) -- (8,0) node[anchor=west] {$n_{i}$};
 \draw[->] (0,0) -- (0,8) node[anchor=south] {$n_{j}$};
 \fill (1,1) circle (0.2);
 \draw (1,1)  -- (1,2) -- (4,2) -- (4,4) -- (5,4) -- (5,6) -- (6,6);
 \draw[dashed] (3,2) -- (3,3) -- (4,3);
 \fill (6,6) circle (0.2);
  \draw(4,4)node[anchor=north west] {$\Gamma^{\prime}$};
\end{tikzpicture}
\end{center}
\caption{Deformation of the discrete curve $\Gamma$.}\label{curve_deformation}
\end{figure}
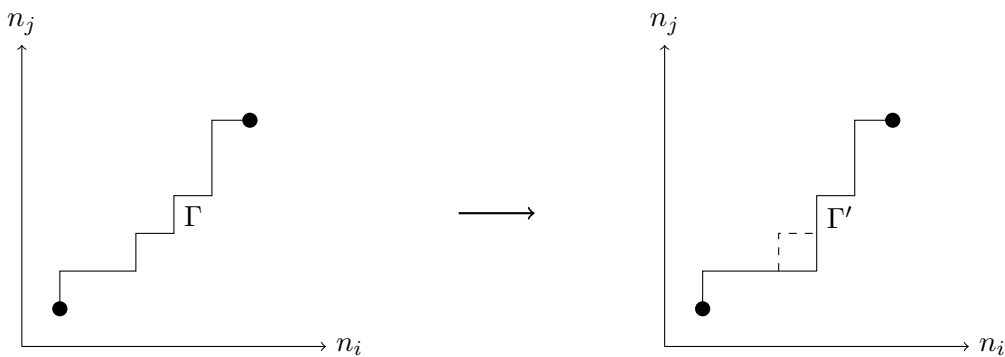

Choosing a discrete curve $\Gamma$ consisting of connected elements $\Gamma_{i}$, we can define an action on the curve by summing up the contributions $\mathcal L_i$ from each of the oriented links $\Gamma_{i}$ in the curve, to get
\begin{equation}\label{eq:3.13}
\mathcal S(\boldsymbol{x}(\textbf{n});\Gamma) = \sum\limits_{\boldsymbol{n} \in \Gamma } {\mathcal L_i(\boldsymbol{x}(\boldsymbol{n}),\boldsymbol{x}(\boldsymbol{n} +\boldsymbol{e}_i ))}. 
\end{equation}
The closure relation \eqref{eq:3.10} is actually equivalent to the invariance of the action under local deformations of 
the curve. To see this, suppose we have an action $\mathcal S$ evaluated on a curve $\Gamma$, and we deform this 
(keeping end points fixed) to get a curve $\Gamma^{\prime}$ on which an action $\mathcal S^{\prime}$ is evaluated,
 such as in Figure \ref{curve_deformation}.

Then $\mathcal S^{\prime}$ is related to $\mathcal S$ by the following:
\begin{eqnarray}\label{eq:3.14ss}
\mathcal S^{\prime} & = & \mathcal S -\mathcal{L}_{i}(\boldsymbol{x}(\boldsymbol{n}+\boldsymbol{e}_{j}),
                                                      \boldsymbol{x}(\boldsymbol{n}+\boldsymbol{e}_{i}+\boldsymbol{e}_{j}))
                                     +\mathcal{L}_{i}(\boldsymbol{x}(\boldsymbol{n}),\boldsymbol{x}(\boldsymbol{n}+\boldsymbol{e}_{i}))\nn\\
                                  && +\mathcal{L}_{j}(\boldsymbol{x}(\boldsymbol{n}+\boldsymbol{e}_{i}),
                                                      \boldsymbol{x}(\boldsymbol{n}+\boldsymbol{e}_{j}+\boldsymbol{e}_{i}))
                                     -\mathcal{L}_{j}(\boldsymbol{x}(\boldsymbol{n}),\boldsymbol{x}(\boldsymbol{n}+\boldsymbol{e}_{j})).
\end{eqnarray}
Equation \eqref{eq:3.14ss} shows that the independence of the action under such a deformation is locally equivalent to the closure relation. 
The invariance of the action under the local deformation is a crucial aspect of the underlying variational principle.
%
\begin{figure}[h]
\begin{center}
\begin{tikzpicture}[scale=0.5]
 \draw[->] (0,0) -- (-8,0) ;
  \draw (-9,0) node[anchor=west] {$n$};
 \draw[->] (0,0) -- (0,8) node[anchor=south] {$m$};
 \draw[thick] (0,0)  --(-1,0)--(-1,1)--(-2,1)--(-2,2)--(-3,2)--(-3,3)
--(-4,3)--(-4,4)--(-5,4)--(-5,5)--(-6,5)--(-6,6);
 \draw[->,thick] (0,0) -- (-.5,0);
 \draw[->,thick] (-1,0) -- (-1,.5);
\draw[->,thick]   (-1,1)--(-1.5,1);
\draw[->,thick]   (-2,1)--(-2,1.5);
\draw[->,thick]   (-2,2)--(-2.5,2);
\draw[->,thick]   (-3,2)--(-3,2.5);
\draw[->,thick]   (-3,3)--(-3.5,3);
\draw[->,thick]   (-4,3)--(-4,3.5);
\draw[->,thick]   (-4,4)--(-4.5,4);
\draw[->,thick]   (-5,4)--(-5,4.5);
\draw[->,thick]   (-5,5)--(-5.5,5);
\draw[->,thick]   (-6,5)--(-6,5.5);
 \draw (-5,-1.5)node[anchor=west] {(a)};
 \draw (-4,4)node[anchor=west] {$\Gamma$};
\draw (0,0)node[anchor=west] {$(n_0,m_0)$};
\draw (-6.5,-0.5)node[anchor=west] {$n_{1}$};
\draw [dashed](-6,5.5)--(-6,0);
\draw[dashed](-6,6)-- (0,6)node[anchor=west] {$m_1$};
\end{tikzpicture}
\begin{tikzpicture}[scale=0.5]
 \draw[white] (0,0) -- (2,0);
 \draw[white] (0,0) -- (0,2);
 \draw[->,thick] (0,4) -- (4,4);
\end{tikzpicture}
\begin{tikzpicture}[scale=0.5]
 \draw[->] (0,0) -- (-8,0) ;
  \draw (-9,0) node[anchor=west] {$\mathsf N$};
 \draw[->] (0,0) -- (0,8) node[anchor=south] {$m$};
 \draw[thick] (0,0) --(-1,0)--(0,1)--(-1,1)--(0,2)--(-1,2)--(0,3)--(-1,3)--(0,4)
--(-1,4)--(0,5)--(-1,5)--(0,6);
 \draw[->,thick] (0,0) -- (-.5,0);
 \draw[->,thick] (-1,0) -- (-.5,.5);
\draw[->,thick]   (0,1)--(-.5,1);
\draw[->,thick]   (-1,1)--(-.5,1.5);
\draw[->,thick]   (0,2)--(-.5,2);
\draw[->,thick]   (-1,2)--(-.5,2.5);
\draw[->,thick]   (0,3)--(-.5,3);
\draw[->,thick]   (-1,3)--(-.5,3.5);
\draw[->,thick]   (-0,4)--(-.5,4);
\draw[->,thick]   (-1,4)--(-.5,4.5);
\draw[->,thick]   (-0,5)--(-.5,5);
\draw[->,thick]   (-1,5)--(-.5,5.5);
\draw (-5,-1.5)node[anchor=west] {(b)};
 \draw (-3,3)node[anchor=west] {$\Gamma^\prime$};
\draw (0,0)node[anchor=west] {$(\mathsf{N}_0,m_0)$};
\draw (-4.5,-.5)node[anchor=west] {$\mathsf{N}_{0}-1=\mathsf N_1$};
\draw (0,6)node[anchor=west] {$m_1$};
\end{tikzpicture}
\end{center}
\caption{The effect of changing variables on the discrete curve I.}\label{curve_deformation_skew}
\end{figure}
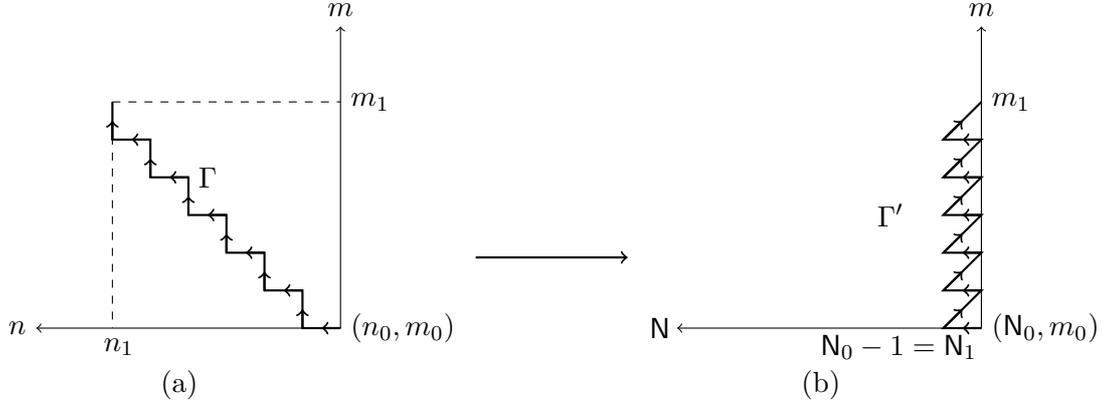
\\

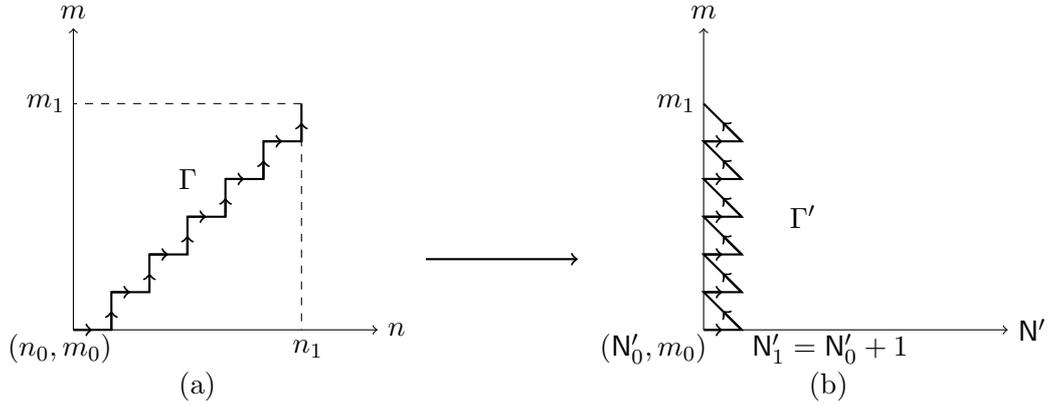
\begin{figure}[h]
\begin{center}
\begin{tikzpicture}[scale=0.5]
 \draw[->] (0,0) -- (8,0) ;
  \draw (8,0) node[anchor=west] {$n$};
 \draw[->] (0,0) -- (0,8) node[anchor=south] {$m$};
 \draw[thick] (0,0)  --(1,0)--(1,1)--(2,1)--(2,2)--(3,2)--(3,3)
--(4,3)--(4,4)--(5,4)--(5,5)--(6,5)--(6,6);
 \draw[->,thick] (0,0) -- (.5,0);
 \draw[->,thick] (1,0) -- (1,.5);
\draw[->,thick]   (1,1)--(1.5,1);
\draw[->,thick]   (2,1)--(2,1.5);
\draw[->,thick]   (2,2)--(2.5,2);
\draw[->,thick]   (3,2)--(3,2.5);
\draw[->,thick]   (3,3)--(3.5,3);
\draw[->,thick]   (4,3)--(4,3.5);
\draw[->,thick]   (4,4)--(4.5,4);
\draw[->,thick]   (5,4)--(5,4.5);
\draw[->,thick]   (5,5)--(5.5,5);
\draw[->,thick]   (6,5)--(6,5.5);
 \draw (2.5,-1.5)node[anchor=west] {(a)};
 \draw (2.5,4)node[anchor=west] {$\Gamma$};
\draw (-2,-.5)node[anchor=west] {$(n_0,m_0)$};
\draw (5.5,-0.5)node[anchor=west] {$n_{1}$};
\draw [dashed](6,5.5)--(6,0);
\draw[dashed](6,6)-- (0,6);
\draw (-1.5,6)node[anchor=west] {$m_1$};
\end{tikzpicture}
\begin{tikzpicture}[scale=0.5]
 \draw[white] (0,0) -- (2,0);
 \draw[white] (0,0) -- (0,2);
 \draw[->,thick] (0,4) -- (4,4);
\end{tikzpicture}
\begin{tikzpicture}[scale=0.5]
 \draw[->] (0,0) -- (8,0) ;
  \draw (8,0) node[anchor=west] {$\mathsf N^\prime$};
 \draw[->] (0,0) -- (0,8) node[anchor=south] {$m$};
 \draw[thick] (0,0) --(1,0)--(0,1)--(1,1)--(0,2)--(1,2)--(0,3)--(1,3)--(0,4)
--(1,4)--(0,5)--(1,5)--(0,6);
 \draw[->,thick] (0,0) -- (.5,0);
 \draw[->,thick] (1,0) -- (.5,.5);
\draw[->,thick]   (0,1)--(.5,1);
\draw[->,thick]   (1,1)--(.5,1.5);
\draw[->,thick]   (0,2)--(.5,2);
\draw[->,thick]   (1,2)--(.5,2.5);
\draw[->,thick]   (0,3)--(.5,3);
\draw[->,thick]   (1,3)--(.5,3.5);
\draw[->,thick]   (0,4)--(.5,4);
\draw[->,thick]   (1,4)--(.5,4.5);
\draw[->,thick]   (0,5)--(.5,5);
\draw[->,thick]   (1,5)--(.5,5.5);
\draw (2.5,-1.5)node[anchor=west] {(b)};
 \draw (2,3)node[anchor=west] {$\Gamma^\prime$};
\draw (-3,-0.5)node[anchor=west] {$(\mathsf{N}^\prime_0,m_0)$};
\draw (1,-.5)node[anchor=west] {$\mathsf N^\prime_1=\mathsf{N}^\prime_{0}+1$};
\draw (-1.5,6)node[anchor=west] {$m_1$};
\end{tikzpicture}
\end{center}
\caption{The effect of changing variables on the discrete curve II.}\label{curve_deformation_skewII}
\end{figure}

Next, it is interesting to investigate further how to derive the discrete Euler-Lagrange equation from the variational principle.
The system given in Fig. (\ref{curve_deformation}) is difficult to work with because of a matter of notation. For general discrete curves it is cumbersome to
implement the variational principle because of the notation it would require. We will, however, demonstrate how the principle works for a few simple cases: 
\textbf{(a)} the curve shown
 in Fig. (\ref{curve_deformation_skew}), \textbf{(b)} the curve shown in Fig. (\ref{curve_deformation_skewII}).
\\
\\
\textbf{(a)}: The curve shown in Fig. (\ref{curve_deformation_skew}): We now introduce a new variable $\mathsf N=n+m$ (which will play an important role in the next section)
together with the change of notation
\[
\boldsymbol{x}(n,m)\mapsto\boldsymbol{x}(\mathsf N,m),\;\;\;\;\;\;\wt{\boldsymbol{x}}:=\boldsymbol{x}(\mathsf N+1,m)\;\;\;\mbox{and}\;\;\;
\wh{\boldsymbol x}:=\boldsymbol{x}(\mathsf N,m+1)\;,
\] 
and so we work with the curve given in
 Fig. (\ref{curve_deformation_skew}b). The action evaluated on this curve can be written in the form
\begin{eqnarray}\label{eq:3.14}
\mathcal S[\boldsymbol{x};\Gamma^\prime]=\sum_{m=m_0}^{m_1-1}-\mathcal{L}_{(\mathsf N)}(\boldsymbol{x}(\mathsf N_0,m),\boldsymbol{x}(\mathsf N_{0}-1,m))
+\sum_{m=m_0}^{m_1-1}\mathcal{L}_{(m)}(\boldsymbol{x}(\mathsf N_{0}-1,m),\boldsymbol{x}(\mathsf N_{0},m+1)),
\end{eqnarray}
where
\begin{eqnarray}\label{eq:3.14a}
\mathcal{L}_{(\mathsf N)}(\boldsymbol{x},\boldsymbol{y})&=&\sum\limits_{\mathop {i,j = 1}
\limits_{i \ne j} }^N\left( \frac{1}{2}\log|y_i-y_j|+\frac{1}{2}\log|{ x}_i-{ x}_j|\right)
-\sum_{i,j=1}^N\log|y_i-x_j|\nn\\
&&+p\sum_{i=1}^N(y_i-x_i)\;,\\
\mathcal{L}_{(m)}(\boldsymbol{x},\boldsymbol{y})&=&\sum\limits_{\mathop {i,j = 1}
\limits_{i \ne j} }^N\left( \frac{1}{2}\log| x_i-x_j|+\frac{1}{2}\log|y_i-y_j|\right)
-\sum_{i,j=1}^N\log|{x}_i-y_j|\nn\\
&&+p\sum_{i=1}^N(x_i-y_i)\;.
\end{eqnarray}
The minus sign in \eqref{eq:3.14} indicates the reverse direction of the Lagrangian $\mathcal{L}_{(\mathsf N)}$ along the horizontal links. 
Performing the variation $\boldsymbol{x}\mapsto\boldsymbol{x}+\delta\boldsymbol{x}$, we have
\begin{eqnarray}
&&\delta\mathcal S=0=\nn\\
&&\sum_{m=m_0}^{m_1-1}\left(-\frac{\partial{\mathcal{L}_{(\mathsf N)}
(\boldsymbol{x}(\mathsf N_0,m),\boldsymbol{x}(\mathsf N_{0}-1,m)}}{\partial{\boldsymbol{x}(\mathsf N_0,m)}}\delta\boldsymbol{x}(\mathsf N_0,m)\right.\nn\\
&&\;\;\;\;\;\;\;\;\;\;\;\;\;\;\;\;\;\;\;\;\;\;\;\;\;\;\;\;\;\;\;\;\;\;\;\;\;\;\;\;\;\;\left.-\frac{\partial{\mathcal{L}_{(\mathsf N)}
(\boldsymbol{x}(\mathsf N_0,m),\boldsymbol{x}(\mathsf N_{0}-1,m)}}{\partial{\boldsymbol{x}(\mathsf N_{0}-1,m)}}\delta\boldsymbol{x}(\mathsf N_{0}-1,m)
\right)\nn\\
&&+\sum_{m=m_0}^{m_1-1}\left(\frac{\partial{\mathcal{L}_{(m)}
(\boldsymbol{x}(\mathsf N_{0}-1,m),\boldsymbol{x}(\mathsf N_{0},m-1)}}{\partial{\boldsymbol{x}(\mathsf N_0,m+1)}}\delta\boldsymbol{x}(\mathsf N_{0},m+1)\right.\nn\\
&&\;\;\;\;\;\;\;\;\;\;\;\;\;\;\;\;\;\;\;\;\;\;\;\;\;\;\;\;\;\;\;\;\;\;\;\;\;\;\left.+\frac{\partial{\mathcal{L}_{(m)}
(\boldsymbol{x}(\mathsf N_{0}-1,m),\boldsymbol{x}(\mathsf N_{0},m+1)}}{\partial{\boldsymbol{x}(\mathsf N_{0}-1,m)}}\delta\boldsymbol{x}(\mathsf N_{0}-1,m)
\right)\;.
\end{eqnarray}
We now obtain the Euler-Lagrange equations
\begin{subequations}
\begin{eqnarray}
-\frac{\partial{\mathcal{L}_{(\mathsf N)}
(\boldsymbol{x}(\mathsf N_0,m),\boldsymbol{x}(\mathsf N_{0}-1,m)}}{\partial{\boldsymbol{x}(\mathsf N_0,m)}}
+\frac{\partial{\mathcal{L}_{(m)}
(\boldsymbol{x}(\mathsf N_{0}-1,m-1),\boldsymbol{x}(\mathsf N_{0},m)}}{\partial{\boldsymbol{x}(\mathsf N_0,m+1)}}=0\;,\\
-\frac{\partial{\mathcal{L}_{(\mathsf N)}
(\boldsymbol{x}(\mathsf N_0,m),\boldsymbol{x}(\mathsf N_{0}-1,m)}}{\partial{\boldsymbol{x}(\mathsf N_{0}-1,m)}}
+\frac{\partial{\mathcal{L}_{(m)}
(\boldsymbol{x}(\mathsf N_{0}-1,m),\boldsymbol{x}(\mathsf N_{0},m+1)}}{\partial{\boldsymbol{x}(\mathsf N_{0}-1,m)}}=0\;,
\end{eqnarray}
\end{subequations}
which produce 
\begin{subequations}
\begin{eqnarray}
p-q&=&\sum_{j=1}^N\left( \frac{1}{\boldsymbol{x}_i(\mathsf N_0,m)-\boldsymbol{x}_i(\mathsf N_0+1,m-1)}-
\frac{1}{\boldsymbol{x}_i(\mathsf N_0,m)-\boldsymbol{x}_i(\mathsf N_0-1,m)}\right),\\
p-q&=&\sum_{j=1}^N\left( \frac{1}{\boldsymbol{x}_i(\mathsf N_0,m)-\boldsymbol{x}_i(\mathsf N_0+1,m)}-
\frac{1}{\boldsymbol{x}_i(\mathsf N_0,m)-\boldsymbol{x}_i(\mathsf N_0-1,m+1)}\right)\;,
\end{eqnarray}
\end{subequations}
which are equivalent to \eqref{eq:con} and \eqref{eq:con2}, respectively.
\\
\\
\\
\textbf{(b)}: The curve shown in Fig. (\ref{curve_deformation_skewII}):  
Introducing the variable $\mathsf N^\prime=n-m$, the corresponding curve is given in Fig. (\ref{curve_deformation_skewII}b).
 The action evaluated on the curve $\Gamma^\prime$ reads
\begin{eqnarray}\label{eq:3.18}
\mathcal S[\boldsymbol{x};\Gamma^\prime]=\sum_{m=m_0}^{m_1-1}\mathcal{L}_{(\mathsf N^\prime)}(\boldsymbol{x}(\mathsf N^\prime_0,m),\boldsymbol{x}(\mathsf N^\prime_{0}+1,m))
+\sum_{m=m_0}^{m_1-1}\mathcal{L}_{(m)}(\boldsymbol{x}(\mathsf N^\prime_{0}+1,m),\boldsymbol{x}(\mathsf N^\prime_{0},m+1)),
\end{eqnarray}
where
\begin{eqnarray}\label{eq:3.19}
\mathcal{L}_{(\mathsf N^\prime)}(\boldsymbol{x},\boldsymbol{y})&=&\sum\limits_{\mathop {i,j = 1}
\limits_{i \ne j} }^N\left( \frac{1}{2}\log|y_i-y_j|+\frac{1}{2}\log|{ x}_i-{ x}_j|\right)
-\sum_{i,j=1}^N\log|{x}_i-y_j|\nn\\
&&+p\sum_{i=1}^N(x_i-y_i)\;,\\
\mathcal{L}_{(m)}(\boldsymbol{x},\boldsymbol{y})&=&\sum\limits_{\mathop {i,j = 1}
\limits_{i \ne j} }^N\left( \frac{1}{2}\log|x_i-x_j|+\frac{1}{2}\log|y_i-y_j|\right)
-\sum_{i,j=1}^N\log|x_i-y_j|\nn\\
&&+p\sum_{i=1}^N(x_i-y_i)\;.
\end{eqnarray}
Performing the variation $\boldsymbol{x}\mapsto\boldsymbol{x}+\delta\boldsymbol{x}$, we have
\begin{eqnarray}
&&\delta\mathcal S=0=\nn\\
&&\sum_{m=m_0}^{m_1-1}\left(\frac{\partial{\mathcal{L}_{(\mathsf N^\prime)}
(\boldsymbol{x}(\mathsf N^\prime_0,m),\boldsymbol{x}(\mathsf N^\prime_{0}+1,m)}}{\partial{\boldsymbol{x}(\mathsf N^\prime_0,m)}}\delta\boldsymbol{x}(\mathsf N^\prime_0,m)\right.\nn\\
&&\;\;\;\;\;\;\;\;\;\;\;\;\;\;\;\;\;\;\;\;\;\;\;\;\;\;\;\;\;\;\;\;\;\;\;\;\;\;\;\;\;\;+\left.\frac{\partial{\mathcal{L}_{(\mathsf N^\prime)}
(\boldsymbol{x}(\mathsf N^\prime_0,m),\boldsymbol{x}(\mathsf N^\prime_{0}+1,m)}}{\partial{\boldsymbol{x}(\mathsf N^\prime_{0}+1,m)}}\delta\boldsymbol{x}(\mathsf N^\prime_{0}+1,m)
\right)\nn\\
&&+\sum_{m=m_0}^{m_1-1}\left(\frac{\partial{\mathcal{L}_{(m)}
(\boldsymbol{x}(\mathsf N^\prime_{0}+1,m),\boldsymbol{x}(\mathsf N^\prime_{0},m+1)}}{\partial{\boldsymbol{x}(\mathsf N^\prime_0,m+1)}}\delta\boldsymbol{x}(\mathsf N^\prime_{0},m+1)\right.\nn\\
&&\;\;\;\;\;\;\;\;\;\;\;\;\;\;\;\;\;\;\;\;\;\;\;\;\;\;\;\;\;\;\;\;\;\;\;\;\;\;\left.+\frac{\partial{\mathcal{L}_{(m)}
(\boldsymbol{x}(\mathsf N^\prime_{0}+1,m),\boldsymbol{x}(\mathsf N^\prime_{0},m+1)}}{\partial{\boldsymbol{x}(\mathsf N^\prime_{0}+1,m)}}\delta\boldsymbol{x}(\mathsf N^\prime_{0}+1,m)
\right)\;.
\end{eqnarray}
We now obtain the Euler-Lagrange equations
\begin{subequations}
\begin{eqnarray}
\frac{\partial{\mathcal{L}_{(\mathsf N^\prime)}
(\boldsymbol{x}(\mathsf N^\prime_0,m),\boldsymbol{x}(\mathsf N^\prime_{0}+1,m)}}{\partial{\boldsymbol{x}(\mathsf N^\prime_0,m)}}
+\frac{\partial{\mathcal{L}_{(m)}
(\boldsymbol{x}(\mathsf N^\prime_{0}+1,m-1),\boldsymbol{x}(\mathsf N^\prime_{0},m)}}{\partial{\boldsymbol{x}(\mathsf N^\prime_0,m+1)}}=0\;,\\
\frac{\partial{\mathcal{L}_{(\mathsf N^\prime)}
(\boldsymbol{x}(\mathsf N^\prime_0,m),\boldsymbol{x}(\mathsf N^\prime_{0}+1,m)}}{\partial{\boldsymbol{x}(\mathsf N^\prime_{0}+1,m)}}
+\frac{\partial{\mathcal{L}_{(m)}
(\boldsymbol{x}(\mathsf N^\prime_{0}+1,m),\boldsymbol{x}(\mathsf N^\prime_{0},m+1)}}{\partial{\boldsymbol{x}(\mathsf N^\prime_{0}+1,m)}}=0\;,
\end{eqnarray}
\end{subequations}
which produce 
\begin{subequations}
\begin{eqnarray}
q-p&=&\sum_{j=1}^N\left( \frac{1}{\boldsymbol{x}_i(\mathsf N^\prime_0,m)-\boldsymbol{x}_i(\mathsf N^\prime_0-1,m+1)}-
\frac{1}{\boldsymbol{x}_i(\mathsf N^\prime_0,m)-\boldsymbol{x}_i(\mathsf N^\prime_0+1,m)}\right),\\
p-q&=&\sum_{j=1}^N\left( \frac{1}{\boldsymbol{x}_i(\mathsf N^\prime_0+1,m)-\boldsymbol{x}_i(\mathsf N^\prime_0+2,m)}-
\frac{1}{\boldsymbol{x}_i(\mathsf N^\prime_0+1,m)-\boldsymbol{x}_i(\mathsf N^\prime_0,m+1)}\right)\;,
\end{eqnarray}
\end{subequations}
which are again equivalent to \eqref{eq:con} and \eqref{eq:con2}, respectively.

By working with the specific type of curves given in Fig. (\ref{curve_deformation_skew}) and Fig. (\ref{curve_deformation_skewII}), we can perform the variational principle
with either the new variables $(\mathsf N,m)$ or $(\mathsf N^\prime,m)$. Even though we have these two possibilities, the natural choice is
to work with $(\mathsf N,m)$. This is because the variable $\mathsf N^\prime$ leads to difficulties in performing the skew limit in the next section. 
This can, in fact, be seen from the
structure of the plane wave solution \eqref{eq:2.4}. 

It is not yet clear whether the processes of varying the curve of the independent variables, from which we get the closure relation, and 
that of varying the dependent variables, from which we get Euler-Lagrange equations, commute. We defer consideration of this point to a future publication.

%
%
%
\section{The semi-continuum limit: skew limit}
\setcounter{equation}{0}
In this section, we study a continuum analogue of a previous construction in Section 2 by considering a particular semi-continuous limit.
Since the semi-discrete KP equation \eqref{eq:2.1} contains two discrete variables $n$ and $m$, we could perform a continuum limit on one of these 
variables separately, while leaving the other discrete variable intact, and thus obtain a semi-continuous equation with one remaining discrete 
and two continuous independent variables. Alternatively, we can first perform a change of independent variables on the lattice and subsequently perform 
the limit on one of the new variables. The advantage of the latter approach over the former is that it often leads in a more direct way to a hierarchy of 
higher order flows. Adopting the latter approach in this section, we use a new discrete variable $\mathsf N :=n+m$, and perform the 
transformation on the dependent variables by setting $u(n,m)\mapsto \mathsf u(\mathsf N,m)=:{\mathsf u}$, which leads to the following expressions for 
the shifted variables:
\begin{eqnarray}\label{eq:4.1a}
u=u(n+1,m) &\mapsto & \mathsf u(\mathsf N+1,m)=:\wt{\mathsf u}\;, \nn\\
\widehat u=u(n,m+1) &\mapsto & \mathsf u(\mathsf N+1,m+1)=:\widehat{\widetilde{ {\mathsf u}}}\;,\nn\\ 
\widetilde u=u(n+1,m+1) &\mapsto & \mathsf u(\mathsf N+2,m+1)=:\widehat{\widetilde{\widetilde {\mathsf u}}}\;.\nn
\end{eqnarray}
Rearranging the discrete exponential factor in \eqref{eq:2.4}, we have
\begin{eqnarray}\label{eq:4.1}
\phi(n,m)  &=& \left( {1 - \frac{1}{k}\sum\limits_{i = 1}^N {\frac{{b_i (n,m)}}{{\xi  - x_i (n,m)}}} } \right)\left( {p + k} \right)^{\mathsf N} \left( {\frac{q + k}{p+k}} \right)^m e^{k\xi } ,\nonumber\\
&=& \left( {1 - \frac{1}{k}\sum\limits_{i = 1}^N {\frac{{b_i (n,m)}}{{\xi  - x_i (n,m)}}} } \right)\left( {p + k} \right)^{\mathsf N} \left( {1-\frac{p-q }{p+k}} \right)^m e^{k\xi }.
\end{eqnarray}
We perform the limit $n\rightarrow -\infty$, $m \rightarrow \infty$, $\varepsilon \rightarrow 0$ while
keeping $\mathsf N$ fixed and setting $\varepsilon=p-q$,
such that $\varepsilon m=\tau$ remains finite. Focusing on the penultimate factor in \eqref{eq:4.1} we have that
\begin{equation}\label{eq:4.2}
\lim\limits_{\mathop {m \rightarrow \infty} \limits_{\mathop {\varepsilon \rightarrow 0}\limits_{\varepsilon
 m \rightarrow \tau}} }\left(1-\frac{\varepsilon}{p+k} \right)^m =\lim\limits_{m \rightarrow \infty}\left( 1-\frac{\tau}{m(p+k)}
\right)^m=e^{-\frac{\tau}{p+k}}, 
\end{equation}
so that the solution to the Lax equations \eqref{eq:2.2} and \eqref{eq:2.3} takes the form
\begin{equation}\label{eq:4.3}
\mathsf\phi(\mathsf N,\tau) =\left( {1 - \frac{1}{k}\sum\limits_{i = 1}^N {\frac{{\mathsf {b}_i (\mathsf N,\tau)}}{{\xi  -\mathsf {x}_i (\mathsf N,\tau)}}} } \right)\left( {p + k} \right)^{\mathsf N}e^{k\xi}e^{-\frac{\tau}{p+k}}.
\end{equation}
\\
We would now like to see the effect of this limit on the semi-discrete KP equation. We rewrite eq. (\ref{eq:2.1}) as
\begin{equation}\label{eq:4.3a}
\partial_\xi\ln(p-q+\widehat { u}-\widetilde{ u})= u-\widehat{ u}-\widetilde { u}+\widehat{\widetilde{ {\mathsf u}}}.
\end{equation}
Taking the limit, we have
\begin{eqnarray}
\partial_\xi\ln(\varepsilon+\widehat{\widetilde{ {\mathsf u}}}-\widetilde{ \mathsf {u}})= \mathsf u-
\widehat{\widetilde{{\mathsf u}}}-\widetilde { \mathsf u}+
\widehat{\widetilde{\widetilde{{ \mathsf u}}}}.
\end{eqnarray}
Setting $\tau=\tau_0+m\varepsilon$, where $\varepsilon_0$ is a background constant of the continuous variable, and applying the Taylor expansion, we get
\begin{equation}\label{eq:4.6}
\mathsf u(\tau+\varepsilon)=\mathsf u(\tau)+\varepsilon\frac{\partial}{\partial \tau}\mathsf u(\tau)+
\frac{1}{2}\varepsilon^2\frac{\partial^2}{\partial {\tau^2}}\mathsf u(\tau)+ \dots
\end{equation}
where $\dot{\mathsf u}=\frac{\partial}{\partial\tau}\mathsf u$. The leading order $\mathcal O(\delta^0)$ gives us
\begin{equation}\label{eq:4.6a}
\partial_\xi\ln(1+\dot {\mathsf u})= \undertilde{\mathsf u}+\widetilde { \mathsf u}-2\mathsf u.
\end{equation}
Eq. \eqref{eq:4.6a}, which contains two continuous and one discrete variable, we call the semi-continuous KP equation (to distinguish it from the semi-discrete KP equation,
which contains one continuous and two discrete variables).
\\
\\
\textsf{\textbf{Remark:} The connection to the Toda system.}

Introducing the variable $1+\dot {\mathsf u}=: \exp({\mathsf v})$ we can write eq. (\ref{eq:4.6a}) in the form
\begin{equation}\label{eq:4.6c}
\mathsf {v}_{\tau \xi}=2\exp({\mathsf v})-\exp({\undertilde{\mathsf{v}}})-\exp({\widetilde{\mathsf{v}}}).
\end{equation}
Introducing next another variable $\mathsf{y}$ through the relation  $\mathsf v=\widetilde {\mathsf y}-\mathsf y$,  we obtain
\begin{equation}\label{eq:4.6d}
\widetilde{\mathsf y}_{\tau\xi}-\mathsf{y}_{\tau\xi}=( \exp(\widetilde{\mathsf y}-\mathsf y)-\exp(\widetilde{\widetilde{\mathsf y}}-\widetilde{\mathsf y}))+ 
( \exp(\mathsf y-\undertilde{\mathsf y})-\exp(\widetilde{\mathsf y}-\mathsf y))\;,
\end{equation}
from which we can identify
\begin{equation}\label{eq:4.6e}
\mathsf{y}_{\tau\xi}=  \exp(\mathsf y-\undertilde{\mathsf y})-\exp(\widetilde{\mathsf y}-\mathsf y)\  . 
\end{equation}
This is the well-known two-dimensional Toda lattice, cf. \cite{Mikh}. A Lagrangian exists for this system is given by 
\begin{equation}\label{eq:TodaLagr}
\mathcal{L}_{2D-Toda}= \frac{1}{2} \mathsf{y}_\tau \mathsf{y}_\xi +e^{\textstyle \wt{\mathsf{y}}-\mathsf{y}}\; 
\end{equation} 
To follow the reductions of the Lagrangian structure for the CM system, it would be desirable to have a Lagrangian for the 
semi-discrete KP equation \eqref{eq:2.1} and for the semi-continuous KP \eqref{eq:4.6a} equation, but these are strangely elusive. 
Obviously, the fully continuous KP equation does possess a Lagrangian structure which is easy to establish by inspection, and 
furthermore the 2D Toda lattice \eqref{eq:4.6e} reduces to the KP equation through some specific continuum limits, see e.g. \cite{F5}. 
Thus it is conceivable that CM lagrangians could be established through the ``Toda route'', but we will not pursue this line 
in the present paper. We would like to mention also Krichever's elliptic analogue of the one-dimensional Toda lattice \cite{Kr2}, which  
remarkably contains a term corresponding to the elliptic discrete-time CM given in \cite{FP}. We will give some results on 
the Lagrangian 1-form structure of the discrete-time elliptic CM model in Appendix {\ref{ET}}, but our main concern in the present paper is to study the 
Lagrangian 1-form structure in the simplest possible case, for the sake of transparency. 
\\
\\
\textbf{The skew limit on the Lax equations}: To obtain the Lax representation for the semi-continuous KP equation 
\eqref{eq:4.6a}, we perform a similar limit on the Lax equations \eqref{eq:2.2}, \eqref{eq:2.3} by making the transformation 
$\phi(n,m,\xi)\;\mapsto\;{\mathsf \phi}(\mathsf{N},m,\xi)$ with similar replacements as above for its lattice shifts, i.e. 
$$ 
\phi(n+1,m,\xi)\;\mapsto\;{\mathsf \phi}(\mathsf{N}+1,m,\xi)=:\wt{\mathsf \phi}\quad,\quad 
\phi(n,m+1,\xi)\;\mapsto\;{\mathsf \phi}(\mathsf{N}+1,m+1,\xi)=:\wh{\wt{\mathsf \phi}}\  . 
$$  
Thus, we obtain 
\begin{eqnarray}
\widetilde{\mathsf \phi}  &=& {\mathsf \phi} _\xi   + (p +\mathsf u - \widetilde {\mathsf u})\phi ,\label{eq:4.4}\\
\widehat{\widetilde {\mathsf \phi}}  &=& {\mathsf \phi}_\xi   + (q + \mathsf u - \widehat{\widetilde {\mathsf u}}){\mathsf \phi} .\label{eq:4.5}
\end{eqnarray}
Using a Taylor expansion as in the previous case, we can write \eqref{eq:4.5} as
\begin{equation}\label{eq:4.7}
\widetilde \phi+\varepsilon\dot{\widetilde {\phi}}+...=\phi_{\xi}+(p-\varepsilon+\mathsf u-\widetilde 
{\mathsf u}-\varepsilon \dot{\widetilde {\mathsf u}}+...)\phi,
\end{equation}
where $\dot{\phi}=\frac{\partial}{\partial\tau}\phi$. Then to leading order $\mathcal{O}(\varepsilon)$ we have
\begin{equation}\label{eq:4.8}
\dot\phi=-(1+\dot {\mathsf u})\undertilde \phi.
\end{equation}
Equation \eqref{eq:4.8} is a mixed equation, with one discrete and one continuous variable. It is easy to check that eq. (\ref{eq:4.6a}) arises 
as the compatibility condition of the Lax pair consisting of (\ref{eq:4.8}) and (\ref{eq:4.4}).

Inserting the special form of the solution, eq. \eqref{eq:4.3}, into eq. \eqref{eq:4.4}, we recover the set of equations 
\eqref{eq:2.5} and \eqref{eq:2.6}, obviously with the replacements of $x_i(n,m)$ by $\mathsf{x}_i(\mathsf N,m)$ and with 
$\widetilde{\mathsf{x}}_i=\mathsf{x}_i(\mathsf{N}+1,m)$. For the sake of selfcontainedness we write the Lax corresponding matrices (which are 
essentially the same as before) in the new notation as follows: 
\begin{eqnarray}
\boldsymbol{\mathsf L} &=& \sum\limits_{i,j = 1}^N {\frac{E_{ii}}{\mathsf x_i-\widetilde {\mathsf x}_j}}  -
 \sum\limits_{\mathop {i,j = 1}\limits_{j \ne i} }^N \frac{E_{ii}+E_{ij}}{\mathsf x_i-\mathsf x_j}\;, \label{eq:4.8a}\\
\boldsymbol{\mathsf M} &=&  - \sum\limits_{i,j = 1}^N {\frac{{E_{ij} }}{{\widetilde {\mathsf x}_i  - 
\mathsf x_j }}} \;.\label{eq:4.8b}
\end{eqnarray}
The compatibility between the matrices $\boldsymbol{\mathsf L}$ and $\boldsymbol{\mathsf M}$ produces the equations of motion
\begin{equation}\label{eq:4.8c}
\sum\limits_{j = 1}^N {\left( {\frac{1}{{\mathsf x_i  - \widetilde {\mathsf x}_j }} + \frac{1}{{\mathsf x_i  - \undertilde {\mathsf{x}_j} }}} \right)}  - 2\sum\limits_{\mathop {j = 1}\limits_{j \ne i} }^N {\frac{1}{{\mathsf x_i  - \mathsf x_j }}=0,} \begin{array}{*{20}c}
   \mbox{~where~} & {i = 1,...,N} , \\
\end{array}
\end{equation}
implying as before the system of equations \eqref{eq:2.11a}, but now viewed as part of a semi-continuous Calogero-Moser system. This set of equations 
will be complemented by equations involving derivatives with respect to the variable $\tau$, which are obtained by  
inserting the solution \eqref{eq:4.3} into eq. \eqref{eq:4.8} and obtaining the complementary equations for $\mathsf{b}_i$. 
Equating to zero the resulting coefficients of $(\xi-\mathsf{x}_i)^{-1}$, $(\xi-\mathsf{x}_i)^{-2}$ and $(\xi-\undertilde{\mathsf{x}_i})^{-1}$, we obtain
%
%
\begin{eqnarray}
(p+k)\dot{\mathsf{b}}_i&=&\mathsf{b}_i+\sum_{j=1}^N\frac{\dot{\mathsf{x}}_i{\hypotilde 0 {\mathsf{b}}}_j}{(\mathsf{x}_i-
{\hypotilde 0 {\mathsf{x}}}_j)^2}\;,\label{eq:4.9}\\
(p+k)\mathsf{b}_i&=&k-\sum_{j=1}^N\frac{{\hypotilde 0 {\mathsf{b}}}_j}{\mathsf{x}_i-\undertilde{\mathsf{x}_j}},\label{eq:4.10}\\
-1&=&\sum_{j=1}^N\frac{\dot{\mathsf{x}}_j}{(\mathsf{x}_j-\undertilde{\mathsf{x}_i})^2}.\label{eq:4.11}
\end{eqnarray}
Introducing the matrix 
\begin{eqnarray}\label{eq:4.12}
\boldsymbol{\mathsf A}&=&\sum_{i,j=1}^N\frac{\dot{\mathsf{x}}_i}{(\mathsf{x}_i-\undertilde{\mathsf{x}_j})^2}E_{ij},\label{eq:4.12a}
\end{eqnarray}
together with the matrix $\boldsymbol{\mathsf M}$ as given above, we can rewrite equations \eqref{eq:4.9} and \eqref{eq:4.10} as
\begin{eqnarray}
(p+k)\dot{\boldsymbol{\mathsf{b}}}&=&\boldsymbol{\mathsf b}+\boldsymbol{\mathsf A}\undertilde{\boldsymbol{\mathsf b}},\label{eq:4.13}\\
(p+k)\widetilde{\boldsymbol{\mathsf b}}&=&k\boldsymbol{e}+\boldsymbol{\mathsf M}\boldsymbol{\mathsf b}.\label{eq:4.14a}
\end{eqnarray}
Note that \eqref{eq:4.13} and \eqref{eq:4.14a} can be directly obtained by taking the
skew limit on \eqref{eq:2.13b} in order $\mathcal O(\delta^0)$ and $\mathcal O(\delta)$ respectively.
The compatibility condition between \eqref{eq:4.13} and \eqref{eq:4.14} leads to
$$ (p+k)\left( \wt{\boldsymbol{\mathsf A}}-\dot{\boldsymbol{\mathsf M}}- \boldsymbol{\mathsf M}\;\boldsymbol{\mathsf A}\;
{\hypotilde 0 {\boldsymbol{\mathsf M}}}^{-1}\right)\; \boldsymbol{\mathsf b} + 
k \left( \boldsymbol{\mathsf I} + \boldsymbol{\mathsf M}\;\boldsymbol{\mathsf A}\;
{\hypotilde 0 {\boldsymbol{\mathsf M}}}^{-1}\right)\;\boldsymbol{e} =0 $$ 
which, using \eqref{eq:4.12} and \eqref{eq:4.8b}, leads to: 
\begin{eqnarray*}
&& \sum_{i,j=1}^N\, \left[ -\frac{E_{ij}}{\wt{x}_i-{\hypotilde 0 x}_j} \left( \sum_{l=1}^N \frac{\dot{x}_l}{(\wt{x}_i-x_l)^2} 
- \sum_{l=1}^N \frac{\dot{x}_l}{({\hypotilde 0 {x}}_i-x_l)^2}\right) {\hypotilde 0 {\boldsymbol{\mathsf M}}}^{-1} \left( (p+k)\boldsymbol{ \mathsf b}-k\boldsymbol{e}\right)
\right. \\  
&& \qquad \left. \frac{k}{N} \left(1+\sum_{l=1}^N \frac{\dot{x}_l}{(\wt{x}_i-x_l)^2} \right) E_{ij} \boldsymbol{e} \right] =0  \;. 
\end{eqnarray*} 
This, using \eqref{eq:4.11}, implies that we also have  
\begin{equation}
1+ \sum_{j=1}^N\frac{\dot{\mathsf{x}}_j}{(\widetilde{\mathsf x}_i-\mathsf {x}_j)^2}=0\;,\label{eq:4.16}
\end{equation} 
and thus, combining these two relations, we get: 
\begin{equation}
0= \sum_{j=1}^N\left( \frac{\dot {\mathsf x}_j}{(\widetilde{\mathsf x}_i-\mathsf{x}_j)^2}-\frac{\dot {\mathsf x}_j}
{(\mathsf{x}_j-\undertilde{\mathsf{x}_l})^2}\right)\;.\label{eq:4.17}
\end{equation}
We observe that if we use \eqref{eq:4.16}, we recover \eqref{eq:4.11} from \eqref{eq:4.17}. Furthermore, we can show that
eq. \eqref{eq:4.17} is a consequence of the skew limit of the eq. \eqref{eq:3.9a} of order $\mathcal O(\delta)$.
On the level of the matrix relations, this implies 
\begin{subequations}
\begin{eqnarray}
(\mathsf I+\widetilde{\boldsymbol{\mathsf A}}-\dot{\boldsymbol{\mathsf M}})\boldsymbol{e}&=&0,\label{eq:4.14}\\
\boldsymbol{\mathsf M}\,\boldsymbol{\mathsf A}\,{\hypotilde 0 {\boldsymbol{\mathsf M}}}^{-1}-\widetilde{\boldsymbol{\mathsf A}}
+\dot{\boldsymbol{\mathsf M}}&=&0.\label{eq:4.15}
\end{eqnarray}
\end{subequations}
\textbf{Remark}: We observe that \eqref{eq:4.11} and \eqref{eq:4.16} can be obtained by taking the skew 
limit on \eqref{eq:con} and \eqref{eq:con2} in order $\mathcal{O}(\delta)$, respectively.
\\
\\
\textbf{The skew limit on the solution}: we finish this section by performing the skew limit on the full solution \eqref{eq:2.15}. 
Let us observe that due to the fact that the matrices $\bL$ and $\bK$ are related linearly through the constraint \eqref{eq:con} 
(this relation amounts to a shift over $(p-q)$ times the unit matrix) we can identify the diagonal matrices of eigenvalues  as follows: 
\begin{equation}\label{eq:w5}
 \boldsymbol{\Lambda}_{L}=p+\boldsymbol{\Lambda}\;,\;\;\mbox{and}\;\;\;\boldsymbol{\Lambda}_{K}=q+\boldsymbol{\Lambda}\;,
\end{equation}
where $\boldsymbol{\Lambda}$ is the matrix which is independent of both $p$ and $q$ (i.e., independent of the direction of the lattice).
%
%
Then the full solution \eqref{eq:2.15} can be expressed in the form
\begin{eqnarray}\label{eq:w7}
 \boldsymbol{Y}(n,m)&=&(p+\boldsymbol{\Lambda})^{-n}(q+\boldsymbol{\Lambda})^{-m}\boldsymbol{Y}(0,0)(q+\boldsymbol{\Lambda}
)^{m}(p+\boldsymbol{\Lambda})^{n}\nn\\
&&-n(p+\boldsymbol{\Lambda})^{-1}-m(q+\boldsymbol{\Lambda})^{-1}\;.
\end{eqnarray}
Using the definitions of the variable $\mathsf N=n+m$ and $p-q=\varepsilon$, we obtain
\begin{eqnarray}\label{eq:w8}
\boldsymbol{\mathsf Y}(\mathsf N,m)&=&
(p+\boldsymbol{\Lambda})^{-\mathsf N}\left(1-\varepsilon(p+\boldsymbol{\Lambda})^{-1}\right)^{-m}\boldsymbol{\mathsf Y}(0,0)
\left(1-\varepsilon(p+\boldsymbol{\Lambda})^{-1}\right)^{m}(p+\boldsymbol{\Lambda})^{\mathsf N}\nn\\
&&-\mathsf{N}(p+\boldsymbol{\Lambda})^{-1}+m\left((p+\boldsymbol{\Lambda})^{-1}-(q+\boldsymbol{\Lambda})^{-1}\right)\;.
\end{eqnarray}
Taking the limit, we have
\begin{eqnarray}\label{eq:w12}
\lim\limits_{\mathop {m \rightarrow \infty} \limits_{\mathop {\varepsilon \rightarrow 0}\limits_{\varepsilon m \rightarrow \tau}} }
\boldsymbol{\mathsf {Y}}(\mathsf{N},m)=\boldsymbol{\mathsf {Y}}(\mathsf {N},\tau)&=&
(p+\boldsymbol{\Lambda})^{-\mathsf{ N}}e^{\tau(p+\boldsymbol{\Lambda})^{-1}}\boldsymbol{\mathsf Y}(0,0)e^{-\tau(p+\boldsymbol{\Lambda}
)^{-1}}(p+\boldsymbol{\Lambda})^{\mathsf N}\nn\\
&&-\mathsf {N}(p+\boldsymbol{\Lambda})^{-1}-\tau(p+\boldsymbol{\Lambda})^{-2}\;.
\end{eqnarray}
This equation represents the full solution after taking the skew limit. The positions of the particles $\mathsf x_i(\mathsf N,m)$ can be
determined by computing the eigenvalues of \eqref{eq:w12}.

\section{Semi-continuous Lagrangian and closure relation}

Having obtained, in the previous section, the skew limit of the Lax equation and the corresponding differential-difference 
system comprising eqs. \eqref{eq:4.11} and \eqref{eq:4.16}, together with eq. \eqref{eq:4.8c}, we now proceed to present the corresponding 
Lagrange form. First, we observe that eq. \eqref{eq:4.8c} can be once again be obtained by implementing the usual variational principle 
on the following action $\mathcal S_{(\mathsf N)}$ given by
\begin{eqnarray}\label{eq:4.17a}
&&\mathcal S_{(\mathsf N)}[\boldsymbol {\mathsf x}(\mathsf N,\tau)]=\sum_{\mathsf N}\mathcal{L}_{\mathsf N}
 =\sum\limits_{\mathsf N} \left( - \sum\limits_{i,j = 1}^N {\log \left| {\mathsf x_i  - \widetilde {\mathsf x}_j } \right|}
  + \sum\limits_{i \ne j}^N \left(\frac{1}{2}{\log \left| {\mathsf x_i  - \mathsf x_j } \right|}
+\frac{1}{2}{\log \left| {\wt{\mathsf x}_i  - \wt{\mathsf x}_j } \right|}\right)\right.\nn\\
&&\;\;\;\;\;\;\;\;\;\;\;\;\;\;\;\;\;\;\;\;\;\;\;\;\;\;\;\;\;\;\;\;\;\;\;\;\;\;\;\;\;\;\;\left.+p\sum_{i=1}^N(x_i-\wt{\mathsf x}_i)\right)
\end{eqnarray}
where now the Lagrangian $\mathcal L_{(\mathsf N)}$ involves variables $\wt{\mathsf x}_i$ shifted in the discrete variable $\mathsf N$ instead of 
the original variable $n$, and the corresponding discrete Euler-Lagrange equation reads: 
\begin{equation}\label{eq:4.20s}
\widetilde{\frac{\partial\mathcal L_{(\mathsf N)}}{\partial {\mathsf x}_i}}+
\left( \frac{\partial\mathcal L_{(\mathsf N)}}{\partial {\widetilde {\mathsf x}_i}}\right)=0 ,
\end{equation}
yielding \eqref{eq:4.8c}.  

Next we observe that the combined differential-difference equation \eqref{eq:4.17} can be obtained from the two-dimensional 
action
\begin{equation}\label{eq:2Dact} 
\mathcal S_{(\tau)}[\boldsymbol{\mathsf x}(\mathsf N,\tau)]
=\sum_{\mathsf N}\int d\tau \left( \sum\limits_{i,j=1}^N\, \frac{\dot {\mathsf x}_i}{\mathsf x_i-\undertilde {\mathsf {x}_j}} \right),
\end{equation}
yielding \eqref{eq:4.17} from the two-dimensional EL equations
\begin{equation}
\frac{\pl {\mathcal L}}{\pl {\mathsf x}_i} + \wt{\frac{\pl {\mathcal L}}{\pl {\hypotilde 0 {\mathsf x}}_i} }  -\frac{\pl}{\pl\tau} \left( 
\frac{\pl {\mathcal L}}{\pl \dot{\mathsf x}_i}\right)=0\  .  
\end{equation}
However, we will argue that this is not the correct point of view to take in the context of the CM systems, because they consist  
after all of ODE's rather than PDE's or differential-difference equations. In fact, this action, containing a summation over the variable 
$\mathsf{N}$ and an integration over the variable $\tau$ can never be obtained as a continuum limit of the one-dimensional discrete 
actions presented in Section 3, which contain one single summation. 
\\
\\
\textbf{The skew limit on the action}:
We note that the actions in \eqref{eq:4.17a} and \eqref{eq:2Dact} are obtained directly from the corresponding equations of motion 
\eqref{eq:4.8c}, \eqref{eq:4.11} and \eqref{eq:4.16}, respectively. An interesting question is whether we can obtain the mixed discrete and
continuous 1-form structure directly by taking the skew limit on the fully discrete 1-form action \eqref{eq:3.13}. This will be rather 
tricky because it will sensitively depend on parametrization of the curve. Furthermore, in order to perform a systematic expansion on the action
functional one would need to employ not only a Taylor expansion on the relevant lattice shifts, but at the same time also the opposite of the 
Taylor expansion on the sums. For the latter we can employ the following ``anti-Taylor'' expansion of the sum of a discrete function 
$f(n)=:F(\tau)$, where $\tau=\tau_0+n\varepsilon$ with $\tau_0$ fixed:
the formula
\[
 \sum_{n=n_1}^{n_2-1}f(n)=\frac{1}{\varepsilon}\int_{\tau_1}^{\tau_2}d\tau\;F(\tau)-\frac{1}{2}\left(F(\tau_2)-F(\tau_1)\right)
+\frac{\varepsilon}{12}\left((F^{\prime}(\tau_2)-F^{\prime}(\tau_1))\right)+ \dots\;,
\]
which guarantees that the analogue of the fundamental theorem of integration holds, namely:  
\[
\sum_{n=n_1}^{n_2-1}(f(n+1)-f(n))=f(n_2)-f(n_1)=F(\tau_2)-F(\tau_1)\;. 
\]
We now would like to pursue this task by considering a simple discrete curve $\Gamma$ given in Fig. (\ref{curve_deformation_skew}a). 
The action corresponding to the curve has been given by \eqref{eq:3.13}.
Performing the skew limit, the curve has been transformed to Fig. (\ref{curve_deformation_skew}b) with the new variables $(\mathsf N,m)$.
%
%
Using the anti-Taylor expansion, the action \eqref{eq:3.14} becomes
\begin{eqnarray}\label{eq:actionskew2}
\mathcal{S}[\boldsymbol {\mathsf x}(\mathsf N,m);\Gamma^\prime]&=& \frac{1}{\varepsilon}\int_{\tau_1}^{\tau_2}d\tau\mathcal{L}_{(m)}
(\boldsymbol{\mathsf x}(\mathsf N_{0}-1,\tau),
\boldsymbol{\mathsf x}(\mathsf N_0,\tau-\varepsilon))-\frac{1}{2}[\mathcal{L}_{(m)}(\boldsymbol{\mathsf x}(\mathsf N_{0}-1,\tau_2),
\boldsymbol{\mathsf x}(\mathsf N_0,\tau_2-\varepsilon))\nn\\
&-&\mathcal{L}_{(m)}(\boldsymbol{\mathsf x}(\mathsf N_{0}-1,\tau_1),\boldsymbol{\mathsf x}(\mathsf N_0,\tau_1-\varepsilon))]+.....\nn\\
&-& \frac{1}{\varepsilon}\int_{\tau_1}^{\tau_2-\varepsilon}d\tau\mathcal{L}_{(\mathsf N)}(\boldsymbol{\mathsf x}(\mathsf N_0,\tau),
\boldsymbol{\mathsf x}(\mathsf N_{0}-1,\tau))+\frac{1}{2}[\mathcal{L}_{(\mathsf N)}(\boldsymbol{\mathsf x}(\mathsf N_0,\tau_2-\varepsilon),
\boldsymbol{\mathsf x}(\mathsf N_{0}-1,\tau_2))\nn\\
&-&\mathcal{L}_{(\mathsf N)}(\boldsymbol{\mathsf x}(\mathsf N_0,\tau_1-\varepsilon),\boldsymbol{\mathsf x}(\mathsf N_{0}-1,\tau_1))]+.....\;\;.
\end{eqnarray}
We also use the fact that the Lagrangian is an antisymmetric function under interchange of the arguments 
\[
\mathcal{L}(\boldsymbol{\mathsf x}(\mathsf N_0,\tau),\boldsymbol{\mathsf x}(\mathsf N_{0}+1,\tau))=
-\mathcal{L}(\boldsymbol{\mathsf x}(\mathsf N_{0}+1,\tau),\boldsymbol{\mathsf x}(\mathsf N_0,\tau))\;,
\]
together with expansion with respect to $\varepsilon$, then the action becomes
\begin{equation}\label{eq:actionskew3}
\mathcal{S}[\boldsymbol {\mathsf x}(\mathsf N,m);\Gamma^\prime]=\int_{\tau_1}^{\tau_2}d\tau\mathcal{L}_{(\tau)}
\left(\boldsymbol{\mathsf{x}}(\mathsf N_{0}-1,\tau),\frac{\partial{\boldsymbol{\mathsf{x}}(\mathsf N_0,\tau)}}{\partial \tau}\right)\;,
\end{equation}
where $\mathcal{L}_{(\tau)}$ is given in \eqref{eq:4.26aaa}.

By using a specific choice of curve given in Fig. (\ref{curve_deformation_skew}), we can perform the skew limit on the discrete action
leading to what is called the ``semi-continuous'' action in \eqref{eq:actionskew3}.  
\\
\\
\textbf{The skew limit on the discrete-time closure relation}: We would like to consider the closure relation of the semi-continuous Lagrangians. Taking, 
as we did in the equations of motion, the skew 
limit on the discrete-time closure relation eq. (\ref{eq:3.10}), implementing a change of variables on the lattice first, we obtain: 
\begin{eqnarray*}
\widehat{\widetilde {\mathcal L}}_{(\mathsf N)}-\mathcal L_{(\mathsf N)}&=&\widetilde {\mathcal L}_{(m)}-\mathcal L_{(m)}\;\quad \Rightarrow \nonumber\\
\left( \widetilde {\mathcal {L}}_{(\mathsf N)}+\varepsilon\widetilde{\mathcal {\dot L}}_{(\mathsf N)} +...\right)-\mathcal L_{(\mathsf N)} 
&=&\left( \widetilde {\mathcal L}_{(\mathsf N)}-\varepsilon\widetilde{\mathcal L}_{(\tau)}+...\right)-\left( \mathcal L_{(\mathsf N)}-
\varepsilon\mathcal L_{(\tau)}\right)\;.  
\end{eqnarray*}
Equating the coefficients of $\varepsilon$ gives us
\begin{eqnarray}\label{eq:4.25}
\frac{\partial \mathcal L_{(\mathsf N)}}{\partial \tau}&=&{\hypotilde 0 {\mathcal L}}_{(\tau)}- {\mathcal L_{(\tau)}}\;,  
\end{eqnarray}
where the Lagrangian $\mathcal L_{(\tau)}$ is obtained by performing the skew limit on the Lagrangian $\mathcal L_{(m)}$ in
eq. (\ref{eq:3.10qb}) to give
\begin{eqnarray}\label{eq:4.26aaa}
\mathcal L_{(\tau)}=\mathsf{Tr}\left(\boldsymbol{\mathsf M}^{-1}\wt{\boldsymbol{\mathsf A}}\right)-\Xi+\wt{\Xi}
=\sum_{i \ne j}^N\frac{\dot{\wt{\mathsf x}}_j}{\wt{\mathsf x}_i-\wt{\mathsf x}_j}+\sum_{i,j=1}^N
\frac{\dot{\wt{\mathsf x}}_j}{\mathsf x_i-\wt{\mathsf x}_j}-\Xi+\wt{\Xi}-p\dot{\wt{\Xi}}\; ,
\end{eqnarray}
which produces the semi-continuous equations of motion. The derivation of \eqref{eq:4.26aaa} is given in Appendix \ref{L}.
\\
\\
Furthermore, the variational derivatives with respect to $\wt{\boldsymbol{\mathsf x}}(\tau)$ and 
$\undertilde{\boldsymbol{\mathsf x}}(\tau)$ yield: 
\begin{eqnarray}
\frac{\partial {\mathcal L}_{(\tau)}}{\partial {\widetilde{\mathsf x}}_i}=0\;\;\;\mbox{and}\;\;\;\
\frac{\partial {\mathcal L}_{(\tau)}}{\partial {\hypotilde 0 {\mathsf x}}_i}=0\ ,\label{eq:4.21a}
\end{eqnarray}
which yield the equations 
\begin{eqnarray}
\frac{\partial {\mathcal L}_{(\tau)}}{\partial {\widetilde{\mathsf x}}_i}&=&\sum_{j=1}^N\frac{\dot{\mathsf x}_j}{(\widetilde{\mathsf x}_i-
\mathsf x_j)^2}+1=0\;,\\\label{eq:4.2d}
\frac{\partial {\mathcal L}_{(\tau)}}{\partial {\hypotilde 0 {\mathsf x}}_i}&=&\sum_{j=1}^N\frac{\dot{\mathsf x}_j}{({\hypotilde 0 
{\mathsf x}}_i-\mathsf x_j)^2}+1=0\;,\label{eq:4.21e}
\end{eqnarray}
which are identical to eqs. \eqref{eq:4.11} and \eqref{eq:4.16}.
\\

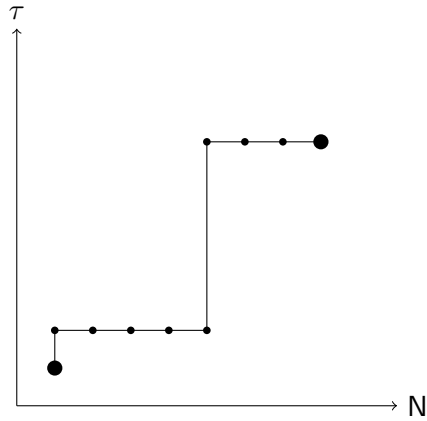
\begin{figure}[h]
\begin{center}
\begin{tikzpicture}[scale=0.5]
 \draw[->] (0,0) -- (10,0) node[anchor=west] {$\mathsf N$};
 \draw[->] (0,0) -- (0,10) node[anchor=south] {$\tau$};
 \fill (1,1) circle (0.2);
 \draw (1,1)  -- (1,2) ;
 \draw (1,2) -- (5,2);
\draw (5,2) -- (5,7);
\draw (5,7) -- (8,7);
 \fill (8,7) circle (0.2);
\fill (1,2) circle (0.1);
\fill (2,2) circle (0.1);
\fill (3,2) circle (0.1);
\fill (4,2) circle (0.1);
\fill (5,2) circle (0.1);
\fill (5,7) circle (0.1);
\fill (6,7) circle (0.1);
\fill (7,7) circle (0.1);
\fill (8,7) circle (0.1);
\end{tikzpicture}
\end{center}
\caption{The semi-continuous curve.}\label{semi-curve44}
\end{figure}

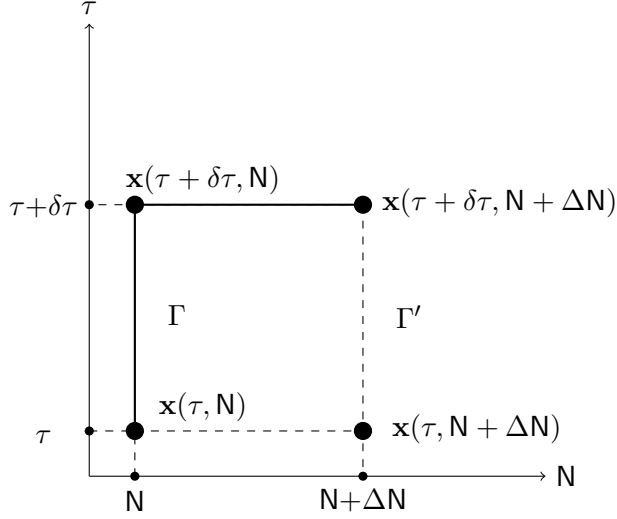
\begin{figure}[h]
\begin{center}
\begin{tikzpicture}[scale=0.6]
 \draw[->] (0,0) -- (10,0) node[anchor=west] {$\mathsf N$};
 \draw[->] (0,0) -- (0,10) node[anchor=south] {$\tau$};
 \fill (1,1) circle (0.2)  ;
 \draw[thick] (1,1) -- (1,6);
 \draw[thick] (1,6) -- (6,6) ;
 \fill (6,6) circle (0.2);
 \fill (1,6) circle (0.2);
 \draw (1.5,4)node[anchor=north west] {$\Gamma$};
 \draw (6.5,4)node[anchor=north west] {$\Gamma^\prime$};
 \draw[dashed] (1,0)--(1,1);
 \draw[dashed] (0,1)--(1,1);
 \fill (1,0) circle (0.1);
\draw (-1,.5) node[anchor=south] {$\tau$};
 \fill (0,1) circle (0.1);
\draw (1,-1) node[anchor=south] {$\mathsf N$};
 \draw (-1,5.5) node[anchor=south] {$\tau$+$\delta\tau$};
 \draw[dashed] (6,6)--(0,6);
 \draw[dashed] (6,6)--(6,0);
\fill (6,0) circle (0.1);
\fill (0,6) circle (0.1);
\draw (6,-1) node[anchor=south] {$\mathsf N$+$\Delta\mathsf N$};
 \draw[dashed] (1,1)--(6,1);
\fill (6,1) circle (0.2);
\draw (8.5,.5) node[anchor=south] {$\textbf x(\tau,\mathsf N+\Delta\mathsf N)$};
\draw (9,5.5) node[anchor=south] {$\textbf x(\tau+\delta\tau,\mathsf N+\Delta\mathsf N)$};
\draw (2.5,1) node[anchor=south] {$\textbf x(\tau,\mathsf N)$};
\draw (2.5,6) node[anchor=south] {$\textbf x(\tau+\delta\tau,\mathsf N)$};
\end{tikzpicture}
\end{center}
\caption{The semi-continuous curve $\Gamma$ and $\Gamma^\prime$.}\label{semi-curve}
\end{figure}

The relation \eqref{eq:4.25} represents a closure relation between the semi-continuous Lagrangian 
$\mathcal L_{(\tau)}$ and the semi-discrete Lagrangian $\mathcal L_{(\mathsf N)}$. 
This points again to a Lagrangian 1-form structure but now of a mixed, part discrete and part continuous, type
where the total action is a combination of a discrete part of the action $\mathcal S_{(\mathsf N)}$ and 
continuous part $\mathcal S_{(\tau)}$ defined on a semi-discrete curve in the space of the discrete independent 
variable $\mathsf N$ and the the continuous variable $\tau$. This is most easily visualized in the case where we 
have a piecewise linear curve as in Fig. \ref{semi-curve44}, where in the horizontal direction we have 
step-wise jumps over units, while vertically the curve is continuous. The closure relation guarantees, nevertheless, 
that any such semi-discrete curve may be ``locally'' deformed without changing the action functional 
consisting of sums of the form \eqref{eq:4.17a} over the horizontal iterates and of the form \eqref{eq:actionskew3} in 
the vertical segments .  
\\
\\
\textbf{Variational principle for a semi-continuous 1-form structure}: In \eqref{eq:4.25}, we obtain the semi-continuous closure relation. It is interesting to 
see how the variational principle works
in this case. We will end this section with a scheme for deriving the closure relation. Now, suppose we have a curve $\Gamma$ living on the semi-continuous
 ($\mathsf N-\tau$)-configuration shown in Fig. (\ref{semi-curve}).
The action associated with the curve $\Gamma$ can be written in the form
\begin{eqnarray}\label{eq:4.26a}
\mathcal {S}(\mathsf x(\mathsf N, \tau); \Gamma)&=&\int_{\tau}^{\tau+\delta\tau}\mathcal L_{(\tau)}(\textbf x
(\tau,\mathsf N),\textbf x(\tau+\delta\tau,\mathsf N))d\tau\nn\\
&&+\mathcal L_{(\mathsf N)}(\textbf {x}(\tau+\delta\tau,\mathsf N)
,\textbf x(\tau+\delta\tau,\mathsf N+\Delta\mathsf N)),
\end{eqnarray}
where the first term is the Lagrangian living on the vertical (continuous) line and the second term corresponds to the Lagrangian living
on the horizontal (discrete) line of the curve $\Gamma$.
\\
\\
We now consider the action along the dashed curve
\begin{eqnarray}\label{eq:4.26b}
\mathcal {S}^{\prime}(\mathsf x(\mathsf N, \tau); \Gamma^\prime)&=&\mathcal L_{(\mathsf N)}(\textbf {x}(\tau,\mathsf N)
,\textbf x(\tau,\mathsf N+\Delta\mathsf N))\nn\\
&&+\int_{\tau}^{\tau+\delta\tau}\mathcal L_{(\tau)}(\textbf x
(\tau,\mathsf N+\Delta\mathsf N),\textbf x(\tau+\delta\tau,\mathsf N+\Delta\mathsf N))d\tau,
\end{eqnarray}
where the first term is the Lagrangian living on the horizontal (discrete) line and the second term corresponds to the Lagrangian living
on the vertical (continuous) line of the dashed curve.
\\
\\
We now impose the condition that $\delta \mathcal S=\mathcal S^{\prime}-\mathcal S=0$, yielding
\begin{eqnarray}\label{eq:4.26c}
\delta \mathcal S&=&\mathcal L_{(\mathsf N)}(\textbf {x}(\tau,\mathsf N)
,\textbf x(\tau,\mathsf N+\Delta\mathsf N))-\mathcal L_{(\mathsf N)}(\textbf {x}(\tau+\delta\tau,\mathsf N)
,\textbf x(\tau+\delta\tau,\mathsf N+\Delta\mathsf N))\nn\\
&+&\int_{\tau}^{\tau+\delta\tau}\left(\mathcal L_{(\tau)}(\textbf x
(\tau,\mathsf N+\Delta\mathsf N),\textbf x(\tau+\delta\tau,\mathsf N+\Delta\mathsf N))-\mathcal L_{(\tau)}(\textbf x
(\tau,\mathsf N),\textbf x(\tau+\delta\tau,\mathsf N))\right)d\tau.\nn
\end{eqnarray}
Using the Taylor expansion with respect to the variable $\tau$, we have
\begin{eqnarray}\label{eq:4.26d}
\delta\mathcal S&=&\delta\tau\left[\mathcal L_{(\tau)}(\textbf x
(\tau,\mathsf N+\Delta\mathsf N),\textbf x(\tau+\delta\tau,\mathsf N+\Delta\mathsf N))-\mathcal L_{(\tau)}(\textbf x
(\tau,\mathsf N),\textbf x(\tau+\delta\tau,\mathsf N))\right.\nn\\
&&-\frac{\partial}{\partial\tau}\left.\mathcal L_{(\mathsf N)}(\textbf {x}(\tau,\mathsf N),\textbf x(\tau,
\mathsf N+\Delta\mathsf N))\right].
\end{eqnarray}
The term inside the bracket is required to be zero according to the condition $\delta\mathcal S=0$, yielding the semi-continuous 
closure relation.

\section{The full continuum limit}
\setcounter{equation}{0}
In the previous section, we took the continuum limit on the discrete variable $m$, leading to a system of
differential-difference equations. The full continuum limit, performed on the remaining discrete variable $\mathsf N$, will lead to a coupled system
of poles in the first instance, from which a hierarchy of ODEs can be retrieved, which is the CM hierarchy. How to perform this limit is inspired by the
structure of the solutions of \eqref{eq:4.3}. Performing the following computation, 
\begin{eqnarray}\label{eq:5.1}
&&(p+k)^{\mathsf N}e^{k\xi-\frac{\tau}{p+k}}=p^{\mathsf N}\exp\left\lbrace k\xi+\mathsf N \ln\left( 1+\frac{k}{p}\right)-
\frac{\tau/p}{1+\frac{k}{p}}  \right\rbrace,\nonumber\\
&=&p^{\mathsf N}\exp\left\lbrace k\xi+\mathsf N\left(\frac{k}{p}-\frac{k^2}{2p^2}+\frac{k^3}{3p^3}+...\right)-\frac{\tau}{p}
\left( 1-\frac{k}{p}+\frac{k^2}{p^2}-...\right) \right\rbrace \nonumber\\
&=&p^{\mathsf N}e^{-\frac{\tau}{p}}e^{kt_1+k^2t_2+k^3t_3+k^4t_4+...}\;,
\end{eqnarray}
we can identify the full continuum variables as the coefficients of the various powers of $k$, namely
\begin{eqnarray}\label{eq:5.2}
t_1=\xi+\frac{\tau}{p^2}+\eta, \mbox{~~} t_2=-\frac{\tau}{p^3}-\frac{\eta}{2p}, \mbox{~~} t_3=\frac{\tau}{p^4}+\frac{\eta}{3p^2},
\mbox{~~}.....\;\;,
\end{eqnarray}
where $\eta=\frac{\mathsf N}{p}$. This reduces the KP eigenfunction given in \eqref{eq:2.4} to the following form.
\begin{eqnarray}\label{eq:5.3}
\phi=\left( 1-\frac{1}{k}\sum\limits_{i=1}^N\frac{\mathsf{b}_i}{t_1-X_i}\right)p^{\mathsf N}
e^{-\frac{\tau}{p}}e^{kt_1+k^2t_2+k^3t_3+...},\;\;\mbox{where}\;\; \mathsf x_i=X_i-\frac{\tau}{p^2}-\eta\;.
\end{eqnarray}
The corresponding reduction to CM removes the centre of mass, which can be identified with the $t_1$ flow, from the system of equations.
Since $X_i=X_i(t_1, t_2, t_3,...)$ and we take the assumption that 
$\frac{\partial X_i}{\partial t_1}=0$, we find that
\begin{eqnarray}\label{eq:5.4}
\dot {\mathsf x}_i&=&\frac{\partial {\mathsf x}_i}{\partial \tau}=-\frac{1}{p^2}+\frac{\partial X_i}{\partial t_1}
\frac{\partial t_1}{\partial \tau}+\frac{\partial X_i}{\partial t_2}\frac{\partial t_2}{\partial \tau}+\frac{\partial X_i}{\partial t_3}
\frac{\partial t_3}{\partial \tau}+...\nonumber\\
&=&-\frac{1}{p^2}-\frac{1}{p^3}\frac{\partial X_i}{\partial t_2}+\frac{1}{p^4}\frac{\partial X_i}{\partial t_3}+...\;.
\end{eqnarray}
%
 We also find that
\begin{eqnarray}\label{eq:5.5}
\widetilde {\mathsf x}_i&=&\mathsf x_i-\frac{1}{p}-\frac{1}{2p^2}\frac{\partial X_i}{\partial t_2}+
\frac{1}{3p^3}\frac{\partial X_i}{\partial t_3}+\frac{1}{8p^4}\frac{\partial^2 X_i}{\partial t_2^2}\nonumber\\
&&-\frac{1}{6p^5}\frac{\partial^2 X_i}{\partial t_2\partial t_3}+\frac{1}{18p^6}\frac{\partial^2 X_i}{\partial t_3^2}+...\;,
\end{eqnarray}
and
\begin{eqnarray}\label{eq:5.6}
{\hypotilde 0 {\mathsf x}}_i&=&\mathsf x_i+\frac{1}{p}+\frac{1}{2p^2}\frac{\partial X_i}{\partial t_2}-
\frac{1}{3p^3}\frac{\partial X_i}{\partial t_3}+\frac{1}{8p^4}\frac{\partial^2 X_i}{\partial t_2^2}\nonumber\\
&&-\frac{1}{6p^5}\frac{\partial^2 X_i}{\partial t_2\partial t_3}+\frac{1}{18p^6}\frac{\partial^2 X_i}{\partial t_3^2}+...\;.
\end{eqnarray}
\textbf{The full limit on the solution}: Next, we would like to perform the full limit on the exact solution. We now rewrite \eqref{eq:w12}
in the form
\begin{eqnarray}\label{eq:w13}
 \boldsymbol{\mathsf {Y}}(\mathsf {N},\tau)&=&
\left(1+\frac{\boldsymbol{\Lambda}}{p}\right)^{-\mathsf{ N}}e^{\frac{\tau}{p}\left(1+\frac{\boldsymbol{\Lambda}
}{p}\right)^{-1}}\boldsymbol{\mathsf Y}(0,0)
e^{-\frac{\tau}{p}\left(1+\frac{\boldsymbol{\Lambda}}{p}\right)^{-1}}
\left(1+\frac{\boldsymbol{\Lambda}}{p}\right)^{\mathsf N}\nn\\
&&-\frac{\mathsf {N}}{p}\left(1+\frac{\boldsymbol{\boldsymbol{\Lambda}}}{p}\right)^{-1}-
\frac{\tau}{p}\left(1+\frac{\boldsymbol{\Lambda}}{p}\right)^{-2}\;\nn\\
&=&e^{-\mathsf{ N}\left(1+\frac{\boldsymbol{\Lambda}}{p}\right)+\frac{\tau}{p}
\left(1+\frac{\boldsymbol{\Lambda}}{p}\right)^{-1}}\boldsymbol{\mathsf Y}(0,0)
e^{\mathsf{ N}\left(1+\frac{\boldsymbol{\Lambda}}{p}\right)-\frac{\tau}{p}\left(1+\frac{\boldsymbol{\Lambda}}{p}\right)^{-1}}\nn\\
&&-\frac{\mathsf {N}}{p}\left(1+\frac{\boldsymbol{\Lambda}}{p}\right)^{-1}-\frac{\tau}{p^2}\left(1+\frac{\boldsymbol{\Lambda}}{p}\right)^{-2}\;.
\end{eqnarray}
Using the definitions of $t_1$, $t_2$, and $t_3$ in \eqref{eq:5.2}, we now have the solution in the form
\begin{eqnarray}\label{eq:w14}
 \boldsymbol{\mathsf {Y}}(t_1,t_2,t_3,...)&=&e^{-\boldsymbol{\Lambda}(t_1-\xi)+\boldsymbol{\Lambda}^2t_2-\boldsymbol{\Lambda}^3t_3+...}
\boldsymbol{\mathsf Y}(0,0,...)
e^{\boldsymbol{\Lambda}(t_1-\xi)-\boldsymbol{\Lambda}^2t_2+\boldsymbol{\Lambda}^3t_3+...}\nn\\
&&-(t_1-\xi)-2\boldsymbol{\Lambda}t_2-3\boldsymbol{\Lambda}^2t_3+....\;\;,
\end{eqnarray}
which is a function of time variables $(t_1,t_2,t_3,..)$. The positions of the particles $X_i(t_1,t_2,t_3,...)$ can be computed by
looking for the eigenvalues of \eqref{eq:w14}. Since, the term $t_1-\xi$ represents the centre of mass flow, the explicit 
expression of the solution for the CM can be obtained from the secular problem for the matrix
\begin{equation}\label{eq:exact}
 \boldsymbol{ X}(0,0)-2\boldsymbol{ L}(0,0)t_2-3\boldsymbol{L}^2(0,0)t_3,
\end{equation}
where $\boldsymbol{ X}(0,0)=\boldsymbol{ U}^{-1}(0,0)\boldsymbol{ Y}(0,0)\boldsymbol{ U}(0,0)$ and 
$\boldsymbol{ L}(0,0)=\boldsymbol{ U}^{-1}(0,0)\boldsymbol{\Lambda}\boldsymbol{ U}(0,0)$. We note that \eqref{eq:exact} is identical to
what was obtained in \cite{OP} up to the time variable $t_2$ (involving a scaling of the variable). The solution \eqref{eq:w14} involves $N$-time flows for 
the CM system. The next solutions in the hierarchy can be generated by pushing further on with the expansion.
\\
\\
\textbf{\textsf{The full limit on the equations of motion}}: We now would like to see what would result from 
taking the limit on the equations of motion. We know that the combination of eqs. (\ref{eq:4.11}) and (\ref{eq:4.16}) gives us the equations of motion
in the skew limit case. We look at the full limit for each of these. 

Using eqs. (\ref{eq:5.4}), (\ref{eq:5.5}) and (\ref{eq:5.6}),
we find that eq. (\ref{eq:4.11}) gives
\begin{subequations}\label{eq:fullcon1}
 \begin{eqnarray}
  \mathcal{O}\left(\frac{1}{p^2}\right)&:&\frac{1}{4}\left(\frac{\partial X_i}{\partial t_2}\right)^2+\frac{1}{3}\frac{\partial X_i}{\partial t_3}
-\sum\limits_{\mathop {j = 1}\limits_{j \ne i} }^N\frac{1}{(X_i-X_j)^2}=0\;,\\
\mathcal{O}\left(\frac{1}{p^3}\right)&:&\frac{1}{4}\frac{\partial^2 X_i}{\partial t_2^2}-\frac{2}{3}\frac{\partial X_i}{\partial t_2}\frac{\partial X_i}{\partial t_3}
-\frac{1}{4}\left(\frac{\partial X_i}{\partial t_2}\right)^3-\frac{1}{2}\frac{\partial X_i}{\partial t_4}\nn\\
&&\;-\sum\limits_{\mathop {j = 1}\limits_{j \ne i} }^N\left( \frac{\frac{\partial X_j}{\partial t_2}}{(X_i-X_j)^2}
-\frac{2}{(X_i-X_j)^3}\right)=0\;,\\
\mathcal{O}\left(\frac{1}{p^4}\right)&:&\frac{1}{3}\left(\frac{\partial X_i}{\partial t_3}\right)^3-\frac{1}{3}\frac{\partial^2 X_i}{\partial t_2\partial t_3}
+\frac{3}{4}\left(\frac{\partial X_i}{\partial t_2}\right)^2\frac{\partial X_i}{\partial t_3}-\frac{1}{8}\frac{\partial X_i}{\partial t_2}\frac{\partial^2 X_i}{\partial t_2^2}\nn\\
&&+\frac{1}{2}\left(\frac{\partial X_i}{\partial t_2}\right)^4+\frac{3}{4}\frac{\partial X_i}{\partial t_2}\frac{\partial X_i}{\partial t_4}
-\sum\limits_{\mathop {j = 1}\limits_{j \ne i} }^N\left(\frac{3}{(X_i-X_j)^4} \right.\nn\\
&&\left.-\frac{\frac{\partial X_j}{\partial t_3}}{(X_i-X_j)^2}-\frac{2\frac{\partial X_j}{\partial t_2}+\frac{\partial X_i}{\partial t_2}}{(X_i-X_j)^3}\right)=0\;,
 \end{eqnarray}
\end{subequations}
and (\ref{eq:4.16}) yields 
\begin{subequations}\label{eq:fullcon2}
 \begin{eqnarray}
  \mathcal{O}\left(\frac{1}{p^2}\right)&:&\frac{1}{4}\left(\frac{\partial X_i}{\partial t_2}\right)^2+\frac{1}{3}\frac{\partial X_i}{\partial t_3}
-\sum\limits_{\mathop {j = 1}\limits_{j \ne i} }^N\frac{1}{(X_i-X_j)^2}=0\;,\\
\mathcal{O}\left(\frac{1}{p^3}\right)&:&-\frac{1}{4}\frac{\partial^2 X_i}{\partial t_2^2}-\frac{2}{3}\frac{\partial X_i}{\partial t_2}\frac{\partial X_i}{\partial t_3}
-\frac{1}{4}\left(\frac{\partial X_i}{\partial t_2}\right)^3-\frac{1}{2}\frac{\partial X_i}{\partial t_4}\nn\\
&&\;-\sum\limits_{\mathop {j = 1}\limits_{j \ne i} }^N\left( \frac{\frac{\partial X_j}{\partial t_2}}{(X_i-X_j)^2}
+\frac{2}{(X_i-X_j)^3}\right)=0\;,\\
\mathcal{O}\left(\frac{1}{p^4}\right)&:&\frac{1}{3}\left(\frac{\partial X_i}{\partial t_3}\right)^3+\frac{1}{3}\frac{\partial^2 X_i}{\partial t_2\partial t_3}
+\frac{3}{4}\left(\frac{\partial X_i}{\partial t_2}\right)^2\frac{\partial X_i}{\partial t_3}+\frac{1}{8}\frac{\partial X_i}{\partial t_2}\frac{\partial^2 X_i}{\partial t_2^2}\nn\\
&&+\frac{1}{2}\left(\frac{\partial X_i}{\partial t_2}\right)^4+\frac{3}{4}\frac{\partial X_i}{\partial t_2}\frac{\partial X_i}{\partial t_4}
-\sum\limits_{\mathop {j = 1}\limits_{j \ne i} }^N\left(\frac{3}{(X_i-X_j)^4} \right.\nn\\
&&\left.-\frac{\frac{\partial X_j}{\partial t_3}}{(X_i-X_j)^2}+\frac{2\frac{\partial X_j}{\partial t_2}+\frac{\partial X_i}{\partial t_2}}{(X_i-X_j)^3}\right)=0\;.
 \end{eqnarray}
\end{subequations}
The difference between \eqref{eq:fullcon1} and \eqref{eq:fullcon2} gives the following:
\\
\\
$\blacklozenge{}$The leading term of order $\mathcal O(1/p^3)$ gives us
\begin{eqnarray}\label{eq:5.8}
\frac{\partial^2X_i}{\partial t_2^2}=-\sum\limits_{\mathop {j = 1}\limits_{j \ne i} }^N\frac{8}{(X_i-X_j)^3}\;,
\end{eqnarray}
which is the equation of the motion for the continuous CM system \cite{F2}.
\\
\\
$\blacklozenge{}$The term of order $\mathcal O(1/p^4)$ gives us
\begin{eqnarray}\label{eq:5.8a}
\frac{2}{3}\frac{\partial^2X_i}{\partial t_2 \partial t_3}+\frac{1}{4}\frac{\partial X_i}{\partial t_2}\frac{\partial^2X_i}{\partial t_2^2}
=\sum\limits_{\mathop {j = 1}\limits_{j \ne i} }^N\frac{2\frac{\partial X_i}{\partial t_2}+4\frac{\partial X_j}{\partial t_2}}{(X_i-X_j)^3}\;.
\end{eqnarray}
We now use eq. (\ref{eq:5.8}) to simplify eq. (\ref{eq:5.8a}), and we get
\begin{eqnarray}\label{eq:5.8b}
\frac{\partial^2X_i}{\partial t_2 \partial t_3}=6\sum\limits_{\mathop {j = 1}\limits_{j \ne i} }^N\frac{\frac{\partial X_i}{\partial t_2}+
\frac{\partial X_j}{\partial t_2}}{(X_i-X_j)^3}\;.
\end{eqnarray}
This equation represents the next equation of motion beyond the usual continuous CM in the hierarchy.
We will stop at this equation, but we can actually get the higher terms of the equation in which the variable $t_4$ and higher 
order time-flows must be taken into account.
\\
\\
Note finally that summing \eqref{eq:fullcon1} and \eqref{eq:fullcon2} gives the following equations
\begin{subequations}\label{eq:fullcon3}
 \begin{eqnarray}
  \mathcal{O}\left(\frac{1}{p^2}\right)&:&\frac{1}{4}\left(\frac{\partial X_i}{\partial t_2}\right)^2+\frac{1}{3}\frac{\partial X_i}{\partial t_3}
-\sum\limits_{\mathop {j = 1}\limits_{j \ne i} }^N\frac{1}{(X_i-X_j)^2}=0\;,\label{eq:fullcon3a}\\
\mathcal{O}\left(\frac{1}{p^3}\right)&:&-\frac{2}{3}\frac{\partial X_i}{\partial t_2}\frac{\partial X_i}{\partial t_3}
-\frac{1}{4}\left(\frac{\partial X_i}{\partial t_2}\right)^3-\frac{1}{2}\frac{\partial X_i}{\partial t_4}
-\sum\limits_{\mathop {j = 1}\limits_{j \ne i} }^N \frac{\frac{\partial X_j}{\partial t_2}}{(X_i-X_j)^2}=0\;,\\
\mathcal{O}\left(\frac{1}{p^4}\right)&:&\frac{1}{3}\left(\frac{\partial X_i}{\partial t_3}\right)^3
+\frac{3}{4}\left(\frac{\partial X_i}{\partial t_2}\right)^2\frac{\partial X_i}{\partial t_3}
+\frac{1}{2}\left(\frac{\partial X_i}{\partial t_2}\right)^4+\frac{3}{4}\frac{\partial X_i}{\partial t_2}\frac{\partial X_i}{\partial t_4}\nn\\
&&-\sum\limits_{\mathop {j = 1}\limits_{j \ne i} }^N\left(\frac{3}{(X_i-X_j)^4} 
-\frac{\frac{\partial X_j}{\partial t_3}}{(X_i-X_j)^2}\right)=0\;.
 \end{eqnarray}
\end{subequations}
These equations are the constraints for the full limit case which will play an important role in the next section. The occurrence of these constraints
seems to be related to the superintegrability of the CM system, cf. \cite{SW}, and their validity can be checked on solutions.
\section{The continuous Lagrangian 1-form and its closure relation}
In the previous section, we obtained directly the equations of motion of the continuous CM from the full limit.
In this section, we obtain the continuous Lagrangians from the full limit as well. We will restrict ourselves to the case of 3-particles involving time flows $t_2$ and $t_3$,
to establish the simplest example of a Lagrangian 1-form for the CM system.
Both the Lagrangians $\mathcal{L}_{(\tau)}$ and $\mathcal{L}_{(\mathsf N)}$ produce the hierarchy of the continuous CM Lagrangians corresponding to the equations of
motion in the previous section. Rather than performing the full limit on the individual Lagrangians, we prefer to take the limit on the action itself. 
\\
\\
\textbf{The full limit on the action}: In Section 5, we performed the skew limit on the discrete action. In the present section, we will continue to perform
the full continuum limit on the semi-continuous action. We now take the action to be of the form
\begin{equation}
 \mathcal{S}[\boldsymbol{x}(\mathsf N,\tau);\Gamma]=\int_{\tau_1}^{\tau_2}d\tau\mathcal{L}_{(\tau)}(\boldsymbol{x}(\mathsf N,\tau),\dot{\boldsymbol{x}}(\mathsf N,\tau))
+\sum_{\mathsf N}\mathcal{L}_{(\mathsf N)}(\boldsymbol{x}(\mathsf N, \tau),\boldsymbol{x}(\mathsf N+1,\tau))\;,
\end{equation}
where the first term belongs to the vertical part and the second term belongs to the horizontal part of the curve given in Fig. (\ref{semi-curve44}).

Using anti-Taylor expansion, the action now becomes
\begin{equation}
 \mathcal{S}[\boldsymbol{x}(\mathsf N,\tau);\Gamma]=\int_{\tau_1}^{\tau_2}d\tau\mathcal{L}_{(\tau)}(\boldsymbol{x}(\mathsf N,\tau),\dot{\boldsymbol{x}}(\mathsf N,\tau))
+p\int_{\eta_1}^{\eta_2}d\eta\mathcal{L}_{(\mathsf N)}(\boldsymbol{x}(\mathsf N, \tau),\boldsymbol{x}(\mathsf N+1,\tau))\;,
\end{equation}
where we do not need to take into account the boundary terms coming from the expansion, because they are constant at the end points and do not contribute 
to the variational process.
\\
We now perform a change of variables $(\tau,\eta)\mapsto (t_2,t_3)$ by using \eqref{eq:5.2} 
\begin{subequations}
\begin{eqnarray}
d\tau&=&2p^3dt_2+3p^4dt_3\;,\\
d\eta&=&-6pdt_2-6p^2dt_3\;,
\end{eqnarray}
\end{subequations}
and also expand the Lagrangians with respect to variable $p$. We obtain 
\begin{eqnarray}\label{eq:5.19}
 \mathcal{S}[\boldsymbol{x}(t_2,t_3);\Gamma]&=&\int_{t_2(1)}^{t_2(2)}dt_2\mathcal{L}_{(t_2)}\left(\boldsymbol{x}(t_2,t_3),
\frac{\partial\boldsymbol{x}(t_2,t_3)}{\partial t_2}\right)\nn\\
&&+\int_{t_3(1)}^{t_3(2)}dt_3\mathcal{L}_{(t_3)}\left(\boldsymbol{x}(t_2,t_3),\frac{\partial\boldsymbol{x}(t_2,t_3)}{\partial t_2},
\frac{\partial\boldsymbol{x}(t_2,t_3)}{\partial t_3}\right)\;,
\end{eqnarray}
where $\mathcal{L}_{(t_2)}$ and $\mathcal{L}_{(t_3)}$ are given by
\begin{eqnarray}\label{eq:5.10}
\mathcal L_{(t_2)}\equiv \mathcal L_{CM}=\sum\limits_{i=1}^N\frac{1}{2}\left( \frac{\partial X_i}{\partial t_2}\right)^2
+\sum\limits_{i \ne j}^N\frac{2}{(X_i-X_j)^2}\;, 
\end{eqnarray}
of which the Euler-Lagrange equation 
\begin{eqnarray}\label{eq:5.10c}
\frac{\partial \mathcal L_{(t_2)}}{\partial X_i}-\frac{\partial}{\partial t_2}\left( \frac{\partial \mathcal
 L_{(t_2)}}{\partial(\frac{\partial X_i}{\partial t_2})}\right)=0\;, 
\end{eqnarray}
gives exactly eq. (\ref{eq:5.8}) and
\begin{eqnarray}\label{eq:5.10a}
\mathcal L_{(t_3)}=\sum\limits_{i=1}^N\left( \frac{\partial X_i}{\partial t_2}\frac{\partial X_i}{\partial t_3}+\frac{1}{4}
\left( \frac{\partial X_i}
{\partial t_2}\right)^3\right)-\sum\limits_{i \ne j}^N\frac{3\frac{\partial X_i}{\partial t_2}}
{(X_i-X_j)^2}\;. 
\end{eqnarray}
We see that the Lagrangian $\mathcal L_{(t_3)}$ contains derivatives with respect to two time flows $t_2$ and $t_3$. 
We observe that the equation of motion (\ref{eq:5.8b}) requires the Euler-Lagrange equation in the form
\begin{eqnarray}\label{eq:5.10d}
\frac{\partial \mathcal L_{(t_3)}}{\partial X_i}-\frac{\partial}{\partial t_3}\left( \frac{\partial \mathcal 
L_{(t_3)}}{\partial(\frac{\partial X_i}{\partial t_3})}\right)=0\;. 
\end{eqnarray}
Furthermore, we find that
\begin{eqnarray}\label{eq:5.10s}
\frac{\partial \mathcal L_{(t_3)}}{\partial(\frac{\partial X_i}{\partial t_2})}
=\frac{\partial X_i}{\partial t_3}+\frac{3}{4}\left(\frac{\partial{X}_i}{\partial t_2}\right)^2-\sum_{j \neq i}^N\frac{3}{(X_i-X_j)^2}=0\;,
\end{eqnarray} 
which is identical to eq. \eqref{eq:fullcon3a}. 

Here we obtained the hierarchy of Lagrangians for the CM-model through the full continuum limit. Obviously, higher Lagrangians in the family
can be generated by pushing further on with the expansion.
\\
\\
\textbf{The full limit on the closure relation}: Using the results regarding the limits of the Lagrangians $\mathcal{L}_{(\tau)}$ and $\mathcal{L}_{(\mathsf N)}$ in the
previous section, and also the relations of eq. (\ref{eq:5.2}), we can write
\begin{equation}\label{eq:5.14}
\frac{\partial\mathcal {L}_{(\mathsf N)}}{\partial \tau}\mapsto -\frac{1}{p^3}
\frac{\partial \mathcal L_{(t_2,t_3,...)}}{\partial t_2}+\frac{1}{p^4}\frac{\partial \mathcal L_{(t_2,t_3,...)}}{\partial t_
3}.
\end{equation}
Substituting eq. (\ref{eq:5.14}) into the closure relation eq. (\ref{eq:4.25}) together with the full limit of 
${\hypotilde 0 {\mathcal {L}}}_{(\tau)}-\mathcal L_{(\tau)}$, we find that the leading term is of order $\mathcal {O}(1/p^3)$, and is
\begin{equation}\label{eq:5.15}
\frac{\partial \mathcal L_{(t_3)}}{\partial t_2}=\frac{\partial \mathcal L_{(t_2)}}{\partial t_3}.
\end{equation}
This equation represents the closure relation for the continuous case of the Lagrangian $1$-form. We are now taking into account 
not only the usual CM Lagrangian, but also a higher-order Lagrangian; these have up until now not been considered. 
The verification of the closure relation \eqref{eq:5.15} is presented in Appendix {\ref{proof}}.
\\
\\

\section{The variational principle for forms}
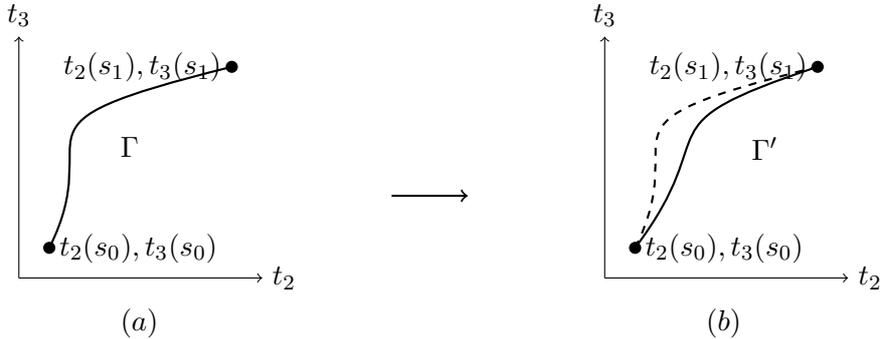
\begin{figure}[h]
\begin{center}
\begin{tikzpicture}[scale=0.4]
 \draw[->] (0,0) -- (8,0) node[anchor=west] {$t_{2}$};
 \draw[->] (0,0) -- (0,8) node[anchor=south] {$t_{3}$};
 \fill (1,1) circle (0.2) node[anchor=west]{$t_2(s_0),t_3(s_0)$};
 \draw (3,5) node[anchor=north west] {$\Gamma$};
 \fill (7,7) circle (0.2) node[anchor=east]{$t_2(s_1),t_3(s_1)$} ;
\draw [thick] (1,1) .. controls (3,5) and (-1,5) .. (7,7);
 \draw (3,-1.5) node[anchor=west] {$(a)$};
\end{tikzpicture}
\begin{tikzpicture}[scale=0.5]
 \draw[white] (0,0) -- (6.5,0);
 \draw[white] (0,0) -- (0,6);
 \draw[->,thick] (2,4) -- (4,4);
\end{tikzpicture}
\begin{tikzpicture}[scale=0.4]
 \draw[->] (0,0) -- (8,0) node[anchor=west] {$t_{2}$};
 \draw[->] (0,0) -- (0,8) node[anchor=south] {$t_{3}$};
 \fill (1,1) circle (0.2)node[anchor=west]{$t_2(s_0),t_3(s_0)$};
 \draw (4.5,5) node[anchor=north west] {$\Gamma^\prime$};
 \fill (7,7) circle (0.2) node[anchor=east]{$t_2(s_1),t_3(s_1)$};
\draw [thick,dashed] (1,1) .. controls (3,5) and (-1,5) .. (7,7);
\draw [thick] (1,1) .. controls (4,5) and (1,5) .. (7,7);
 \draw (3,-1.5) node[anchor=west] {$(b)$};
\end{tikzpicture}
\end{center}
\caption{The deformation of the continuous curve $\Gamma$ .}\label{con-curve_deformation}
\end{figure}

In the previous section, we obtained the Lagrangian hierarchy of the CM system along with the corresponding Euler-Lagrange
equations. In this section, we are interested in investigating the variational principle for continuous
Lagrangian $1$-forms. Here we limit ourselves to the first two time flows in the hierarchy, $t_2$ and $t_3$, corresponding to
a 3-particle system in order to exhibit the simplest example. 
Let us define the time variables to be functions of a new variable $s$, so that $t_i=t_i(s)$, 
associated to the curve 
$\Gamma:(t_2(s),t_3(s))$ as in
Fig. (\ref{con-curve_deformation}a). We now
define the Lagrangians to be
\begin{subequations}
\begin{eqnarray}
\mathcal{L}_{(t_2)}&=&\mathcal{L}_{(t_2)}(\textbf{x}(t_2,t_3),\textbf x_{t_2}(t_2,t_3),\textbf x_{t_3}(t_2,t_3))\;,\\
\mathcal{L}_{(t_3)}&=&\mathcal{L}_{(t_3)}(\textbf x(t_2,t_3),\textbf x_{t_2}(t_2,t_3),\textbf x_{t_3}(t_2,t_3))\;,
\end{eqnarray}
\end{subequations}
where
\[
 \textbf{x}_{t_2}(t_2,t_3)=\frac{\partial \textbf{x}}{\partial t_2}\;\;\mbox{and}\;\; \textbf{x}_{t_3}(t_2,t_3)=\frac{\partial \textbf{x}}{\partial t_3}\;.
\]
\\
The action evaluated on the curve $\Gamma$ can be defined as
\begin{eqnarray}
\mathcal{S}(\textbf x(t_2,t_3);\Gamma)&=&\int_{\Gamma}\left( \mathcal{L}_{(t_2)}dt_1+\mathcal{L}_{(t_3)} dt_2\right)\;,\nonumber\\
&=&\int_{s_0}^{s_1}\left( \mathcal{L}_{(t_2)}(t_2(s),t_3(s))\frac{dt_2}{ds}+\mathcal{L}_{(t_3)}(t_2(s),t_3(s))\frac{dt_3}{ds}\right)ds.
\end{eqnarray}
%

We now perform variation on the curve of time variables $t_2$ and $t_3$ in the following way. Suppose we have an action $\mathcal S$ defined on a 
curve $\Gamma$, 
and the curve is deformed $t_2 \mapsto t_2+\delta t_2$ and $t_3 \mapsto t_3+\delta t_3 $ in Fig. (\ref{con-curve_deformation}b),
keeping the end points fixed, so we have the conditions
\begin{eqnarray}
\delta t_2(s_0)=\delta t_2(s_1)=0,\;\;\;\delta t_3(s_0)=\delta t_3(s_1)=0.
\end{eqnarray}
The action $\mathcal S^{\prime}$ is defined as
\begin{eqnarray}\label{eq:5.20}
\mathcal{S}^{\prime}\equiv\mathcal{S}(\textbf x(t_2+\delta t_2,t_3+\delta t_3);\Gamma^\prime)
\end{eqnarray}
Extremizing the action, we have $\delta\mathcal{S}=0$, where $\delta\mathcal{S}$ is defined by
\begin{eqnarray}\label{eq:5.21}
\delta\mathcal{S}=\mathcal{S}(\textbf x(t_2+\delta t_2,t_3+\delta t_3);\Gamma^\prime)-\mathcal{S}(\textbf x(t_2,t_3);\Gamma).
\end{eqnarray}
Using the Taylor expansion, we obtain
\begin{eqnarray}\label{eq:5.22}
\delta\mathcal{S}&=&\int_{s_0}^{s_1}ds\left[\left(\frac{\partial \mathcal L_{(t_2)}}{\partial t_2}\delta t_2 +\frac{\partial \mathcal L_{(t_2)}}
{\partial t_3}\delta t_3\right)
\frac{dt_2}{ds}+\mathcal L_{(t_2)}\frac{d\delta t_2}{ds}\right.\nonumber\\
&&+\left.\left(\frac{\partial \mathcal L_{(t_3)}}{\partial t_2}\delta t_2 +\frac{\partial\mathcal L_{(t_3)}}{\partial t_3}\delta t_3\right)
\frac{dt_3}{ds}+\mathcal L_{(t_3)}\frac{d\delta t_3}{ds}
\right].\nonumber\\
&=&\mathcal L_{(t_2)}\delta t_2(s_0)-\mathcal L_{(t_2)}\delta t_2(s_1)+\mathcal L_{(t_3)}\delta t_3(s_0)-\mathcal L_{(t_3)}\delta t_3(s_1)\nonumber\\
&&+\int_{s_0}^{s_1}ds\left[\left(\frac{\partial\mathcal L_{(t_2)}}{\partial t_2}\delta t_2 +\frac{\partial\mathcal  L_{(t_2)}}{\partial t_3}\delta t_3\right)
\frac{dt_2}{ds}-\frac{d\mathcal L_{(t_2)}}{ds}\delta t_2\right.\nonumber\\
&&+\left.\left(\frac{\partial \mathcal L_{(t_3)}}{\partial t_2}\delta t_2 +\frac{\partial \mathcal L_{(t_3)}}{\partial t_3}\delta t_3\right)\frac{dt_3}{ds}
-\frac{d\mathcal L_{(t_3)}}{ds}\delta t_3\right].
\end{eqnarray}
Using the fact that the end points are fixed, the first line of eq. \eqref{eq:5.22} is zero. We can also use 
\begin{subequations}
\begin{eqnarray}\label{eq:5.23}
\frac{d\mathcal L_{(t_2)}}{ds}&=&\frac{\partial \mathcal L_{(t_2)}}{\partial t_2}\frac{dt_2}{ds}+\frac{\partial \mathcal L_{(t_2)}}{\partial t_3}\frac{dt_3}{ds}\;,\\
\frac{d\mathcal L_{(t_3)}}{ds}&=&\frac{\partial\mathcal L_{(t_3)}}{\partial t_2}\frac{dt_2}{ds}+\frac{\partial \mathcal L_{(t_3)}}{\partial t_3}\frac{dt_3}{ds}\;,
\end{eqnarray}
\end{subequations}
to obtain
\begin{eqnarray}\label{eq:5.24}
\delta\mathcal{S}&=&\int_{s_0}^{s_1}ds\left[\delta t_2\left(\frac{\partial\mathcal L_{(t_2)}}{\partial t_3} -
\frac{\partial\mathcal L_{(t_3)}}{\partial t_2}
\right)\frac{dt_2}{ds}
+\delta t_3\left(\frac{\partial\mathcal L_{(t_3)}}{\partial t_2} -\frac{\partial \mathcal L_{(t_2)}}{\partial t_3}
\right)\frac{dt_2}{ds}\right]\;.
\end{eqnarray}
Here we must have two copies of the closure relation 
\begin{eqnarray}\label{eq:5.24ss}
 \frac{\partial\mathcal L_{(t_3)}}{\partial t_2} =\frac{\partial \mathcal L_{(t_2)}}{\partial t_3}\;,
\end{eqnarray}
in order that $\delta \mathcal S=0$. The closure relation \eqref{eq:5.24ss} guarantees that any such continuous curve may be locally deformed without
 changing the value of the action functional corresponding
to the time variables $t_2$ and $t_3$. 
\\

\begin{figure}[h]
\begin{center}
\begin{tikzpicture}[scale=0.4]
 \draw[->] (0,0,0) -- (9,0,0) node[anchor=west] {$t_{2}$};
 \draw[->] (0,0,0) -- (0,7,0) node[anchor=south] {$\textbf x$};
 \draw[->] (0,0,0) -- (0,0,9) node[anchor=south] {$t_{3}$};
 \fill (1,0,1) circle (0.2) node[anchor=west]{$t_2(s_0),t_3(s_0)$};
 \draw (3,0,3) node[anchor=north west] {$\Gamma$};
 \fill (7,0,7) circle (0.2) node[anchor=west]{$t_2(s_1),t_3(s_1)$} ;
\draw[dashed] (1,0,1)--(1,5,1);
\draw[dashed] (7,0,7)--(7,5,7);
\fill (1,5,1) circle (0.2) node[anchor=west]{$\textbf x(t_2(s_0),t_3(s_0))$};
\fill (7,5,7) circle (0.2) node[anchor=west]{$\textbf x(t_2(s_1),t_3(s_1))$};
 \draw (3,5,3) node[anchor=north west] {$\mathcal{E}_{\Gamma}$};
 \draw (0,-5,-1.5) node[anchor=west] {$(a)$};
\draw [thick] (1,0,1) .. controls (4,0,5) and (1,0,5) .. (7,0,7);
\draw [thick] (1,5,1) .. controls (4,4,5) and (2,4,5) .. (7,5,7);
\end{tikzpicture}
\begin{tikzpicture}[scale=0.4]
 \draw[->] (0,0,0) -- (9,0,0) node[anchor=west] {$t_{2}$};
 \draw[->] (0,0,0) -- (0,7,0) node[anchor=south] {$\textbf x$};
 \draw[->] (0,0,0) -- (0,0,9) node[anchor=south] {$t_{3}$};
 \fill (1,0,1) circle (0.2) node[anchor=west]{$t_2(s_0),t_3(s_0)$};
 \draw (3,0,3) node[anchor=north west] {$\Gamma$};
 \fill (7,0,7) circle (0.2) node[anchor=west]{$t_2(s_1),t_3(s_1)$} ;
\draw[dashed] (1,0,1)--(1,5,1);
\draw[dashed] (7,0,7)--(7,5,7);
\fill (1,5,1) circle (0.2) node[anchor=west]{$\textbf x(t_2(s_0),t_3(s_0))$};
\fill (7,5,7) circle (0.2) node[anchor=west]{$\textbf x(t_2(s_1),t_3(s_1))$};
 \draw (3,5.5,3) node[anchor=north west] {$\mathcal{E}_{\Gamma}^\prime$};
 \draw (0,-5,-1.5) node[anchor=west] {$(b)$};
\draw [thick] (1,0,1) .. controls (4,0,5) and (1,0,5) .. (7,0,7);
\draw [thick,dashed] (1,5,1) .. controls (4,4,5) and (2,4,5) .. (7,5,7);
\draw [thick] (1,5,1) .. controls (4,6,5) and (2,5.5,5) .. (7,5,7);
\end{tikzpicture}
\end{center}
\caption{The curves $\Gamma$ and $\mathcal{E}_{\Gamma}$ in the $\textbf x-\textbf t$ configuration.}\label{x-curve}
\end{figure}
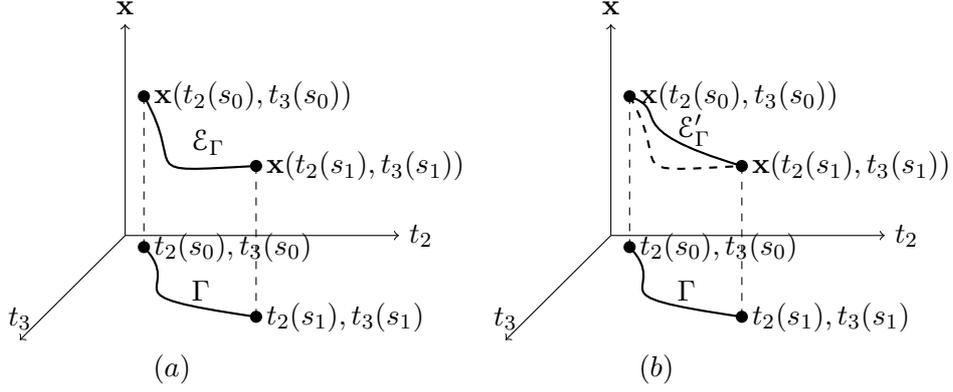

Next, we would like to perform a variation on the field variables $\textbf x$. We now fix the curve $\Gamma$ on the $t_2-t_3$ plane and examining
the corresponding curve $\mathcal{E}_{\Gamma}$ of the variables $\textbf x$ living in the $\textbf x-t_2-t_3$ space as in Fig. (\ref{x-curve}a).
%
The action can be expressed in the form
%
\begin{eqnarray}\label{eq:5.28}
&&\mathcal{S}(\textbf x,\textbf y_2,\textbf y_3;\mathcal{E}_{\Gamma})\nn\\
&=&\int_{\Gamma}\left(\mathcal{L}_{(t_3)}(\textbf x(t_2,t_3),\textbf x_{t_2}(t_2,t_3),\textbf x_{t_3}(t_2,t_3))dt_3
+\mathcal{L}_{(t_2)}\left(\textbf{x}(t_2,t_3),\textbf x_{t_2}(t_2,t_3),\textbf x_{t_3}(t_2,t_3)\right)dt_2\right).\nonumber\\
&=&\int_{s_0}^{s_1}ds\left(\mathcal{L}_{(t_3)}(\textbf x(t_2(s),t_3(s)),\textbf x_{t_2}(t_2(s),t_3(s)),\textbf x_{t_3}(t_2(s),t_3(s)))\frac{dt_3}{ds}\right.\nn\\
&&\;\;\;\;\;\;\;\;\;\;\;\;\;\;\;\;\;+\left.\mathcal{L}_{(t_2)}\left(\textbf{x}(t_2(s),t_3(s)),\textbf x_{t_2}(t_2(s),t_3(s)),\textbf x_{t_3}(t_2(s),t_3(s)\right)\frac{dt_2}{ds}\right)\;.
\end{eqnarray}
%
%
We now introduce 
\begin{subequations}\label{x-y}
 \begin{eqnarray}
  \frac{d}{d s}(\delta \textbf{x})&=&\delta \textbf{x}_{t_2}\frac{dt_2}{ds}+\delta \textbf{x}_{t_3}\frac{dt_3}{ds}\;,\\
  \delta \textbf{y}&=&\delta \textbf{x}_{t_2}\frac{dt_2}{ds}-\delta \textbf{x}_{t_3}\frac{dt_3}{ds}\;,
 \end{eqnarray}
\end{subequations}
where $\textbf y$ is a new independent variable, and the combination of these two equations yields
\begin{subequations}\label{comxy}
 \begin{eqnarray}
  \frac{1}{2}\left(\frac{d}{d s}(\delta \textbf{x})+\delta \textbf{y}\right)&=&\delta \textbf{x}_{t_2}\frac{dt_2}{ds}\;,\\
   \frac{1}{2}\left(\frac{d}{d s}(\delta \textbf{x})-\delta \textbf{y}\right)&=&\delta \textbf{x}_{t_3}\frac{dt_3}{ds}\;.
 \end{eqnarray}
\end{subequations}
Performing the variation on the action, we have 
\begin{eqnarray}\label{eq:5.31}
\delta \mathcal S&=&\int_{s_0}^{s_1}ds\left(\left[ \delta\textbf x \frac{\partial \mathcal{L}_{(t_3)}}{\partial\textbf x}
+\delta\textbf{x}_{t_3}\frac{\partial \mathcal{L}_{(t_3)}}{\partial \textbf x_{t_3}}
+\delta \textbf{x}_{t_2}\frac{\partial \mathcal{L}_{(t_3)}}{\partial \textbf{x}_{t_2}}\right]\frac{dt_3}{ds}\right.\nonumber\\
&&+\left.\left[ \delta\textbf x \frac{\partial \mathcal{L}_{(t_2)}}{\partial\textbf x}
+\delta\textbf{x}_{t_2}\frac{\partial \mathcal{L}_{(t_2)}}{\partial \textbf x_{t_2}}
+\delta \textbf{x}_{t_3}\frac{\partial \mathcal{L}_{(t_2)}}{\partial \textbf{x}_{t_3}}\right]\frac{dt_2}{ds}\right)\;.\nonumber\\
\end{eqnarray}
Using \eqref{comxy}, \eqref{eq:5.31} becomes
\begin{eqnarray}\label{eq:5.31as}
\delta \mathcal S&=&\int_{s_0}^{s_1}ds\left(\left[  \frac{\partial \mathcal{L}_{(t_2)}}{\partial\textbf x}\frac{dt_2}{ds}
+\frac{\partial \mathcal{L}_{(t_3)}}{\partial\textbf x}\frac{dt_3}{ds}
\right]\delta\textbf x\right.\nonumber\\
&&+\frac{1}{2}\frac{d}{ds}(\delta\textbf x )\left.\left[ \frac{\partial \mathcal{L}_{(t_2)}}{\partial\textbf x_{t_2}}
+\frac{\partial \mathcal{L}_{(t_3)}}{\partial\textbf x_{t_3}}
+\frac{\partial t_2/\partial s}{\partial t_3/\partial s}\frac{\partial \mathcal{L}_{(t_2)}}{\partial \textbf x_{t_3}}
+\frac{\partial t_3/\partial s}{\partial t_2/\partial s}\frac{\partial \mathcal{L}_{(t_3)}}{\partial \textbf x_{t_2}}\right]\right.\nonumber\\
&&+\frac{1}{2}\delta\textbf y \left.\left[ \frac{\partial \mathcal{L}_{(t_2)}}{\partial\textbf x_{t_2}}
-\frac{\partial \mathcal{L}_{(t_3)}}{\partial\textbf x_{t_3}}
-\frac{\partial t_2/\partial s}{\partial t_3/\partial s}\frac{\partial \mathcal{L}_{(t_2)}}{\partial \textbf x_{t_3}}
+\frac{\partial t_3/\partial s}{\partial t_2/\partial s}\frac{\partial \mathcal{L}_{(t_3)}}{\partial \textbf x_{t_2}}\right]\right)\;.
\end{eqnarray}
Integrating by parts the second line in \eqref{eq:5.31as} with the boundary conditions
\[
\delta \textbf x(t_2(s_0))=\delta \textbf x(t_3(s_0))=0\;,\;\;\delta \textbf x(t_2(s_1))=\delta \textbf x(t_3(s_1))=0\;,
\]
we obtain
\begin{eqnarray}\label{eq:5.31qw}
\delta \mathcal S&=&\int_{s_0}^{s_1}ds\left(\delta\textbf x\left[  \frac{\partial \mathcal{L}_{(t_2)}}{\partial\textbf x}\frac{dt_2}{ds}
+\frac{\partial \mathcal{L}_{(t_3)}}{\partial\textbf x}\frac{dt_3}{ds}
\right.\right.\nonumber\\
&&-\frac{1}{2}\frac{d}{ds}\left.\left( \frac{\partial \mathcal{L}_{(t_2)}}{\partial\textbf x_{t_2}}
+\frac{\partial \mathcal{L}_{(t_3)}}{\partial\textbf x_{t_3}}
+\frac{\partial t_2/\partial s}{\partial t_3/\partial s}\frac{\partial \mathcal{L}_{(t_2)}}{\partial \textbf x_{t_3}}
+\frac{\partial t_3/\partial s}{\partial t_2/\partial s}\frac{\partial \mathcal{L}_{(t_3)}}{\partial \textbf x_{t_2}}\right)\right]\nonumber\\
&&+\frac{1}{2}\delta\textbf y \left.\left[ \frac{\partial \mathcal{L}_{(t_2)}}{\partial\textbf x_{t_2}}
-\frac{\partial \mathcal{L}_{(t_3)}}{\partial\textbf x_{t_3}}
-\frac{\partial t_2/\partial s}{\partial t_3/\partial s}\frac{\partial \mathcal{L}_{(t_2)}}{\partial \textbf x_{t_3}}
+\frac{\partial t_3/\partial s}{\partial t_2/\partial s}\frac{\partial \mathcal{L}_{(t_3)}}{\partial \textbf x_{t_2}}\right]\right)\;.
\end{eqnarray}
The condition
$\delta \mathcal S=0$ is satisfied once
\begin{eqnarray}
&& \frac{\partial \mathcal{L}_{(t_2)}}{\partial\textbf x}\frac{dt_2}{ds}
+\frac{\partial \mathcal{L}_{(t_3)}}{\partial\textbf x}\frac{dt_3}{ds}\nn\\
&&-\frac{1}{2}\frac{d}{ds}\left( \frac{\partial \mathcal{L}_{(t_2)}}{\partial\textbf x_{t_2}}
+\frac{\partial \mathcal{L}_{(t_3)}}{\partial\textbf x_{t_3}}
+\frac{\partial t_2/\partial s}{\partial t_3/\partial s}\frac{\partial \mathcal{L}_{(t_2)}}{\partial \textbf x_{t_3}}
+\frac{\partial t_3/\partial s}{\partial t_2/\partial s}\frac{\partial \mathcal{L}_{(t_3)}}{\partial \textbf x_{t_2}}\right)=0\;,\label{eq:5.32as}\\
&&\frac{\partial \mathcal{L}_{(t_3)}}{\partial \textbf x_{t_2}}\left(\frac{dt_3}{ds}\right)^2
+\left(\frac{\partial \mathcal{L}_{(t_2)}}{\partial\textbf x_{t_2}}
-\frac{\partial \mathcal{L}_{(t_3)}}{\partial\textbf x_{t_3}}\right)\frac{dt_2}{ds}\frac{dt_3}{ds}
-\frac{\partial \mathcal{L}_{(t_2)}}{\partial \textbf x_{t_3}}\left(\frac{dt_2}{ds}\right)^2=0.\label{eq:5.32bs}
\end{eqnarray}
We consider eq. \eqref{eq:5.32as} to be a generalized Euler-Lagrange equation containing two separate Euler-Lagrange equations, and
the equation \eqref{eq:5.32bs} is the constraint resulting from the variational principle.
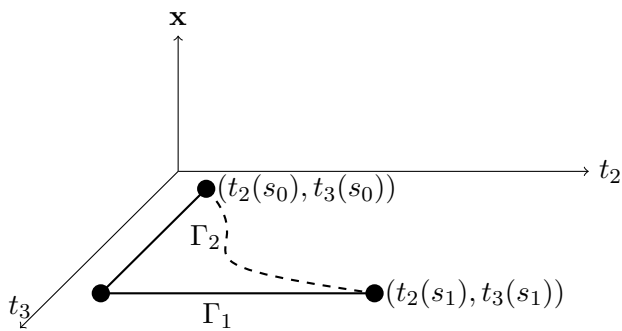
\begin{figure}[h]
\begin{center}
\begin{tikzpicture}[scale=0.6]
 \draw[->] (0,0,0) -- (9,0,0) node[anchor=west] {$t_{2}$};
 \draw[->] (0,0,0) -- (0,3,0) node[anchor=south] {$\textbf x$};
 \draw[->] (0,0,0) -- (0,0,9) node[anchor=south] {$t_{3}$};
 \fill (1,0,1) circle (0.2) node[anchor=west]{$(t_2(s_0),t_3(s_0))$};
 \draw (1,0,2.5) node[anchor=north west] {$\Gamma_2$};
 \draw (3,0,7) node[anchor=north west] {$\Gamma_1$};
 \draw[thick](1,0,1)--(1,0,7)--(7,0,7);
 \fill (1,0,7) circle(0.2);
 \fill (7,0,7) circle (0.2) node[anchor=west]{$(t_2(s_1),t_3(s_1))$} ;
%
\draw [thick,dashed] (1,0,1) .. controls (4,0,5) and (1,0,5) .. (7,0,7);
%
\end{tikzpicture}
\end{center}
\caption{The deformation of the curve $\Gamma$.}\label{deformed-x-curve}
\end{figure}
\\
To summarize, performing the variation on the curve $\Gamma$ with respect to the time variables on the $t_2-t_3$ plane, the computation leads to the closure relation \eqref{eq:5.24ss}.
By fixing the curve $\Gamma$ and varying the curve $\mathcal{E}_{\Gamma}$, the computation leads to the Euler-Lagrange equations corresponding to
the Lagrangian 1-form. However, the closure relation allows the deformation of the curve $\Gamma$ to a simpler curve, as in Fig. (\ref{deformed-x-curve}). If we now are 
on the curve $\Gamma_2$ where
the time variable $t_2$ is a constant leading to $\frac{dt_2}{ds}=0$, we obtain 
\begin{subequations}\label{EL23}
\begin{eqnarray}
&&\frac{\partial{\mathcal{L}_3}}{\partial{\textbf{x}}}
-\frac{\partial}{\partial t_3}\left(\frac{\partial{\mathcal{L}_3}}{\partial \textbf{x}_{t_3}}\right)=0\;,\\\label{eq:5.32a}
&&
\frac{\partial{\mathcal{L}_3}}{\partial {\textbf{x}_{t_2}}}=0.\label{eq:5.32b}
\end{eqnarray}
\end{subequations}
If we are on the curve $\Gamma_1$, where the time variable is a constant leading to $\frac{dt_3}{ds}=0$, we obtain exactly the same equation \eqref{EL23} with interchange of
indices $2\leftrightarrow 3$. This is entirely consistent with the results obtained in the previous section.

The Euler-Lagrange equations \eqref{eq:5.32as}, the constraint equations \eqref{eq:5.32bs} and the closure relation \eqref{eq:5.24ss} together form a consistent set of 
equations, representing the multidimensional consistency of the system.

\section{Concluding remarks}
In this paper, we present what we believe to be the treatment of the Lagrangian structure for the Calogero-Moser system which best
captures the essential integrability characteristics. It is well known that the continuous CM system is Liouville integrable \cite{C3} and
also superintegrable \cite{SW} and this remains the case on the discrete-time level \cite{VUj,F2}. However, in this paper, we focused on another
aspect of the integrability, namely multidimensional consistency, meaning in this context of finite-dimensional systems the 
existence of commuting flows. We believe that in the present paper,
we have given convincing arguments for the assertion that the proper Lagrangian structure should be the one in terms of Lagrange 1-forms. Intriguingly, this is most
manifest on the discrete-time level where it follows directly from the relevant Lax equations. In the continuous case, the relevant Lagrangians (apart from the
obvious Lagrangian for the second order CM flow) are more difficult to establish, but they follow through systematic continuum limits performed on the discrete Lagrangian.
However, the latter is quite a subtle computation, and it is here that the connection between the CM system and the KP system (the former arising from the pole reduction
of the latter) is essential. The construction in this paper starts from the semi-discrete KP equation, and it is the dependence of this equation on (lattice) parameters
which guides the choice of continuum limits. We believe that the CM system forms a first important example for the study of the new variational principle which applies
to the case of Lagrangian 1-forms, cf. \cite{SF1,SF2,SFQ,XNL}, as preparation for similar structures in the case of higher Lagrangian multiforms.
\\
\\
We finish with a few observations.
\begin{itemize}
 \item Since the construction is based on the connection with a semi-discrete KP equation, it would have been more satisfactory if a Lagrangian for
that equation was at our disposal. However, at this stage, such a Lagrangian seems to be elusive, even though it exists for the bilinear discrete
KP case \cite{SFQ}.
 \item Another issue is the connection with the Hamiltonian hierarchy associated with the Calogero-Moser system. Usually, 
the connection between the Hamiltonian and the Lagrangian is given through the Legendre 
transformation, which for the second flow in the CM hierarchy is the standard one. However, when we move to the higher Hamiltonians in the hierarchy, these
are no longer of Newtonian type, and hence lead to complicated expressions when implementing the Legendre transformation. What the treatment in of Section 7 reveals
, however, is that the higher flows fit naturally in the 1-form structure through mixed, but polynomial, Lagrangians in terms of the higher-time derivative.
\item The natural framework for quantization of the discrete-time CM model seems to be that of the path integral, which can be interpreted as a quantum version
of the least-action-principle. One may conjecture that the quantization scheme exploiting the new Lagrange structure could lead to a rigorous approach to
constructing the relevant path integral. 
\end{itemize} 
Our motivation in this paper was to present the most simple example of a Lagrange 1-form structure, hence our restriction to the rational case. However,
most of our results can be readily expanded to the trigonometric and hyperbolic and also elliptic cases. The formulae for those cases are presented in Appendix \ref{ET}.
A natural question is how our results extend to the relativistic case, i.e., the Ruijsenaars-Schneider (RS) model. We intend to address that problem in a future paper.

\appendix
\section{The construction of the exact solution}\label{Solution}
\numberwithin{equation}{section}
In this appendix we review the construction of the exact solution \eqref{eq:2.12}, which follows a procedure 
similar to the continuous case, cf. e.g. \cite{OP}. The basic relations following from the Lax pair, i.e., 
\eqref{eq:2.10} together with the definitions \eqref{eq:2.7} and \eqref{eq:2.8}, lead to: 
\begin{subequations}\label{eq:w1}
\begin{eqnarray}
(\widetilde{\boldsymbol{L}}-\boldsymbol{M})\;\boldsymbol E&=&0\;,\\
\boldsymbol{E}\;(\boldsymbol{L}-\boldsymbol{M})&=&0\;,\\
\widetilde{\boldsymbol{X}}\boldsymbol{M}-\boldsymbol{M}\boldsymbol{X}&=&-\boldsymbol{E}\;,\\
\boldsymbol{X}\boldsymbol{L}-\boldsymbol{L}\boldsymbol{X}&=&\boldsymbol{I}-\boldsymbol{E}\;,
\end{eqnarray}
\end{subequations}
where $\boldsymbol{X}=\boldsymbol{X}(n,m)=\sum_{i=1}^N x_i(n,m) E_{ii}$ is the diagonal matrix of particle positions. 
We concentrate first on the part of the Lax pairs associated with the dynamics in terms of the variable $n$. 
From the Lax equation \eqref{eq:2.10}, we may write
\begin{equation}
 \boldsymbol{M}=\widetilde{\boldsymbol U}\boldsymbol{U}^{-1}\;,\;\;\;\mbox{and}\;\;\;
 \boldsymbol{L}=\boldsymbol{U}\boldsymbol{\Lambda}_{L}\boldsymbol{U}^{-1},
\end{equation}
where $\boldsymbol{U}=\boldsymbol{U}(n,m)$ is the matrix used to diagonalize $\boldsymbol{L}$. Then \eqref{eq:w1} becomes
\begin{subequations}\label{eq:w2}
\begin{eqnarray}
\widetilde{\boldsymbol{U}}^{-1}\boldsymbol{E}&=&\boldsymbol{\Lambda}_{L}^{-1}\boldsymbol{U}^{-1}\boldsymbol{E}\;,\\
\boldsymbol{E}\;\widetilde{\boldsymbol{U}}&=&\boldsymbol{E}\;\boldsymbol{U}\boldsymbol{\Lambda}_{L}\;,\\
\widetilde{\boldsymbol{Y}}-\boldsymbol{Y}&=&-\widetilde{\boldsymbol{U}}^{-1}\boldsymbol{E}\;\boldsymbol{U}\;,\\
\boldsymbol{Y}\boldsymbol{\Lambda}_{L}-\boldsymbol{\Lambda}_{L}\boldsymbol{Y}&=
&\boldsymbol{U}^{-1}(\boldsymbol{I}-\boldsymbol{E})\boldsymbol{U}\;,
\end{eqnarray}
\end{subequations}
where $\boldsymbol{Y}=\boldsymbol{U}^{-1}\boldsymbol{X}\boldsymbol{U}$, which leads to the following expression
\begin{equation}\label{eq:w3}
\boldsymbol{Y}\boldsymbol{\Lambda}_{L}- \boldsymbol{\Lambda}_{L}\widetilde{\boldsymbol{Y}}=\boldsymbol{I}\;.
\end{equation}
Noting the invariance of $\boldsymbol{\Lambda}_{L}$ under the discrete-time shift, we can easily solve \eqref{eq:w3}, and its general
solution is
\begin{equation}\label{eq:w4}
\boldsymbol{Y}(n,m) =\boldsymbol{\Lambda}_{L}^{-n} \boldsymbol{Y}(0,m)\boldsymbol{\Lambda}_{L}^{n}-n\boldsymbol{\Lambda}_{L}^{-1} \; ,
\end{equation}
with $\boldsymbol{Y}(0,m)$ determined from the initial data $\boldsymbol{X}(0,m)$. A similar analysis can be applied to create the solution
associated with the ``\;\;$\wh{}$\;\;'' shift.

Conversely, we can start from a given $N\times N$ diagonal matrix $\bLam$ with distinct entries, and an initial value matrix $\boldsymbol{Y}(0,0)$ 
subject to the condition that 
\begin{equation}
[\boldsymbol{Y}(0,0)\,,\,\boldsymbol{\Lambda}]=\boldsymbol{I} + {\rm rank}\ \ 1 \   , 
\end{equation}
where $[\,,\,]$ denotes the matrix commutator bracket. Let $U^{-1}(0,0)$ be the matrix that diagonalizes $\bsY(0,0)$, i.e., such that 
\begin{equation}\label{eq:Ydiag} 
\bsY(0,0)=\bU^{-1}(0,0)\,\bsX(0,0)\,\bU(0,0))\quad,\quad \bsX(0,0)={\rm diag}(x_1(0,0),\dots, x_N(0,0))\  .  
\end{equation}
If the eigenvalues of $\bsY(0,0)$ are distinct (which we can take as an assumption on the initial condition) then $\bU^{-1}(0,0)$ is 
determined up to multiplication from the right by a diagonal matrix times a permutation matrix of the columns. (Fixing an ordering of the 
eigenvalues $x_i(0,0)$, $\bU^{-1}(0,0)$ unique only up to multiplication by a diagonal matrix from the right).  We can fix $\bU^{-1}(0,0)$ 
up to an overall multiplicative factor by demanding that 
\begin{equation}
[\boldsymbol{Y}(0,0)\,,\,\boldsymbol{\Lambda}]=\boldsymbol{I} -\bU^{-1}(0,0)\,\bE\,\bU(0,0) \   .  
\end{equation}
\\
Next we consider the matrix function given by 
\begin{equation}\label{eq:Ynmdiag} 
\bsY(n,m)=\bLam_L^{-n}\bLam_K^{-m}\bsY(0,0)\bLam_L^n\bLam_K^m -n\bLam_L^{-1}-m\bLam_K^{-1}\  , 
\end{equation}
with $\bLam_L=p\bI+\bLam$ and $\bLam_K=q\bI+\bLam$. Let $\bU^{-1}(n,m)$ be the matrix diagonalizing 
$\bsY(n,m)$; by an appropriate choice of an overall factor (as a function of $n$ and $m$) this matrix can be fixed 
such that it obeys: 
\begin{equation}
\bU^{-1}(n,m)\bE=\bLam_L^{-n}\bLam_K^{-m}\bU^{-1}(0,0)\bE\quad,\quad \bE\,\bU(n,m)=\bE\,\bU(0,0)\bLam_L^n\bLam_K^m\   ,   
\end{equation}
and
\begin{equation}\label{eq:Ycomm}
[\bsY(n,m)\,,\,\bLam]=\bI - U^{-1}(n,m)\bE\,\bU(n,m)\   . 
\end{equation}
From the expression \eqref{eq:Ynmdiag} we can now derive the relations 
\begin{subequations}\label{eq:LamY}\begin{eqnarray}
&& \bLam_L\,\wt{\bsY}-\bsY\,\bLam_L=-\bI\  , \\ 
&& \bLam_K\,\wh{\bsY}-\bsY\,\bLam_K=-\bI\  ,
\end{eqnarray}\end{subequations}
with the usual notation for the shifts in $n$ and $m$ over one unit. Together with the relation \eqref{eq:Ycomm} 
this subsequently yields:
\begin{equation}\label{eq:YY}
\wt{\bsY}-\bsY=-\wt{\bU}^{-1}\bE\,\bU\quad,\quad \wh{\bsY}-\bsY=-\wh{\bU}^{-1}\bE\,\bU\   . 
\end{equation}
Reversing these relations by rewriting them in terms of $\bsX(n,m)=\bU(n,m)\,\bsY(n,m)\,\bU^{-1}(n,m)$ and now 
\textit{defining} the Lax matrices by  
\begin{equation}\label{eq:LKdefs}
\bL:= \bU\,\bLam_L\,\bU^{-1}\quad,\quad \bK:=\bU\,\bLam\,\bU^{-1}\   , 
\end{equation}
together with 
\begin{equation}\label{eq:MNdefs}
\bM:= \wt{\bU}\,\bU^{-1}\quad,\quad \bN:=\wh{\bU}\,\bU^{-1}\   , 
\end{equation}
we recover the relations:
\begin{equation}\label{eq:XMN}
[ \bsX\,,\,\bL]=[\bsX\,,\,\bK]=\bI-\bE\quad,\quad \wt{\bsX}\,\bM-\bM\,\bsX=\wt{\bsX}\,\bN-\bN\,\bsX=-\bE\  , 
\end{equation} 
which determine the matrices $\bM$ and $\bN$ as functions of the $x_i(n,m)$ as well as the off-diagonal 
parts of the matrices $\bL$ and $\bK$. From the definitions of $\bL$ and $\bK$ we have 
that 
\[ \bL-\bK=(p-q)\bI\  . \]
Furthermore, from \eqref{eq:LamY} we obtain
\begin{equation}
\wt{\bL}\,\wt{\bsX}\,\bM-\bM\,\bsX\,\bL=-\bM\quad,\quad  \wh{\bK}\wt{\bsX}\,\bN-\bN\,\bsX\,\bK=-\bN\  , 
\end{equation} 
which, when combined with the latter relations of \eqref{eq:XMN}, yield
\begin{equation}\label{eq:LM}
\left(\wt{\bL}\,\bM-\bM\,\bL\right)(\bsX-\bE)=0 
\quad,\quad  \left(\wt{\bK}\,\bN-\bN\,\bK\right)(\bsX-\bE)=0 \  . 
\end{equation} 
On the other hand, using the first of the relations \eqref{eq:XMN} we also obtain
\begin{equation}\label{eq:XLM}
\wt{\bsX}\left(\wt{\bL}\,\bM-\bM\,\bL\right)=\bE\,(\bL-\bM) 
\quad {\rm and}\quad  \wh{\bsX}\left(\wt{\bK}\,\bN-\bN\,\bK\right)=\bE\,(\bK-\bN)\  ,  
\end{equation} 
as well as
\begin{equation}\label{eq:LMX}
\left(\wt{\bL}\,\bM-\bM\,\bL\right)\bsX=(\wt{\bL}-\bM)\,\bE 
\quad {\rm and}\quad  \left(\wh{\bK}\,\bN-\bN\,\bK\right)\bsX=(\wh{\bK}-\bN)\,\bE \  .   
\end{equation} 
From the relations \eqref{eq:LM}, \eqref{eq:XLM} and \eqref{eq:LMX} it follows that the Lax equations hold 
and their form is determined up to the diagonal part of the matrices $\bL$ and $\bK$.

\section{The elliptic and trigonometric discrete-time Calogero-Moser}\label{ET}
\numberwithin{equation}{section}
In this section, we show that there is a connection between the time-part Lax matrix and the Lagrangian 
for the trigonometric/hyperbolic and elliptic cases of the discrete-time CM system, similar to that which we have established for the rational 
case. 
\subsection{The elliptic case}
The Lax matrices in this case read \cite{FP} 
\begin{equation}\label{eq:a.1}
\boldsymbol{L}_{ell}=\sum_{i=1}^N \left(\sum_{j=1}^N \zeta(q_i-\widetilde {q}_j)-\sum_{j \neq i}^N\zeta(q_i-q_j)
\right)E_{ii} -\sum_{i \neq j}^N \Phi_{\kappa}(q_i-q_j)E_{ij}\;,
\end{equation}
and
\begin{equation}\label{eq:a.2}
\boldsymbol{M}_{ell}=\sum_{i,j=1}^N \Phi_{\kappa}(\wt q_i-q_j)E_{ij}\;,
\end{equation}
where $\zeta(x)$ is the Weierstrass zeta function and $\Phi_{\kappa}(x)=\frac{\sigma(x+\kappa)}{\sigma(\kappa)\sigma(x)}$, where $\sigma(x)$ is the Weierstrass sigma
function. The Lax equation $\widetilde{\boldsymbol L}\boldsymbol M=\boldsymbol M\boldsymbol L$ produces the equation of motion
\begin{equation}\label{eq:a.3}
\sum_{j=1}^N\left( \zeta(q_i-\widetilde{q}_j)+\zeta(q_i-\undertilde{q_j})\right)-2\sum_{j\neq i}^N\zeta(q_i-q_j)=0\;,
\end{equation}
and the corresponding Lagrangian is
\begin{equation}\label{eq:a.4}
\mathcal{L}_{ell}=-\sum_{i,j=1}^N\log\left| \sigma(q_i-\widetilde q_j)\right|+\sum_{i \neq j}^N\log\left| \sigma(q_i-q_j)\right|\;.
\end{equation}
The determinant of the matrix $\boldsymbol M_{elliptic}$ is
\begin{equation}\label{eq:a.5}
\det\left| \boldsymbol M_{ell}\right|=\Phi_{\kappa}(\Sigma)\sigma(\Sigma)\frac{\prod_{i<j}\sigma(q_i-q_j)
\sigma(\widetilde q_i-\widetilde q_j)}{\prod_{ij}\sigma(\widetilde q_i-q_j)}\;,
\end{equation}
where $\Sigma=\sum_{i=1}(q_i-\widetilde q_i)$, and then we can also write
\begin{eqnarray}\label{eq:a.6}
\log\left|\det\left( \boldsymbol M_{ell}\right)\right|&=&\log\left|\Phi_{\kappa}(\Sigma)\sigma(\Sigma)\right|+\sum_{i<j}^N(\log\left|\sigma(q_i-q_j)\right|+\log\left|\sigma(\widetilde q_i-\widetilde q_j)\right|)\nonumber\\
&-&\sum_{ij}^N\log\left|\sigma(\widetilde q_i-q_j)\right|\;.
\end{eqnarray}
The action of the system can be expressed by considering the chain product of the matrix $\boldsymbol M_{ell}$
\begin{eqnarray}\label{eq:a.7}
\mathcal{S}_{ell}&=&\log\left|\det\left(\prod_{n=+\infty}^{\curvearrowleft}\boldsymbol {M}_{ell}(n)\right)\right|,\nonumber\\
&=&\sum_n\left(\log\left|\Phi_{\kappa}(\Sigma)\sigma(\Sigma)\right|+\sum_{i\neq j}^N\log\left|\sigma(q_i-q_j)\right|\right.
-\left.\sum_{ij}^N\log\left|\sigma(\widetilde q_i-q_j)\right|\right).\nonumber\\
&=&\sum_n\left(\mathcal{L}_{ell}+ \log\left|\Phi_{\kappa}(\Sigma)\sigma(\Sigma)\right|\right)\;.
\end{eqnarray}
The last term in eq. (\ref{eq:a.7}) is related to the motion of the centre of mass of the system and
it can be separated from the relative motion.
\subsection{The trigonometric case}
The Lax matrices in this case read \cite{FP} 
\begin{equation}\label{eq:a.8}
\boldsymbol{L}_{trig}=\sum_{i=1}^N \left(\sum_{j=1}^N \coth(q_i-\widetilde {q}_j)-\sum_{j \neq i}^N\coth(q_i-q_j)\right)E_{ii} -\sum_{i \neq j}^N 
\frac{\sinh(q_i-q_j+\kappa)E_{ij}}{\sinh(q_i-q_j)\sinh(\kappa)}\;,
\end{equation}
and
\begin{equation}\label{eq:a.9}
\boldsymbol{M}_{trig}=\sum_{i,j=1}^N \frac{\sinh(\widetilde q_i-q_j+\kappa)E_{ij}}{\sinh(\widetilde q_i-q_j)\sinh(\kappa)}\;.
\end{equation}
The Lax equation $\widetilde{\boldsymbol L}\boldsymbol M=\boldsymbol M\boldsymbol L$ produces the equation of motion
\begin{equation}\label{eq:a.10}
\sum_{j=1}^N\left( \coth(q_i-\widetilde{q}_j)+\coth(q_i-\undertilde{q_j})\right)-2\sum_{j\neq i}^N\coth(q_i-q_j)=0\;,
\end{equation}
and the corresponding Lagrangian is
\begin{equation}\label{eq:a.11}
\mathcal{L}_{trig}=-\sum_{i,j=1}^N\log\left| \sinh(q_i-\widetilde q_j)\right|+\sum_{i \neq j}^N\log\left| \sinh(q_i-q_j)\right|\;.
\end{equation}
The determinant of the matrix $\boldsymbol M_{trig}$ is
\begin{equation}\label{eq:a.12}
\det\left| \boldsymbol M_{trig}\right|=\frac{\sinh(\Sigma+\kappa)}{\sinh(\kappa)}\frac{\prod_{i<j}\sinh(q_i-q_j)
\sinh(\widetilde q_i-\widetilde q_j)}{\prod_{ij}\sinh(\widetilde q_i-q_j)}\;,
\end{equation}
and then we can also write
\begin{eqnarray}\label{eq:a.13}
\log\left|\det\left( \boldsymbol M_{trig}\right)\right|&=&\log\left|\frac{\sinh(\Sigma+\kappa)}{\sinh(\kappa)}\right|+
\sum_{i<j}^N(\log\left|\sinh(q_i-q_j)\right|+\log\left|\sinh(\widetilde q_i-\widetilde q_j)\right|)\nonumber\\
&-&\sum_{ij}^N\log\left|\sinh(\widetilde q_i-q_j)\right|\;.
\end{eqnarray}
The action of the system can be expressed by considering the chain product of the matrix $\boldsymbol M_{trig}$
\begin{eqnarray}\label{eq:a.14}
\mathcal{S}_{trig}&=&\log\left|\det\left(\prod_{n=+\infty}^{\curvearrowleft}\boldsymbol {M}_{trig}(n)\right)\right|,\nonumber\\
&=&\sum_n\left(\log\left|\frac{\sinh(\Sigma+\kappa)}{\sinh(\kappa)}\right|+\sum_{i\neq j}^N\log\left|\sin(q_i-q_j)\right|\right.
-\left.\sum_{ij}^N\log\left|\sigma(\widetilde q_i-q_j)\right|\right).\nonumber\\
&=&\sum_n\left(\mathcal{L}_{trig}+ \frac{\sinh(\Sigma+\kappa)}{\sinh(\kappa)}\right)\;.
\end{eqnarray}
Again the last term in eq. (\ref{eq:a.14}) is related to the center of mass motion of the system, and it is not
 going to effect the equations of the relative motion.
\section{The derivation of the Lagrangian $\mathcal{L}_{(\tau)}$}\label{L}
In this section, we show the derivation of the Lagrangian $\mathcal{L}_{(\tau)}$ \eqref{eq:4.26aaa} from the skew limit of the Lagrangian $\mathcal{L}_{(m)}$ \eqref{eq:3.10qb}. We recall the 
Lagrangian $\mathcal{L}_{(m)}$ again here
\begin{equation}\label{L1}
 \mathcal{L}_{(m)}=\log|\det(\boldsymbol N)|+q(\Xi-\wh{\Xi})\;.
\end{equation}
Performing the skew limit, we obtain
\begin{eqnarray}\label{L2}
 \mathcal{L}_{(m)}&\Rightarrow& \log|\det(\boldsymbol {\mathsf M}-\varepsilon\wt{\boldsymbol {\mathsf A}})|+(p-\varepsilon)(\Xi-\wt{\Xi}-\varepsilon\dot{\wt{\Xi}})\nn\\
&\Rightarrow&\log|\det(\boldsymbol{\mathsf M})|+\log|\det(\boldsymbol{\mathsf I}-\varepsilon\boldsymbol{\mathsf M}^{-1}\boldsymbol{\mathsf A})|+...+p(\Xi-\wt{\Xi})
-\varepsilon(\Xi-\wt{\Xi})-p\varepsilon\dot{\wt{\Xi}}\nn\\
&\Rightarrow&\mathcal{L}_{(\mathsf N)}+\log|1-\varepsilon\mathsf{Tr}\left(\boldsymbol {\mathsf M}^{-1}\wt{\boldsymbol {\mathsf A}}\right)+...|-
\varepsilon(\Xi-\wt{\Xi})-p\varepsilon\dot{\wt{\Xi}}\nn\\
&\Rightarrow&\mathcal{L}_{(\mathsf N)}+\varepsilon\mathcal{L}_{(\tau)}+....
\end{eqnarray}
where
\begin{equation}\label{L3}
\mathcal{L}_{(\tau)} =-\mathsf{Tr}\left(\boldsymbol {\mathsf M}^{-1}\wt{\boldsymbol {\mathsf A}}\right)-\Xi+\wt{\Xi}-p\dot{\wt{\Xi}}\;.
\end{equation}
The inverse of the matrix $\boldsymbol{\mathsf M}$ is given by
\begin{equation}\label{L4}
\boldsymbol{\mathsf M}^{-1}=-\sum_{ij=1}^N\frac{\wt{\Psi}(\mathsf x_i)\Psi(\wt{\mathsf x}_j)}{\mathsf x_i-\wt{\mathsf x}_j}E_{ij}\;,
\end{equation}
where
\begin{subequations}\label{L5}
\begin{eqnarray}
\wt{\Psi}(\mathsf x_i)&=&\frac{\prod_{l=1}^N(\mathsf x_i-\wt{\mathsf x_l})}{\prod_{l \ne i}^N(\mathsf x_i-\mathsf x_l)}\;,\\
\Psi(\wt{\mathsf x}_j)&=&\frac{\prod_{l=1}^N(\wt{\mathsf x}_j-\mathsf x_l)}{\prod_{l \ne i}^N(\wt{\mathsf x}_i-\wt{\mathsf x}_l)}\;.
\end{eqnarray}
\end{subequations}
We now can show that
\begin{eqnarray}\label{L6}
\mathsf{Tr}\left(\boldsymbol{\mathsf M}^{-1}\wt{\boldsymbol{\mathsf A}}\right)
=\sum_{i,j=1}^N\frac{\wt{\Psi}(\mathsf x_i)\Psi(\wt{\mathsf x}_j)\dot{\wt{\mathsf x}}_j}{(\wt{\mathsf x}_j-\mathsf x_i)^3}\;.
\end{eqnarray}
We now consider the identity
\begin{eqnarray}\label{L7}
\prod_{l=1}\frac{\xi-\wt{\mathsf x}_l}{\xi-\mathsf x_l}=1+\sum_{l=1}^N\frac{\wt{\Psi}(\mathsf x_l)}{\xi-\mathsf x_l}\;,
\end{eqnarray}
which we differentiate with respect to $\xi$ at a given value $\xi=\wt{\mathsf x}_j$, yielding 
\begin{eqnarray}\label{L8}
-\frac{1}{\Psi(\wt{\mathsf x}_j)}=\sum_{l=1}^N\frac{\wt{\Psi}(\mathsf x_l)}{(\wt{\mathsf x}_j-\mathsf x_l)^2}\;.
\end{eqnarray}
We now differentiate \eqref{L8} with respect to $\wt{\mathsf x}_j$, yielding
\begin{eqnarray}\label{L9}
\frac{\partial}{\partial \wt{\mathsf x}_j}\frac{1}{\Psi(\wt{\mathsf x}_j)}=\sum_{l=1}^N\frac{\wt{\Psi}(\mathsf x_l)}{(\wt{\mathsf x}_j-\mathsf x_l)^3}\;.
\end{eqnarray}
Using \eqref{L9}, we can rewrite \eqref{L6} in the form
\begin{eqnarray}\label{L10}
\mathsf{Tr}\left(\boldsymbol{\mathsf M}^{-1}\wt{\boldsymbol{\mathsf A}}\right)
&=&\sum_{j=1}^N\dot{\wt{\mathsf x}}_j\Psi(\wt{\mathsf x}_j)\frac{\partial}{\partial \wt{\mathsf x}_j}\frac{1}{\Psi(\wt{\mathsf x}_j)}\;,\nn\\
&=&-\sum_{j=1}^N\dot{\wt{\mathsf x}}_j\frac{\partial}{\partial \wt{\mathsf x}_j}\ln|\Psi(\wt{\mathsf x}_j)|\;,
\end{eqnarray}
then the Lagrangian $\mathcal{L}_{(\tau)}$ can be expressed in the form
\begin{eqnarray}\label{L11}
\mathcal{L}_{(\tau)}=\sum_{i \ne j}^N\frac{\dot{\wt{\mathsf x}}_j}{\wt{\mathsf x}_i-\wt{\mathsf x}_j}+\sum_{i,j=1}^N
\frac{\dot{\wt{\mathsf x}}_j}{\mathsf x_i-\wt{\mathsf x}_j}-\Xi+\wt{\Xi}-p\dot{\wt{\Xi}}\;.
\end{eqnarray}

\section{The direct proof of the closure relation \eqref{eq:5.15}}\label{proof}
\numberwithin{equation}{section}
In this section, we show that the closure relation \eqref{eq:5.15} holds on the equations of motion. Here we recall the equations of motion 
for the time variables $t_2$ and $t_3$
\begin{eqnarray}
\frac{\partial^2X_i}{\partial t_2^2}&=&-\sum\limits_{\mathop {j = 1}\limits_{j \ne i} }^N\frac{8}{(X_i-X_j)^3},\label{eq:B1}\\
\frac{\partial^2X_i}{\partial t_2 \partial t_3}&=&6\sum\limits_{\mathop {j = 1}\limits_{j \ne i} }^N\frac{\frac{\partial X_i}{\partial t_2}+
\frac{\partial X_j}{\partial t_2}}{(X_i-X_j)^3}.\label{eq:B2}
\end{eqnarray}
The Lagrangians corresponding to \eqref{eq:B1} and \eqref{eq:B2} take the form
\begin{eqnarray}
\mathcal L_{(t_2)}&=&\sum\limits_{i=1}^N\frac{1}{2}\left( \frac{\partial X_i}{\partial t_2}\right)^2+\sum\limits_{i \ne j}^N\frac{2}{(X_i-X_j)^2}, \label{eq:B3}\\
\mathcal L_{(t_3)}&=&\sum\limits_{i=1}^N\left( \frac{\partial X_i}{\partial t_2}\frac{\partial X_i}{\partial t_3}+\frac{1}{4}\left( \frac{\partial X_i}
{\partial t_2}\right)^3\right)-\sum\limits_{i \ne j}^N\frac{\frac{\partial X_j}{\partial t_2}+2\frac{\partial X_i}{\partial t_2}}{(X_i-X_j)^2}\;,\label{eq:B4}
\end{eqnarray}
respectively.
\\
\\
We find that
\begin{eqnarray}\label{eq:B5}
\frac{\partial {\mathcal L_{(t_2)}}}{\partial t_3}=\sum\limits_{i \ne j}^N\frac{6(\frac{\partial{X}_i}{\partial t_2})^2-
8\frac{\partial{X}_i}{\partial t_3}}{(X_i-X_j)^3},
\end{eqnarray}
and
\begin{eqnarray}\label{eq:B6}
\frac{\partial {\mathcal L_{(t_3)}}}{\partial t_2}=\sum\limits_{i \ne j}^N\frac{6(\frac{\partial{X}_i}{\partial t_2})^2-8\frac{\partial{X}_i}{\partial t_3}}{(X_i-X_j)^3}
-3\sum\limits_{i \ne j}^N\frac{\frac{\partial^2 X_i}{\partial t_2}}{(X_i-X_j)^2}.
\end{eqnarray}
We see that the first terms of eq. \eqref{eq:B5} and eq. \eqref{eq:B6} are identical. The remaining work is to show that the last term in eq. \eqref{eq:B6} is zero.
Using eq. \eqref{eq:B1}, we may rewrite the last term in eq. \eqref{eq:B6} with the help of the identities:
\begin{eqnarray}\label{eq:B7}
\frac{1}{8}\sum\limits_{i \ne j}^N\frac{\frac{\partial^2 X_i}{\partial t_2}}{(X_i-X_j)^2}&=&\sum\limits_{i \ne j}^N\sum\limits_{\mathop {k = 1}\limits_{k \ne i} }^N
\frac{1}{(X_i-X_j)^2(X_i-X_k)^3}\nonumber\\
&=&\sum\limits_{i \ne j}^N\frac{1}{(X_i-X_j)^5}+\sum\limits_{i \ne j \ne k}^N\frac{1}{(X_i-X_j)^2(X_i-X_k)^3}.
\end{eqnarray}
We see that the first term is an antisymmetric function and hence vanishes, while the second term
\begin{eqnarray}\label{eq:B8}
&&\sum\limits_{i \ne j \ne k}^N\frac{1}{(X_i-X_j)^2(X_i-X_k)^3}\nn\\
&&=\sum\limits_{i \ne j \ne k}^N\left( \frac{1}{X_i-X_j}-\frac{1}{X_i-X_k}\right)^2
\frac{1}{(X_j-X_k)^2}\frac{1}{X_i-X_k}\nn\\
&&=\sum\limits_{i \ne j \ne k}^N\left(\frac{1}{(X_i-X_j)^2(X_j-X_k)^2(X_i-X_k)}\right.\nn\\
&&+\left.\frac{1}{(X_j-X_k)^2(X_i-X_k)^3}-
\frac{2}{(X_i-X_j)(X_i-X_k)^2)(X_j-X_k)^2} \right)\;.
\end{eqnarray}
The first and third terms are the antisymmetric functions, hence vanish leaving the middle term which is actually the opposite of the term on the left hand side, i.e.,
\begin{eqnarray}\label{eq:B9}
\sum\limits_{i \ne j \ne k}^N\frac{1}{(X_i-X_j)^2(X_i-X_k)^3}=-\sum_{i \ne j \ne k}^N\frac{1}{(X_k-X_j)^2(X_k-X_i)^3}=0\;.
\end{eqnarray}
%
%
%

\begin{acknowledgements}

S. Yoo-Kong was supported by the Royal Thai Government and King Mongkut's University of Technology Thonburi. 
S.B. Lobb was supported by the UK Engineering and Physical Sciences Research Council (EPSRC).
\end{acknowledgements}


\end{document}